\documentclass[sigconf]{acmart}
\renewcommand\footnotetextcopyrightpermission[1]{} 
\def\fullversion{1}

\usepackage{enumitem}
\usepackage[algo2e,ruled,lined,linesnumbered,noend]{algorithm2e}
\usepackage{cleveref}
\usepackage{soul}
\theoremstyle{definition}
\newtheorem{definition}{Definition}[section]

\ifx\nocomment\undefined
\newcommand{\guy}[1]{{\color{brown}{\bf Guy:} #1}}
\newcommand{\yan}[1]{{\color{violet}{\bf Yan:} #1}}
\newcommand{\yihan}[1]{{\color{blue}{\bf Yihan:} #1}}
\newcommand{\laxman}[1]{{\color{purple}{\bf Laxman:} #1}}
\newcommand{\magdalen}[1]{{\color{magenta}{\bf Magdalen:} #1}}
\newcommand{\zheqi}[1]{{\color{cyan}{\bf Zheqi:} #1}}
\else
\newcommand{\guy}[1]{{}}
\newcommand{\yan}[1]{{}}
\newcommand{\yihan}[1]{{}}
\newcommand{\laxman}[1]{{}}
\newcommand{\magdalen}[1]{{}}
\newcommand{\zheqi}[1]{{}}
\fi

\newcommand{\hide}[1]{}

\newcommand{\mV}{\mathcal{V}}
\newcommand{\mP}{\mathcal{P}}
\newcommand{\mK}{\mathcal{K}}
\newcommand{\mQ}{\mathcal{Q}}
\newcommand{\la}{\leftarrow}

\providecommand{\norm}[1]{\lVert#1\rVert}

\newcommand{\defn}[1]{{\emph{\textbf{#1}}}}

\SetCommentSty{mycommfont}

\makeatletter

\usepackage{cleveref}
\crefname{section}{Sec.}{Sec.}
\crefname{theorem}{Thm.}{Thm.}
\crefname{lemma}{Lem.}{Lem.}
\crefname{corollary}{Col.}{Col.}
\crefname{table}{Tab.}{Tab.}
\crefname{algorithm}{Alg.}{Alg.}
\crefname{figure}{Fig.}{Fig.}
\crefname{fact}{Fact}{Fact}
\Crefname{table}{Tab.}{Tab.}
\crefname{problem}{Problem}{Problem}

\acmConference[]{}{}{}
\acmYear{}
\acmISBN{} 
\acmDOI{} 
\startPage{1}

\setcopyright{none}

\usepackage{booktabs}   
\usepackage{subcaption} 

\begin{document}

\title{Range Retrieval with Graph-Based Indices}         



\author{Magdalen Dobson Manohar}
\affiliation{
	\institution{Carnegie Mellon University and Microsoft Azure}           
	\country{USA}          
}
\email{mmanohar@microsoft.com}         

\author{Taekseung Kim}
\affiliation{
	\institution{Carnegie Mellon University}    
	\country{USA}       
}
\email{taekseuk@andrew.cmu.edu}         

\author{Guy E. Blelloch}
\affiliation{
	\institution{Carnegie Mellon University and Google}  
	\country{USA}     
}
\email{guyb@cs.cmu.edu}

\begin{abstract}
Retrieving points based on proximity in a high-dimensional vector space is a crucial step in information retrieval applications. The approximate nearest neighbor search (ANNS) problem, which identifies the $k$ nearest neighbors for a query, has been extensively studied in recent years. However, comparatively little attention has been paid to the related problem of finding all points within a given distance of a query, the \textit{range retrieval} problem, despite its applications in areas such as duplicate detection, plagiarism checking, and facial recognition. In this paper, we present new techniques for range retrieval on graph-based vector indices, which are known to achieve excellent performance on ANNS queries. Since a range query may have anywhere from no matching results to thousands of matching results in the database, we introduce a set of range retrieval algorithms based on modifications of the standard graph search that adapt to terminate quickly on queries in the former group, and to put more resources into finding results for the latter group. Due to the lack of existing benchmarks for range retrieval, we also undertake a comprehensive study of range characteristics of existing embedding datasets, and select a suitable range retrieval radius for eight existing datasets with up to 1 billion points in addition to one existing benchmark. We test our algorithms on these datasets, and find up to 100x improvement in query throughput over a standard graph search and the FAISS-IVF range search algorithm. We also find up to 10x improvement over a previously suggested modification of the standard beam search, and strong performance up to 1 billion data points.
\end{abstract}



\maketitle

\section{Introduction}\label{sec:intro}

Similarity search within databases of high-dimensional vectors has become increasingly important over the last decade due to the rise of semantic embeddings generated by neural networks and large language models (LLMs). Similarity search is a foundational building block of applications such as search, recommendations, advertising, and retrieval augmented generation (RAG). Correspondingly, there has been an explosion of work on efficient algorithms for approximate similarity search~\cite{jegou2010product,malkov2018efficient,subramanya2019diskann}, as exact similarity search is prohibitively expensive due to the curse of dimensionality.

The object of approximate similarity search, or approximate nearest neighbor search (ANNS) is finding the top-$k$ most similar embeddings. These top-$k$ results can then be post-processed for the desired application. However, there are some applications for which retrieving top-$k$ embeddings for a fixed $k$ is a poor fit. Applications such as duplicate detection, plagiarism checking, and facial recognition require instead to retrieve all results within a certain \textit{radius} of a query, with some post-processing after to verify whether a match exists out of the retrieved items~\cite{douze2021image,schroff2015facenet,simsearchnet}. Furthermore, in real-world search applications some queries may have tens of thousands of matches while others have none or very few. In these cases, search with a small radius, or range retrieval, may be used to both differentiate these query types~\cite{szilvasy2024vector}. Range search can also be a useful subroutine in applications using nearest neighbor graphs, such as clustering or graph learning~\cite{Grale20,li2020density}.

Range search differs from top-$k$ search by the diversity in size of the ground truth solutions for a set of queries. In practice, ground truth for a set of queries follows a skewed, Pareto-like distribution where the majority of queries tend to have no results within the chosen radius, while a smaller fraction have a small number of results, and a few outliers have thousands to tens of thousands of results. Queries in the middle category are well served by existing similarity search algorithms, which are already highly optimized and efficient. Thus, a good range search algorithm will match the efficiency of top-$k$ search on the middle category of queries, while quickly terminating on queries with no results, and efficiently finding results for those few queries with thousands of results.

Despite the many applications of range search, there is very little work on designing algorithms specifically for the task of high-dimensional range search. From a practical perspective, it would be ideal if indices for top-$k$ search could be reused or lightly adapted for range search, enabling an existing index to serve both types of queries. Indices for top-$k$ search typically fall into one of two categories---partition-based indices, which partition the data set into cells and exhaustively search a small number of cells at query time~\cite{jegou2010product,indyk1998towards}, and graph-based indices, which construct a proximity graph over the data points and use a variant of greedy search to answer queries~\cite{fu2019nsg,subramanya2019diskann,malkov2018efficient,wang2024starling}. Graph-based indices are widely acknowledged as achieving equal or better similarity search performance than partition-based indices~\cite{wang2021comprehensive}.  However, there is almost no work has studied the question of adapting graph-based indices for range retrieval. Furthermore, top-$k$ search with IVF indices naturally explores thousands of candidates per query by exhaustively checking the distances between the query point and all the points in each cell, while the number of nodes a graph-based search typically explores is only a small multiple of $k$. It is thus a much less trivial algorithmic task to effectively adapt graph-based search to serve such queries (as we will show in Section~\ref{sec:algorithms}, naive adaptations of existing search methods are not adequate for the task). The graph structure also suggests potential to terminate queries with no results after examining just a handful of vertices. It is natural then to ask whether graph-based indices can be adapted to efficiently serve range queries. 

\paragraph{Our Contributions} In this paper, we present a set of algorithms designed for approximate high-dimensional range search on graph-based ANNS indices. Our techniques are modifications of the standard graph search algorithm, meaning that we enable one data structure to efficiently serve both top-$k$ queries and range queries. We devise an early-stopping heuristic that is capable of predicting whether a query has no results after only a few hops in the graph search. For queries with many results, we design an algorithm that extends the search path to return more results while minimizing wasted work, and compare and contrast situations where each algorithm dominates the other.

For our experimental evaluation, since there is only one publicly released dataset specifically for range search, we evaluate eight state-of-the-art publicly available metric embedding datasets with up to 1 billion points and select a suitable radius for each one. We also perform additional analysis of the characteristics of different range search datasets.

\begin{figure*}
	\includegraphics[scale=.35]{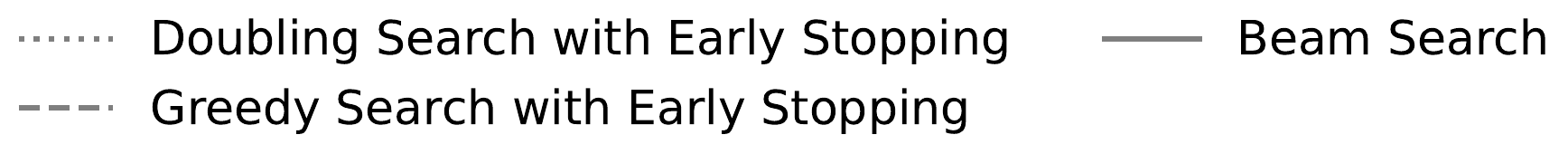} \\
	\includegraphics[scale=.35]{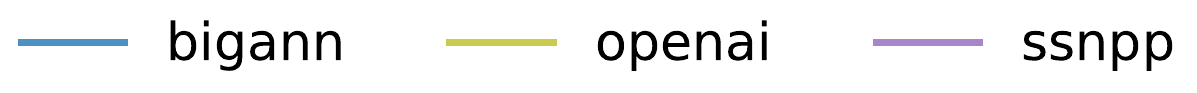} 
	\includegraphics[scale=.35]{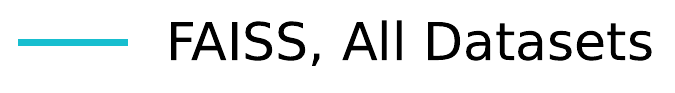}\\
	\begin{subfigure}[t]{.32\textwidth}
			\includegraphics[scale=.32]{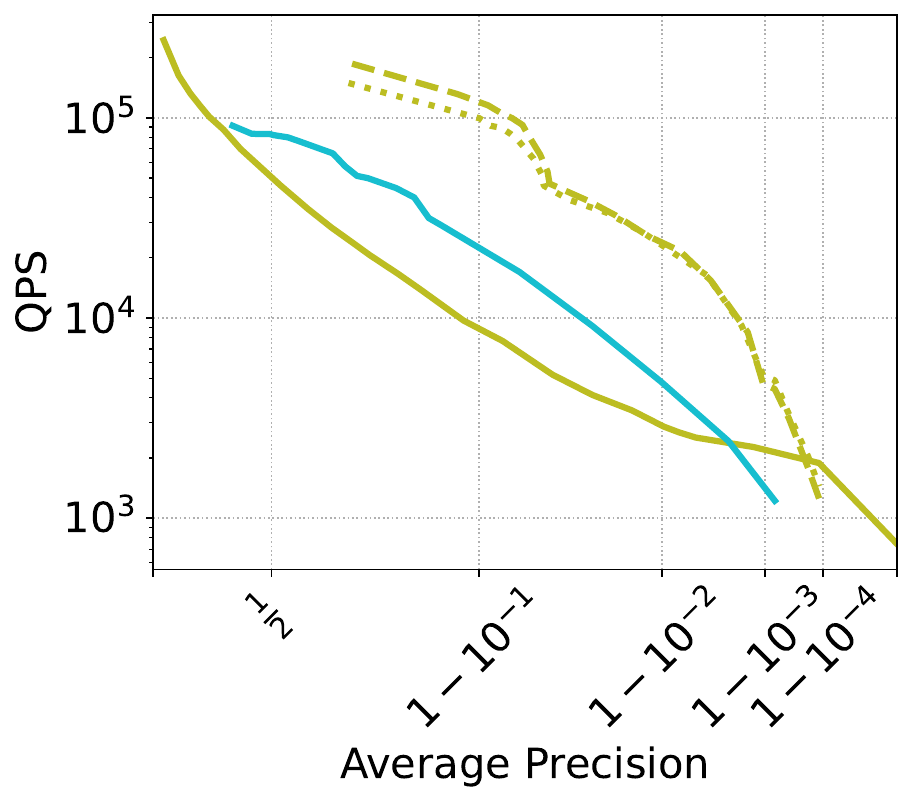}
			\caption{OpenAI-1M}
	\end{subfigure}\hfil
		\begin{subfigure}[t]{.32\textwidth}
			\includegraphics[scale=.32]{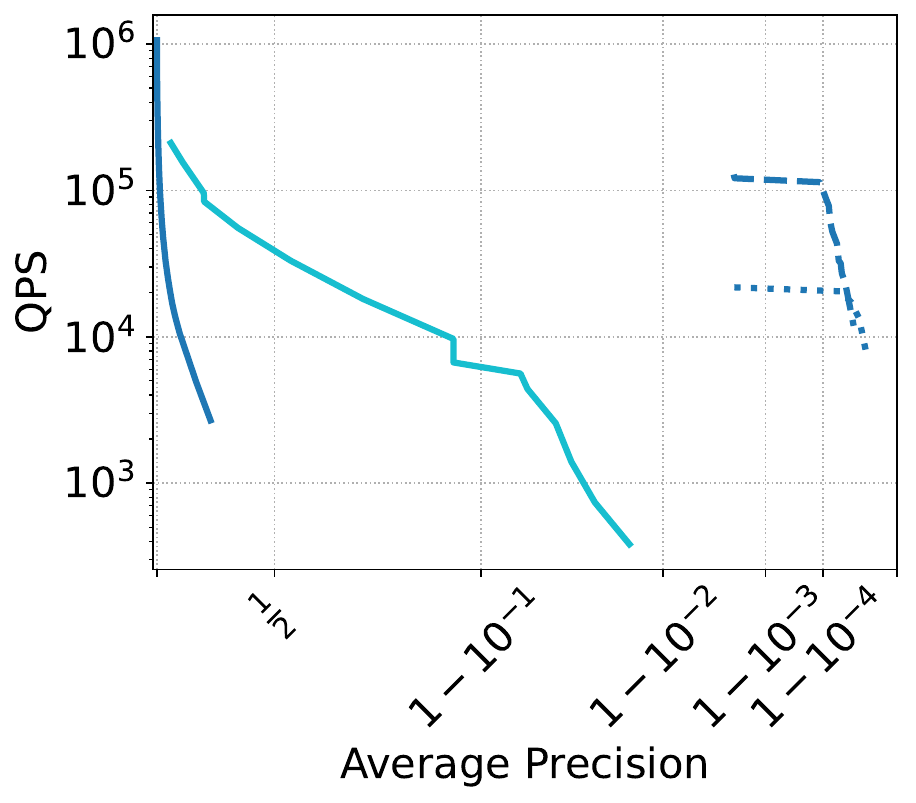}
		\caption{BIGANN-100M}
	\end{subfigure}\hfil
		\begin{subfigure}[t]{.32\textwidth}
			\includegraphics[scale=.32]{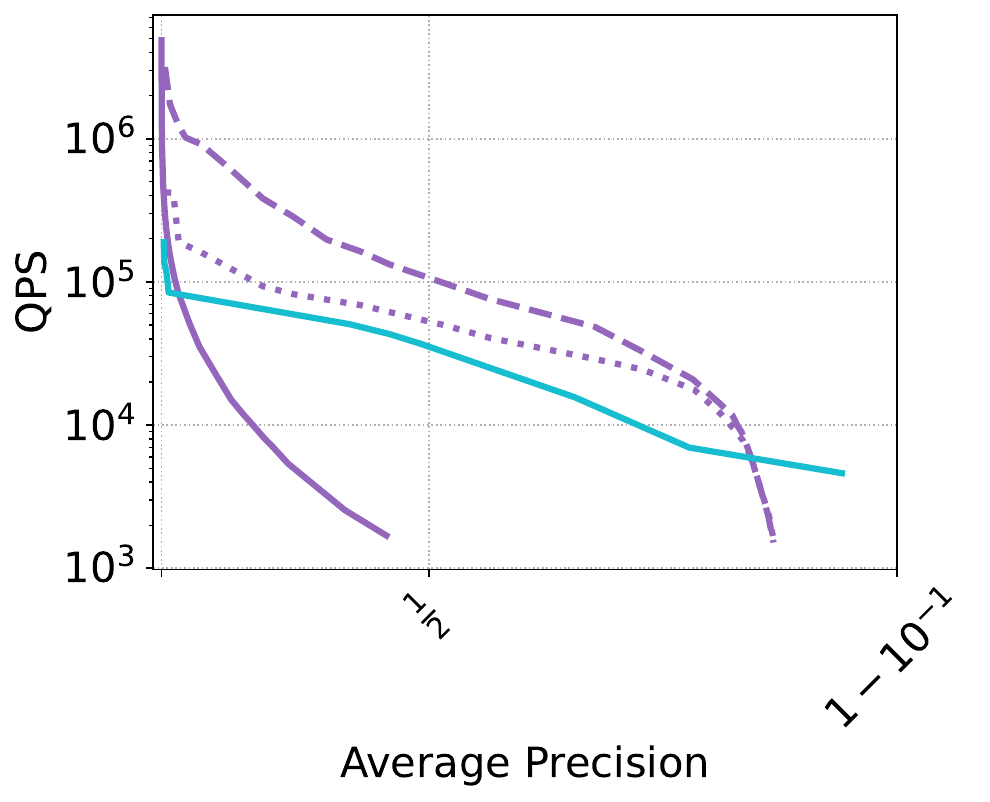}
		\caption{SSNPP-1B}
	\end{subfigure}

	\caption{A preview of our experimental results on three datasets. The solid line shows the beam search baseline, while the dotted and dashed lines show our two new algorithms. Datasets and algorithms are described in detail in Sections~\ref{sec:rangedata} and~\ref{sec:algorithms}, respectively.}
	\label{fig:teaser}
\end{figure*}

We evaluate our algorithms on the set of nine datasets (eight contributed by us, and one public benchmark) against both a simple adaptation of top-$k$ search, an in-memory variant of the range search algorithm used in the Starling system~\cite{wang2024starling} which we refer to as \textit{doubling search}, and the range search algorithm from FAISS-IVF. We find that our range retrieval algorithms are capable of up to 100x speedup over the top-$k$ and FAISS-IVF baselines, and up to 10x speedup over the doubling search. We additionally find that the speedup and scalability of our algorithms extends up to datasets with 1 billion points. Figure~\ref{fig:teaser} shows a preview of our experimental results on three embedding datasets of sizes one million to 1 billion points.

\paragraph{Outline} Sections~\ref{sec:relatedwork} and~\ref{sec:prelim} cover related work and preliminaries, respectively. In Section~\ref{sec:rangedata} we explore the characteristics of range search datasets and provide heuristics for choosing an acceptable radius, which we use to adapt eight public embedding datasets for range searching. In Section~\ref{sec:algorithms} we present our algorithms for range search on graph-based indices, and in Section~\ref{sec:experiments} we present experimental results.

\subsection{Related Work}~\label{sec:relatedwork}

\paragraph{Work on Range Searching} Some earlier work addresses range retrieval in high dimensions from a theoretical standpoint~\cite{chazelle2008approximate}, or with only minimal experiments that do not extend to the large embeddings used today~\cite{wang2013pltree}. Theory results on nearest neighbor search using locality sensitive hashing (LSH) also apply to range search, since their approximation guarantees are of the form of finding neighbors within a $(1+\epsilon)$ radius of the distance to the true top-$k$ result~\cite{indyk1998towards}.

Meta AI released a range retrieval dataset aimed towards detecting misinformation~\cite{simsearchnet}, which was used as one of the competition datasets in the NeurIPS 2021 Big ANN Benchmarks Challenge~\cite{simhadri2021results}, but competitors solved this task using naive adaptations of top-$k$ search rather than novel range searching algorithms. A recent work by Szilvasy, Mazare, and Douze~\cite{szilvasy2024vector} addresses the combined task of retrieving range search results with one metric and then re-ranking them using a more sophisticated model. They design a range search metric specifically for \textit{bulk} search---that is, range search over a group of queries with a fixed search budget---and benchmark its results on an image search application. They use an inverted file (IVF) index, a common data structure for similarity search~\cite{douze2024faiss}, to serve their database queries. The well-known FAISS-IVF library includes an option for range search on an inverted file index~\cite{douze2024faiss}.

The Starling system~\cite{wang2024starling} is an SSD-resident graph-based ANNS index optimized for I/O efficiency. Their work includes benchmarks on range search metrics, but the focus of their paper is not on comprehensive benchmarking of range search, and since their implementation is a disk-resident index and ours is fully in-memory, our experiments are not directly comparable to theirs. In Section~\ref{sec:algorithms} we describe the range search algorithm used in the Starling system, and how we adapted it for the in-memory case and made some small experimental improvements before using it as a baseline.

\paragraph{Other Related Work} Some variants on the high-level ideas behind our range search algorithms---that is, early termination of queries with few results, and extensions of beam search for queries with many results---have been used before in the context of nearest neighbor search. Two works~\cite{li2020improving, chatzakis2025darth} use machine learning models to predict when a query can terminate early and still maintain high accuracy. Another work~\cite{xu2021twostage} uses graph search with two phases (essentially, an initial cheaper phase and a later, more compute intensive phase to gather additional candidates). 

Not to be confused with range search, the \textit{range filtering} problem addresses vectors with numerical labels such as timestamps, and asks to retrieve all vectors within a certain range on the labels (e.g., finding all vectors added to the database within a certain time range). This problem, which deals with external metadata on the vectors rather than finding all vectors within a ball of radius $r$, has been the subject of some recent work~\cite{zuo2024serf,engels2024approximate,xu2024irangegraph}.

\section{Preliminaries}\label{sec:prelim}
 In this work,
we study a set $\mP\subseteq \mathbb{R}^{d}$ of $n$ points (vectors)
in $d$ dimensions.
We denote the \defn{distance} between two points $p,q\in
\mathbb{R}^{d}$ as
 $d(p,q)$.  Smaller distance indicates greater
similarity.

Commonly-used distance functions include Euclidean distance ($L_2$
norm), and maximum inner product search, which uses negative inner product as a distance function~\cite{manohar2024parlayann}.

\begin{definition}[Range Search]
	Given a set of points $\mP$ in $d$-dimensions, a query point $q$ and radius $r$, the range search problem finds a set $\mK_q\subseteq \mP$
	such that $\max_{p\in \mK_q} d(p,q) \le r$.
\end{definition}

Next we define average precision, a commonly used metric that captures the accuracy of a search algorithm. Note that the measure weights a point's contribution to average precision proportionately to its size. 

\begin{definition}[Average Precision]
	Let $\mP$ be a set of points in $d$-dimensions and $\mQ$ a query point set. For $q \in \mQ$ , let $\mK_q$ be the true range neighbors of $q$ in $\mP$. Let $\mK'_q \subset \mP$ be an output of a range search algorithm, which is only allowed to consist of results that are truly within radius $r$ of the query point. Then \defn{Average Precision} is defined as $\frac{\sum\limits_{q \in \mQ}|\mK_q \cap \mK'_q|}{\sum\limits_{q \in \mQ}|\mK_q|}$ 
\end{definition}

Throughout this work, we will use the notation $N_{out \in G}(p)$ to refer to the set of directed edges $(p, v) | v \in G$, where $G$ is a directed graph.

\section{Characteristics of Range Search Datasets}\label{sec:rangedata}

In this section, due to the fact that there is only one existing high-dimensional range search benchmark, we take eight existing embedding datasets from top-$k$ retrieval benchmarks and show how to use them for range search by finding a suitable radius. We evaluate the characteristics of each dataset; alongside, we also evaluate the one existing range search benchmark. 

For a dataset to be a meaningful range search benchmark, the same distance $\epsilon$ must consistently indicate a match for each query point. This is not always true for embedding datasets and is difficult to verify without full access to the source material. Thus, we assess existing benchmarks to determine suitability and select a radius accordingly. Real-world query sets typically follow a power-law distribution, where most points have no matches, but a few have thousands~\cite{szilvasy2024vector,simsearchnet}. Another key factor is whether a dataset has a ``robust'' radius—small perturbations should not drastically alter match distribution. Our experiments show that more robust datasets tend to outperform those sensitive to radius changes. See Table~\ref{fig:datasets} for a description of each dataset used in our benchmarks.

\begin{figure*}
	\centering
	\begin{tabular}{|c|c|c|c|c|c|c|}
		\hline
		Dataset & Metric & Dimension & \begin{tabular}{c} Query \\ Size \end{tabular}  & \begin{tabular}{c} Selected \\ Radius \end{tabular} & Source & \begin{tabular}{c}
			Embedding \\ Model
		\end{tabular}   \\
		\hline
		BIGANN & L2 & 128 (uint8)& 10000 & 10000 &  Images~\cite{jegou2010product} & SIFT descriptors \\
		\hline
		DEEP & L2 & 96 (float)& 10000 & .02 &  Images~\cite{Deep1B} & \begin{tabular}{c} The last fully-connected layer \\ of the GoogLeNet model \end{tabular} \\
		\hline
		MSTuring & L2 & 100 (float) & 100000 & .3  & Bing queries~\cite{msturing} &  Turing AGI v5\\
		\hline
		GIST & L2 & 968 (float) & 10000 & .5 &  Images~\cite{jegou2010product} & GIST descriptors \\
		\hline
		SSNPP & L2 & 200 (uint8) & 100000 & 96237 &  Images~\cite{simsearchnet} & SimSearchNet++\\
		\hline
		OpenAI & L2 & 1536 (float) & 10000 & .2 & \begin{tabular}{c} ArXiv articles~\cite{openai} \end{tabular}  & \begin{tabular}{c} OpenAI text-embedding- \\ada-002 model \end{tabular}\\
		\hline
		Text2Image & IP & 200 (float) & 100000 & -.6 &  Images~\cite{Deep1B}   & Se-ResNext-101   \\
		\hline
		Wikipedia & IP & 768 (float) & 5000 & -10.5 &  \begin{tabular}{c}
			Wikipedia articles~\cite{wikipedia}
		\end{tabular}  & \begin{tabular}{c}  cohere.ai \\ multilingual 22-12 \end{tabular} \\
		\hline
		\begin{tabular}{c}
			MSMARCO- \\
			WebSearch
		\end{tabular}
		& IP & 768 (float) & 9374 & -62 & \begin{tabular}{c}
			Clicked document-query  \\ pairs from the ClueWeb22  \\ website corpus~\cite{msmarco} 
		\end{tabular}  & SimANS \\
		\hline
	\end{tabular}
	\captionof{table}{Descriptions of each dataset used in our experiments, along with the radius used for range search. Every dataset is publicly available~\cite{bigann,siftgist}. The SSNPP dataset is a range search benchmark, while the other datasets were originally published as top-$k$ benchmarks. Unless otherwise indicated, the reference in the ``Source'' column covers all attributes of the dataset.}\label{fig:datasets}
\end{figure*}

\begin{figure}
	\centering
	\includegraphics[scale=.28]{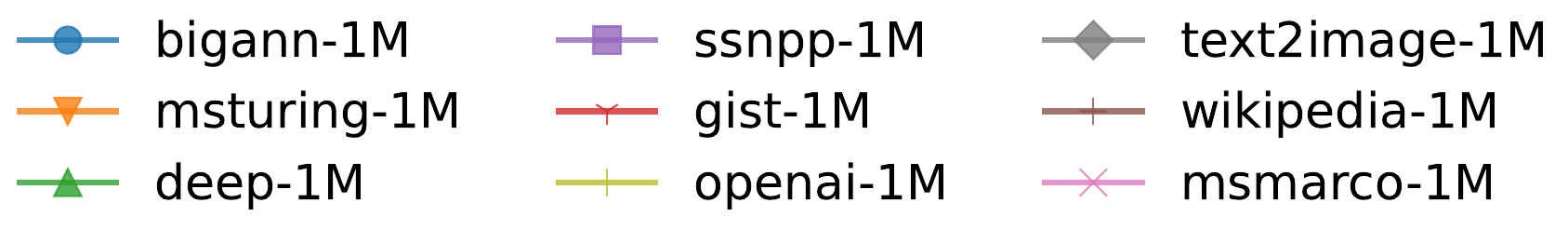}
	
	\includegraphics[scale=.24]{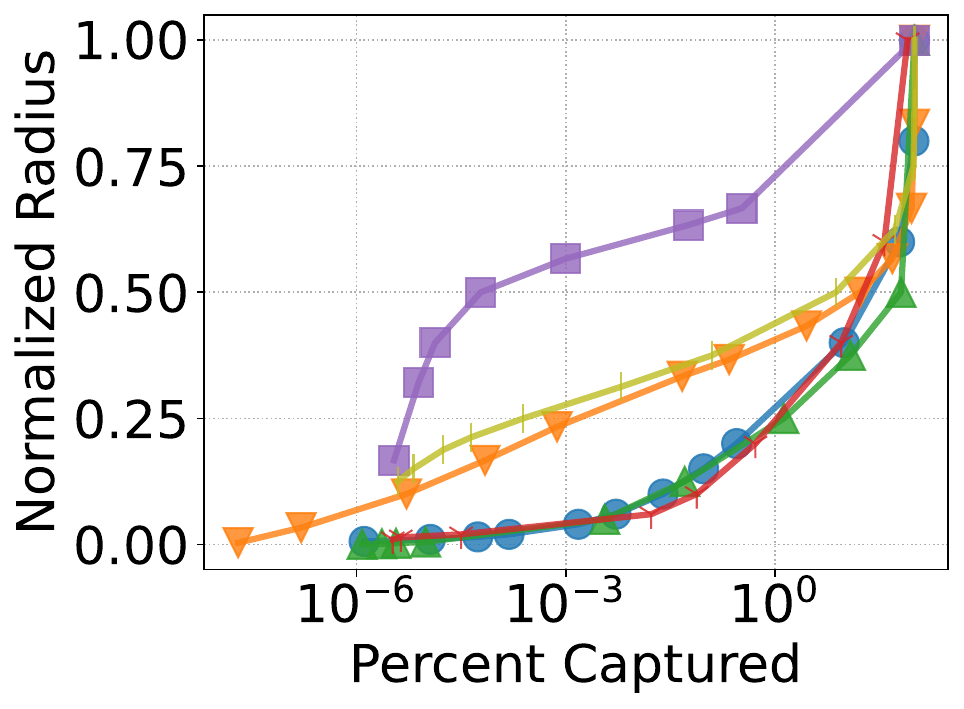}
	\includegraphics[scale=.24]{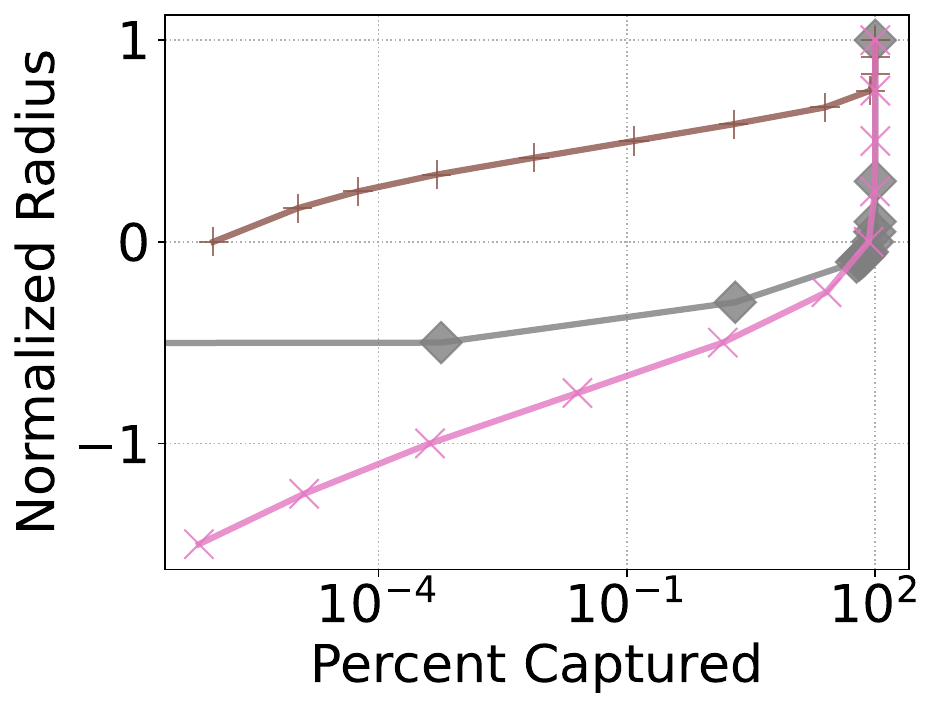}
	
	\caption{Plots of radius versus percent captured for each dataset: Euclidean datasets on the left and inner product datasets on the right. Radius is normalized to enable displaying multiple datasets on one plot. Note that inner product values can be less than zero, so negative ranges of radius are present on the right-hand figure.}
	\label{fig:pctcaptured}
\end{figure}

\subsection{Density of Matches by Dataset}
We begin our investigation by computing a measure of density of each dataset, in order to both find a suitable radius and determine whether the choice of radius is ``robust" in the sense that small changes to the radius do not result in drastically more points in each ball around a query point. To do this, we take each dataset and vary the radius, recording what we refer to as the ``percent captured'' for each radius---that is, for each query point $q$ and radius $r$, what fraction of the dataset is captured in the ball of radius $r$ around $q$?

Figure~\ref{fig:pctcaptured} presents data for each dataset, separately plotting Euclidean and inner product results from the largest radius with 0\% capture to the smallest achieving 100\% capture. As shown, dataset responses to radius perturbations vary widely. BIGANN, GIST, DEEP, Wikipedia, and MSMARCOWebSearch are the most robust in the relevant range (around $10^{-6}$ to $10^{-5}$ for million-size datasets, where most query points yield no results). In these datasets, fewer points lie near the boundary of radius $r$, reducing the effort needed to distinguish between those just inside or outside. Experimental results in Section~\ref{sec:experiments} illustrate this effect. Using insights from this analysis, we select an appropriate radius for each dataset, with real (non-normalized) values shown in Figure~\ref{fig:datasets}

\subsection{Frequency Distribution of Matches}

Given a choice of radius for each dataset, we now evaluate the frequency distribution of the number of range results for each query point. The full distributions are found in Table~\ref{fig:freqdistr}. We observe a general pattern of most queries having no results and a few large outliers, which is more pronounced in BIGANN, DEEP, MSTuring, and Text2Image, and less pronounced in SSNPP, OpenAI, Wikipedia, and MSMARCOWebSearch. GIST is unusual among the datasets in having many extremely large outliers. For BIGANN, SSNPP, and Wikipedia, we also show the frequency distribution for a larger version of the same base dataset. Since these are effectively larger samples from the same distribution, we observe an increase in density of range results. 

\begin{figure}
	\centering
		\addtolength{\tabcolsep}{-0.2em}
	\begin{tabular}{|c|c|c|c|c|c|c|}
		\hline
		Dataset & $0$ & $\leq 10^1$ & $\leq 10^2$ & $\leq 10^3$ & $\leq 10^4$ & $\leq 10^5$  \\
		\hline
		BIGANN-1M & 9728 & 143& 84  & 45 & 0 & 0 \\
		\hline
		BIGANN-10M & 9590 & 173 & 99 & 93 & 45 & 0 \\
		\hline
		BIGANN-100M & 9413 & 212 & 136 & 101 & 93 & 45 \\
		\hline
		DEEP-1M & 9923 & 56 & 19  & 2 & 0 & 0 \\
		\hline
		MSTuring-1M & 95716 & 2443 & 20 & 21 & 0 & 0 \\
		\hline
		GIST-1M & 8487 & 830 & 160  & 143 & 134  & 246 \\
		\hline
		SSNPP-1M & 97422 & 2424 &  254 & 0 & 0 & 0 \\
		\hline
		SSNPP-10M & 91575 & 7310 & 954  & 161 & 0 & 0 \\
		\hline
		SSNPP-100M & 80795 & 13719 & 4357  & 971 & 158  & 0\\
		\hline
		OpenAI-1M & 7030 & 2564 & 372 & 34 & 0 & 0\\
		\hline
		Text2Image-1M & 99327 & 669 & 4 & 0 & 0 & 0 \\
		\hline
		Wikipedia-1M & 4445 & 482 & 73 & 0 & 0 & 0 \\
		\hline
		Wikipedia-10M & 3385 &  1328 & 277 & 10 & 0 & 0 \\
		\hline
		\begin{tabular}{c}
			MSMARCO- \\ WebSearch-1M
		\end{tabular} & 7022 & 2199 & 152  & 3 & 0 & 0 \\
		\hline
	\end{tabular}
	\captionof{table}{Table showing the distribution of result sizes for each dataset. Note that not all datasets have the same number of query points; see Figure~\ref{fig:datasets} for the number of queries and corresponding radius for each dataset. No data point had more than $10^5$ results.}\label{fig:freqdistr}
\end{figure}

\section{Range Search on Graph-Based ANNS Indices}\label{sec:algorithms}

\begin{figure*}
	\centering
	\includegraphics[scale=.35]{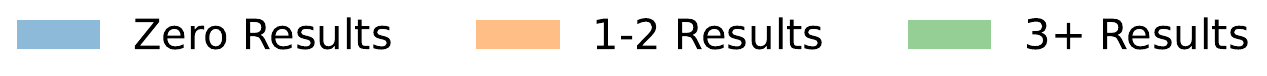} \\
	\begin{subfigure}{\textwidth}
		\centering
		\includegraphics[scale=.24]{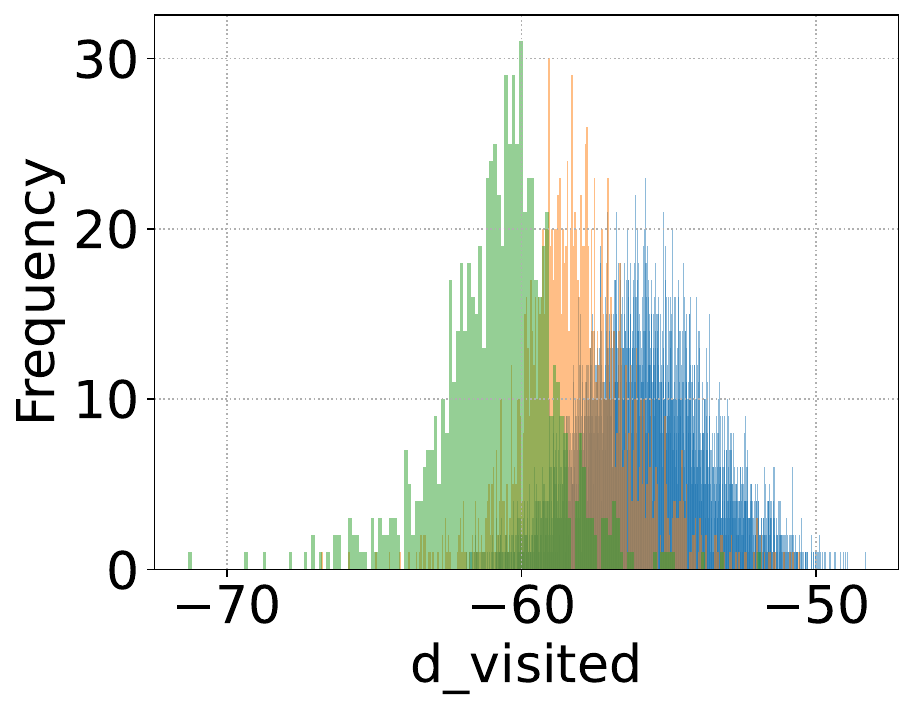}
		\includegraphics[scale=.24]{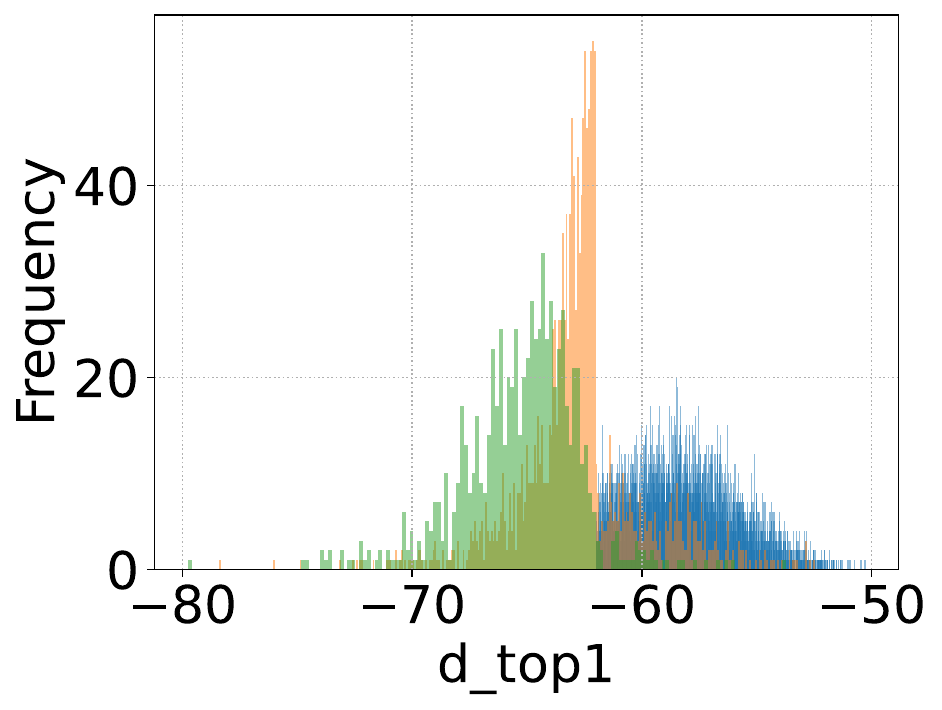}
		\includegraphics[scale=.24]{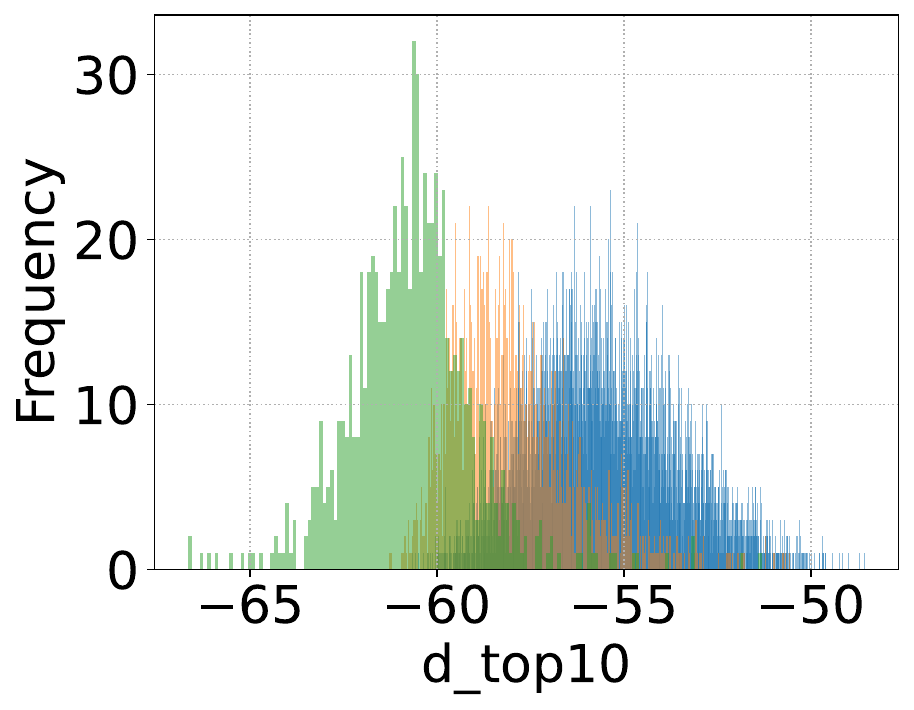}
		\includegraphics[scale=.24]{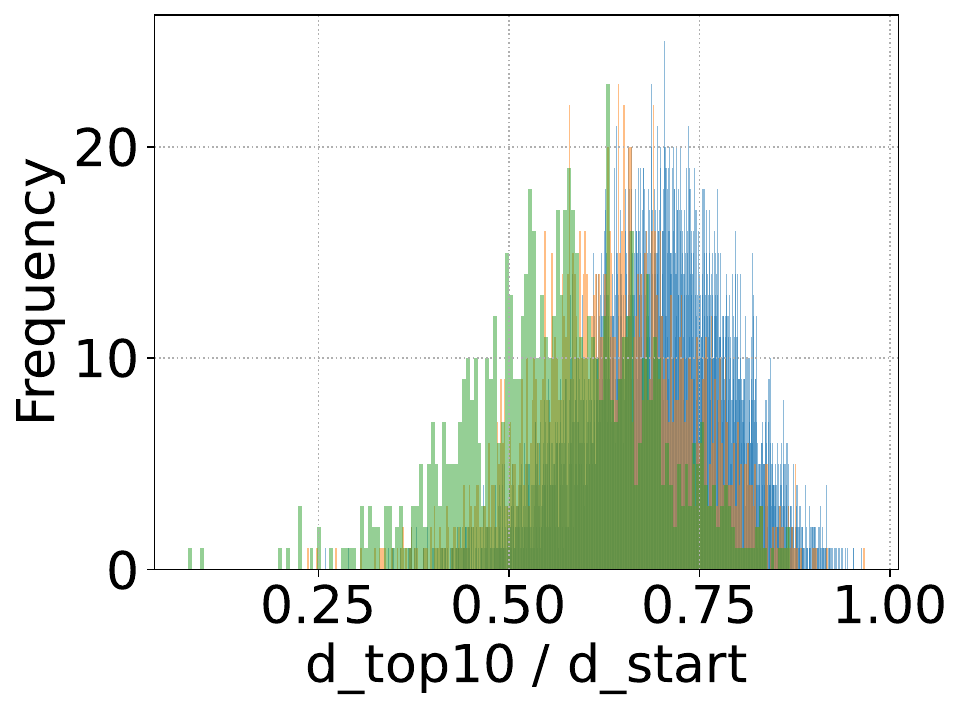}
		\caption{Histograms plotting selected early stopping metrics for all query points using dataset MSMARCO-1M.}
		\label{fig:msmarco_metrics}
	\end{subfigure}
	
	\begin{subfigure}{\textwidth}
		\centering
		\includegraphics[scale=.24]{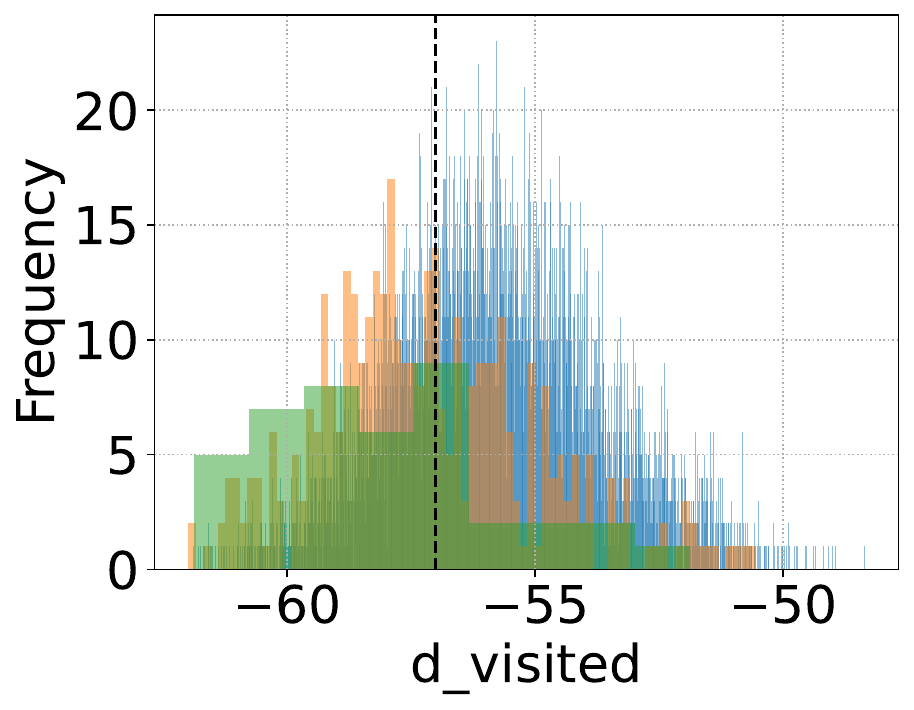}
		\includegraphics[scale=.24]{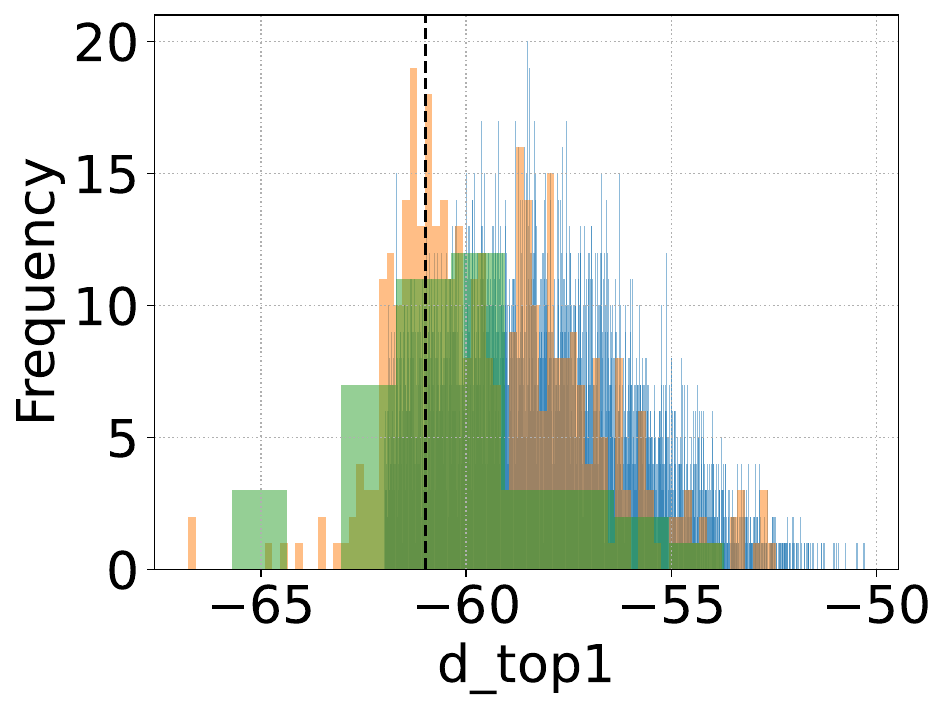}
		\includegraphics[scale=.24]{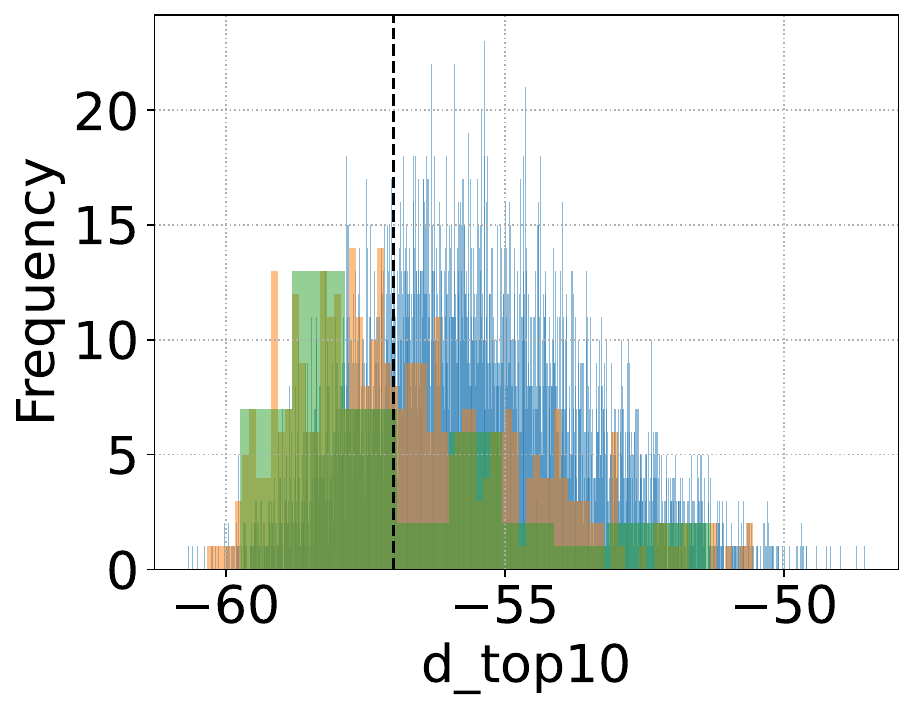}
		\includegraphics[scale=.24]{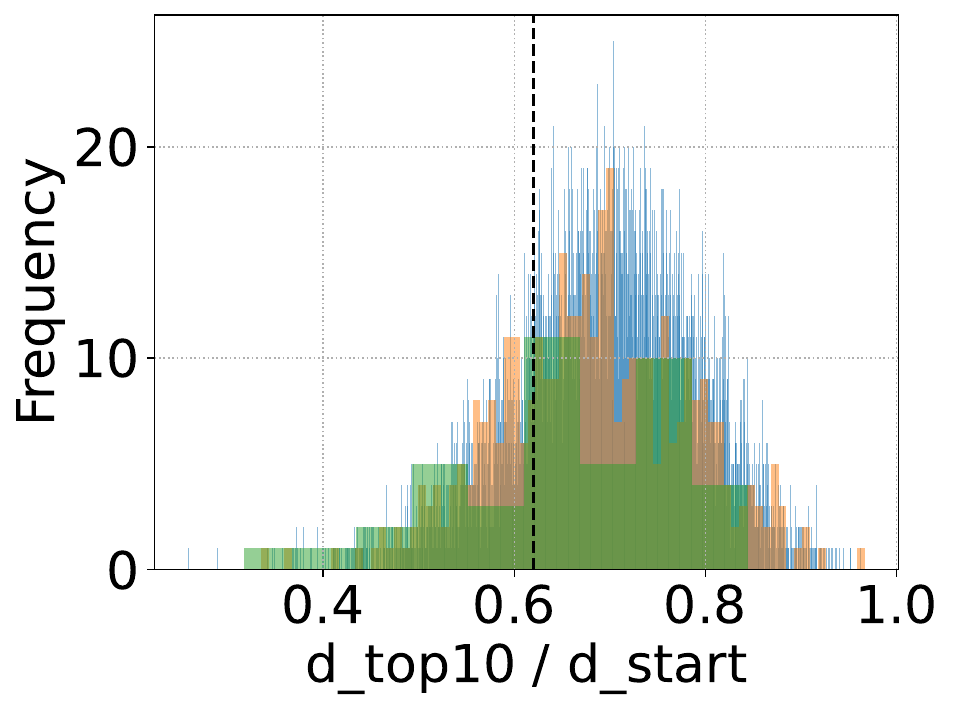}
		\caption{Histograms plotting selected early stopping metrics on dataset MSMARCO-1M, excluding query points which have already found a candidate within the radius.}
		\label{fig:msmarco_withexclusions}
	\end{subfigure}
	
	\begin{subfigure}{\textwidth}
		\centering
		\includegraphics[scale=.24]{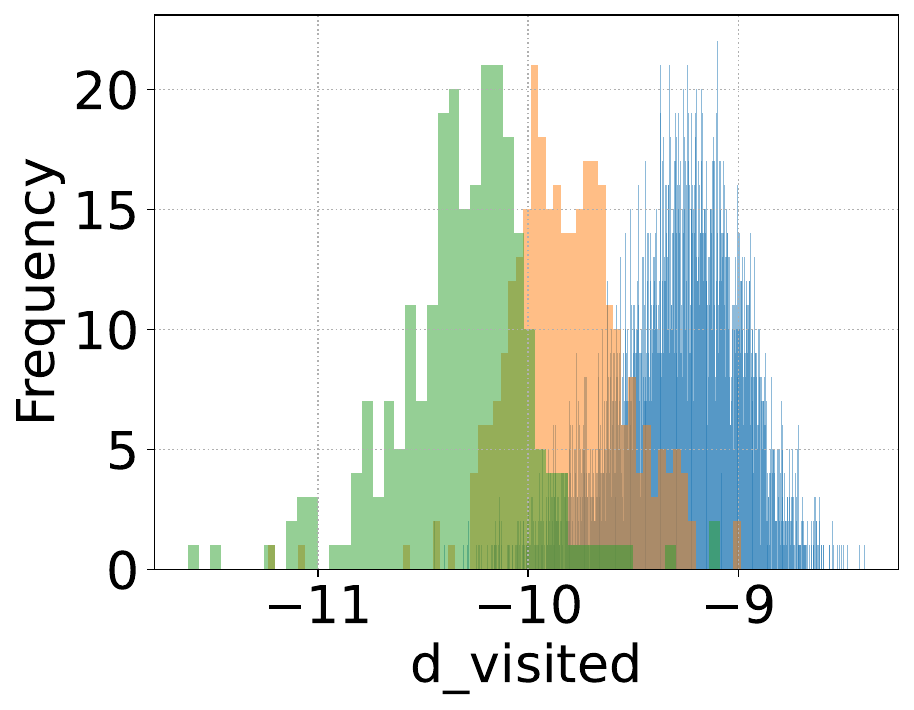}
		\includegraphics[scale=.24]{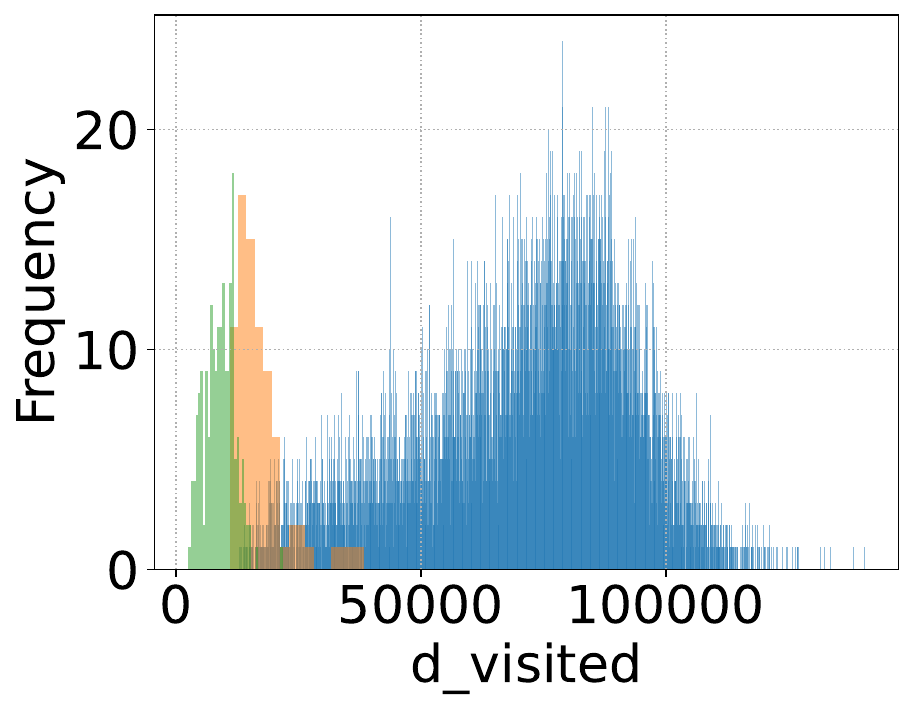}
		\includegraphics[scale=.24]{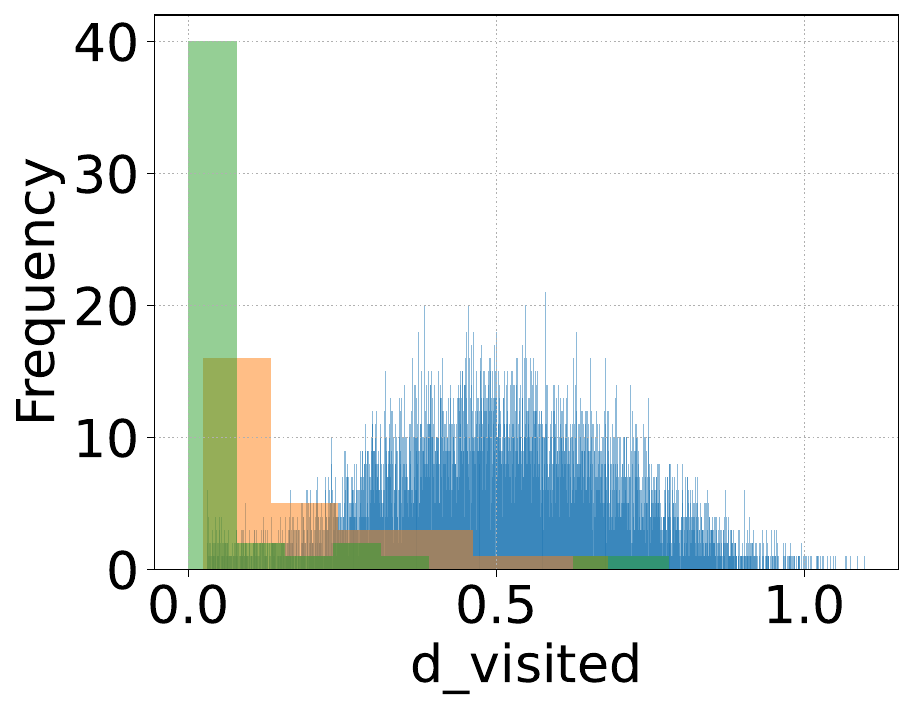}
		\includegraphics[scale=.24]{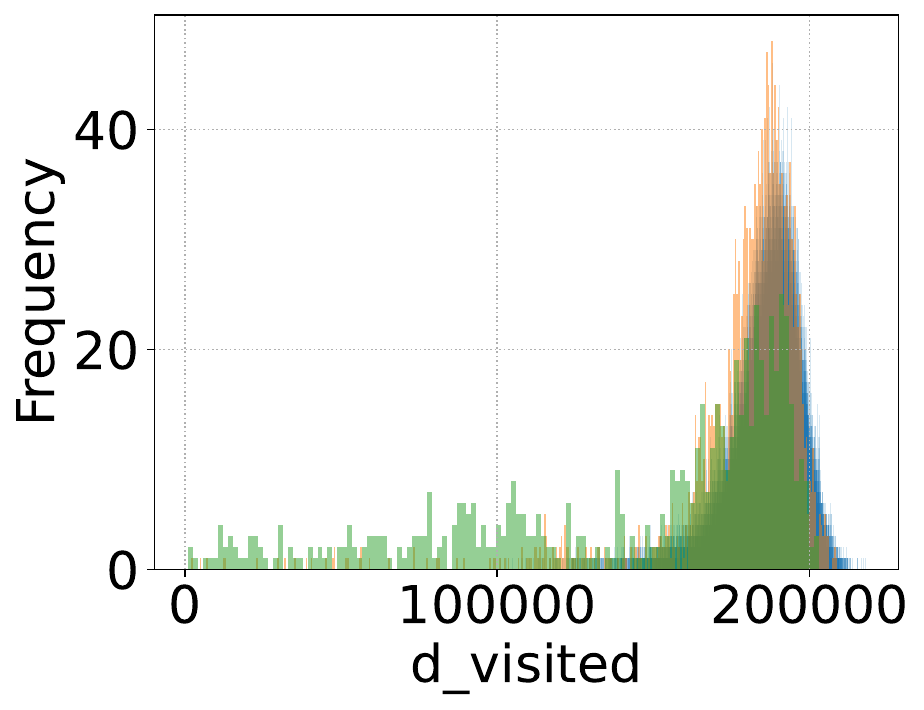}
		\caption{Histograms showing the value of $d_{\text{visited}}$ on more selected datasets. From left to right: Wikipedia-1M, BIGANN-1M, DEEP-1M, and SSNPP-1M.}
		\label{fig:visited_datasets}
	\end{subfigure}
	
	\caption{Frequency distributions of selected metrics broken down by the number of matched points. All metric values were taken at step 20 of a beam search with beam 100. Queries are separated by color based on the number of range results.}

\end{figure*}

This section introduces our range search algorithms, starting with the data structure and baseline beam search. Since beam search performs well for top-$k$ queries, some range queries with an intermediate number of results can be efficiently handled using a small $k$. Thus, we focus on improving queries at the extremes—those with many results and those with none. Intuitively, both require dynamic adjustments: terminating early for queries with no results and dynamically extending the beam for those with many.

\subsection{Data Structure and Baseline Algorithms}

As the aim of this paper is to investigate whether ANNS graphs can be effectively reused for range search, the data structure used for range queries is an ANNS graph. There are many examples of ANNS graphs in the literature~\cite{subramanya2019diskann,malkov2018efficient}. For the experiments in this paper we use the in-memory construction of DiskANN, specifically the ParlayANN library~\cite{manohar2024parlayann}, but the algorithms we present should work on any single-layer ANNS graph that uses the standard beam search algorithm, and should be modifiable to multi-layer graphs such as HNSW with little effort.

For our baseline algorithms, due to lack of existing algorithms for range search on graph-based indices, we use standard beam search on an ANNS graph as one baseline. This algorithm is described in detail in many publications~\cite{manohar2024parlayann,subramanya2019diskann}, and we also show pseudocode in Algorithm~\ref{alg:beamsearch}. We also build on the doubling search baseline suggested in the Starling system~\cite{wang2024starling}, re-implementing it within the ParlayANN library for the in-memory setting for the fairest possible comparison.

\subsection{Improving Range Search for Queries with Many Results}

We first investigate how to approach graph-based range search for queries that have many results. A standard beam search can only capture as many points as are contained within the beam of size $B$, and thus any queries with significantly more than $B$ results cannot achieve high accuracy. The natural question is whether it is possible to identify such queries with more than $B$ results and expand the set of returned candidates for only those queries. In this section we discuss two algorithmic approaches to this idea.

\paragraph{Doubling Beam Search} The first, and perhaps most natural, algorithm extending beam search to queries with variable number results is a dynamic beam search: that is, run the beam search with some starting beam $B$, then either terminate or double the beam and continue searching for more candidates. This algorithm was first proposed in the paper introducing the Starling system~\cite{wang2024starling}. The algorithm terminates if fewer than some fraction $\lambda$ of returned candidates are valid range candidates, and continues otherwise. In practice, we found that the value of $\lambda$ did not change the results very much, so we set it to $1$ as a rule (this finding seems to differ from that of the Starling authors, which may be attributed either to implementation distances or the in-memory setting). That algorithm is shown using beam search as a subroutine in Algorithm~\ref{alg:doubling}. In addition to rewriting the doubling search within the in-memory ParlayANN library, we made some minor optimizations such as using a faster approximate hash table for visited set lookups. For the fairest possible comparison, we also augment it with the early stopping heuristic introduced in Section~\ref{sec:earlystopping} during our experimental evaluation, but we note that the version in their paper contains no such heuristic.

\paragraph{Greedy Search} Another approach to dynamically resizing the queue length is to run a beam search with \textit{unbounded} queue length, but much stricter criteria on which nodes are explored. Candidates are first generated by an initial beam search with beam $B$. If the number of candidates within the radius falls above a certain threshold (again, some fraction $\lambda$ of the candidates returned), we next call our so-called greedy search routine, which has an unbounded queue but only adds nodes to the queue if they are within the ball of radius $r$ around the query point for the specified radius. This approach takes advantage of the fact that nodes in the same ball of radius $r$ are likely to be in connected clusters, and avoids the overhead of the doubling beam search by performing very few distance comparisons to points which are not valid results. See Algorithm~\ref{alg:greedy} in the supplementary material for a complete description, and Section~\ref{sec:experiments} for experimental analysis.

\subsection{Improving Range Search for Queries with No Results}~\label{sec:earlystopping}

This section introduces an algorithm to improve range search for queries with no results. The search begins with a fixed beam $B$, continuing with Algorithm~\ref{alg:greedy} or Algorithm~\ref{alg:doubling} if only candidates within the radius are returned. Ideally, the initial beam search should terminate early when no results are found, using terms that are inexpensive to compute. A 2020 paper by Li et al.~\cite{li2020improving} explores similar termination criteria for top-$k$ search, identifying several predictive terms: query point coordinates, distance to the current top-$k$ neighbor for fixed $k$, and the ratio of that distance to the query point’s distance from the start node.

To apply these ideas to range search, the key question is whether any of these metrics are effective at distinguishing queries with zero results from queries with one or more result. To investigate this, we look at each dataset and separate each query set into three groups: those with no results, those with 1-2 results, and those with more than three results. Then, we histogram the values of these metrics at different steps in the beam search, distinguishing between these three groups. We evaluate the following metrics, the last three of which are proposed in ~\cite{li2020improving}. 

\begin{itemize}[noitemsep,nosep]
	\item $d_{\text{visited}}$: distance from query point $q$ to the point being visited at step $i$ of a beam search.
	\item $d_{\text{top1}}$: distance from query point $q$ to the closest known neighbor of $q$ at step $i$ of a beam search.
	\item $d_{\text{top10}}$: distance from query point $q$ to the tenth-closest known neighbor at step $i$ of a beam search.
	\item $d_{\text{top10}} / d_{\text{start}}$: ratio of $d_{\text{top10}}$ to the distance from query point $q$ to the starting point of the beam search.
\end{itemize}

A natural and general approach to an early stopping condition for range search, using a variety of metrics, is to apply the early stopping metric after a certain number of steps in the beam search. The search does not terminate early if it has already found at least one candidate in the specified range.  Figure~\ref{fig:visited_datasets} shows one choice of metric---$d_{\text{visited}}$ on four different datasets. Algorithm~\ref{alg:earlystopping} shows formally how this condition can be applied.

Based on the histogram data of early stopping conditions, a larger set of which is shown in the supplemental material, we come to three conclusions: first, some datasets show potential for significant improvement using early stopping, while others do not. Second, if a dataset shows potential for improvement using one metric, it also tends to show improvement for all other metrics. Third, using $d_{\text{visited}}$ seems to be the best early stopping metric. We hypothesize that $d_{\text{visited}}$ is the best metric as it is the most frequently updated and contains the most granular information, but the positive results from other metrics suggests that approaches using machine learning such as those in the Li et al. paper~\cite{li2020improving} might also be useful. We present experimental analysis of this algorithm in Section~\ref{sec:experiments}. See Algorithm~\ref{alg:earlystoppingexample} for an example of how to apply the $d_{\text{visited}}$ metric during a search, and Algorithm~\ref{algo: rangesearch} for an example of the end-to-end algorithm using greedy search. 

\begin{algorithm2e}[t]
	\caption{BeamSearch($q, G, \mP, S, r,b , \mathcal{M}$).\protect}
	\label{alg:beamsearch}\small
	\SetKwBlock{ParDo}{do in parallel}{end}
	\SetKwBlock{ParFor}{parallel for}{end}
	\SetKwComment{tcp}{//}{}
	\SetKw{KwRet}{return}
	\SetAlgoLined
	\DontPrintSemicolon
	\KwIn{Query point $q$, graph $G$, point set $\mP$, starting points $S$, radius $r$, beam size $b$, early stopping metric $\mathcal{M}$}
	\KwOut{Set $\mathcal{B}$ of closest points, set $\mathcal{V}$ of visited points}

	$\mathcal{B} \gets S, \mathcal{V} \leftarrow \emptyset$\;
	
	\While{$\mathcal{B}\setminus \mV \neq \emptyset$}{
		$p^* \; \la \; \arg\min_{(p \in \mathcal{B} \setminus \mV)} \norm{p, q}$

		\tcp{Early stopping conditions}
		\lIf{$\mathcal{M} (q, \mathcal{B},\mV , r)$}{$\mathbf{Break}$} 
		
		$\mathcal{V} \leftarrow \mathcal{V} \cup \{ p^*\}$
		$\mathcal{B} \leftarrow  \mathcal{B} \cup (N_{out\in G}(p^*) \setminus \mathcal{V})$\;
		\lIf{$ | \mathcal{B} | > b$}{
			trim $\mathcal{B}$ to size $b$, keeping the closest points to $q$
		}
	}
	\KwRet $(\mathcal{B}, \mathcal{V})$
	\end{algorithm2e}

	\begin{algorithm2e}[t]
\caption{GreedySearch($q, G, \mP, S, r$).\protect}
\label{alg:greedy}\small
\SetKwBlock{ParDo}{do in parallel}{end}
\SetKwBlock{ParFor}{parallel for}{end}
\SetAlgoLined
\DontPrintSemicolon
\KwIn{Query point $q$, graph $G$, point set $\mP$, starting points $S$, radius $r$}
\KwOut{Set $\mathcal{V}$ of visited points}

$\mathcal{F} \gets S, $\ 
$\mathcal{V} \gets \emptyset $

\While{$\mathcal{F} \setminus \mV \neq \emptyset$}{
	$p^* \; \la \; \arg\min_{(p \in \mathcal{F} \setminus \mV)} \norm{p, q}$\;
	$\mathcal{V} \gets \mathcal{V} \cup \{p^*\}$
	
	\ForEach{$f \in (N_{out\in G}(p^*)\setminus \mV) $}{ 
		\lIf{$\|f, q\| \leq r$}{
			$\mathcal{F} = \mathcal{F} \cup \{f\} $\
		}
	}
}
\KwRet{$\mV$}\;
\end{algorithm2e}

\begin{algorithm2e}[t]
\caption{DoublingSearch($q, G,\mP, S, r, b, \mathcal{M}$).\protect}
\label{alg:doubling}\small
\SetKwBlock{ParDo}{do in parallel}{end}
\SetKwFor{ParFor}{parallel for}{do}{end}
\SetKw{KwRet}{return}
\SetAlgoLined
\DontPrintSemicolon
\SetKwProg{myfunc}{Function}{}{}
\KwIn{Query point $q$, graph $G$ , point set $\mP$, starting points $S$, radius $r$, beam size $b$, early stopping metric $\mathcal{M}$}
\KwOut{Set of neighbors $\mathcal{N}$ witin range }

\While{\textbf{true}}{
	$\mathcal{N} \gets \emptyset$

	\tcp{Early stopping is done inside the beam search}
	
	$(\mathcal{B}, \mV) \gets BeamSearch(q, G,\mP, S, r, \mathcal{M}, b)$\;
	
	$S \gets S \cup \mV$\;

	\For{$n \in \mathcal{B}$}{
		\lIf{$\| n, q \| \leq r$}{
			$\mathcal{N} \gets \mathcal{N} \cup \{n\}$
		}
	}

	\lIf{$ | \mathcal{N} | < b$}{ 
		$\mathbf{Break}$ 
	}
	$b \gets 2 \times b$\;
}

\KwRet $\mathcal{N}$\;
\end{algorithm2e}

\begin{algorithm2e}
\caption{EarlyStopping($q, \mathcal{B}, \mV, \mathcal{M}, \mathcal{C}$).\protect}
\label{alg:earlystopping}\small
\SetKwBlock{ParDo}{do in parallel}{end}
\SetKwBlock{ParFor}{parallel for}{end}
\SetAlgoLined
\DontPrintSemicolon
\KwIn{Query point $q$, current beam $\mathcal{B}$, visited set $\mV$, early stopping metric $\mathcal{M}$, cutoff parameter set $\mathcal{C}$}
\KwOut{Bool value for early stopping. True if conditions are met.}
\lIf{$\mathcal{M}(q, \mathcal{B},\mV, \mathcal{C})$}{\KwRet{$\mathbf{True}$}}
\end{algorithm2e}

\begin{algorithm2e}
\caption{EarlyStoppingExample($q, \mathcal{B}, \mV, v, \mathcal{C} = \{r, vl, esr\}$).\protect}
\label{alg:earlystoppingexample}\small
\SetKwBlock{ParDo}{do in parallel}{end}
\SetKwBlock{ParFor}{parallel for}{end}
\SetKwComment{tcp}{//}{}
\SetAlgoLined
\DontPrintSemicolon
\KwIn{Query point $q$, current beam $\mathcal{B}$, visited set $\mV$, number of visited points $v$, radius $r$, limit of number of visits $vl$, early stopping radius $esr$}
\KwOut{Bool value for early stopping.}
\RestyleAlgo{ruled}
\tcp{When the closest point in the beam is not within the radius}
\textbf{return} ($ \arg\min_{p \in \mathcal{B}} \|p,q\| > r , \mathbf{and}$\;
\tcp{When the number of visited points is larger than the given cutoff}
\hspace*{.34in}$v \geq vl, \mathbf{and}$\;
\tcp{When the current point visited is not within the early stopping radius}
\hspace*{.34in}$\arg\min_{p^* \in \mathcal{V}\setminus \mathcal{B}} \|p^*,q\| > esr$)\; 

\end{algorithm2e}

\begin{algorithm2e}[t]
\caption{GreedyRangeSearch($Q,G,\mP, S, r,b , \mathcal{M}$).\protect}
\label{algo: rangesearch}\small
\SetKwBlock{ParDo}{do in parallel}{end}
\SetKwFor{ParFor}{parallel for}{do}{end}
\SetAlgoLined
\DontPrintSemicolon
\SetKwProg{myfunc}{Function}{}{}
\KwIn{Query points $Q$, graph $G$, point set $\mP$, starting points $S$,   radius $r$, beam size $b$, early stopping metric $\mathcal{M}$}
\KwOut{Set of neighbors $\mathcal{V}$ within range  for each of the query points}

\ParFor{$q \in Q$}{
	$\mathcal{N}\gets \emptyset$
	
	$(\mathcal{B}, \mathcal{V}) \gets BeamSearch(q, G, \mP, S, r, b, \mathcal{M})$\; 
	\tcp{Early stopping is done inside the beam search}
	
	\ForEach{$n \in \mathcal{B}$}{
		\lIf{$\| n, q \| \leq r$}{
			$\mathcal{N} \gets \mathcal{N} \cup \{n\}$
		}
	}
	
	\lIf{$|\mathcal{N} | < b$}{
		$\mathbf{V}[q] \gets \mathcal{N}$
	}\Else{
		\tcp{Add results from beam search as starting points for greedy search}
		$\mathcal{V}[q] \gets GreedySearch(Q, G, \mP, \mathcal{N}, r)$\;
	}
}

\KwRet $\mathcal{V}$\;
\end{algorithm2e}

\section{Experimental Results}\label{sec:experiments}

We evaluate our algorithms on the nine datasets covered in Section~\ref{sec:rangedata}.

\paragraph{Experimental Setup and Code} The algorithms introduced in Section~\ref{sec:algorithms} extend the ParlayANN~\cite{parlayann} library, implemented in C++ with ParlayLib~\cite{parlay} for fork-join parallelism and standard parallel primitives such as sorting and filtering. ParlayANN features a top-ranked DiskANN implementation, currently among the top five on ANN Benchmarks~\cite{annbench}. The code is publicly available at \url{https://github.com/cmuparlay/ParlayANN/tree/rangetest2}. Most experiments ran on an Azure Standard\_L32s\_v3 VM with a 3rd Gen Intel® Xeon® Platinum 8370C processor, 32 vCPUs, and 256 GB memory, utilizing all vCPUs unless noted. Billion-scale tests were conducted on a 72-core Dell R930 with 4x Intel® Xeon® E7-8867 v4 processors, each with 18 cores, 2.4GHz, and 45MB L3 cache, and 1 TB main memory.

To produce a curve of QPS versus average precision, the starting beamwidth is varied on a number of searches and the Pareto frontier is reported. The build parameters used for the DiskANN graphs, as well as the early stopping threshold for each dataset, are reported in the supplemental material. We use scalar quantization to 8 bits for all floating-point datasets, with re-ranking to ensure that all reported results are within the radius.

\begin{figure*}
	\centering
	\includegraphics[scale=.35]{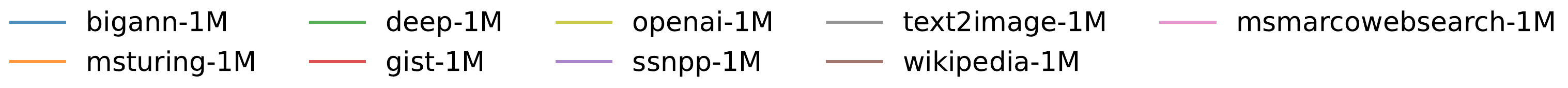}
	\includegraphics[scale=.35]{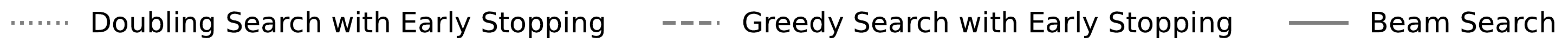} \\
	\includegraphics[scale=.35]{figures/legends/FAISS_all_legend.pdf}\\
	\begin{subfigure}[t]{.32\textwidth}
		\includegraphics[scale=.32]{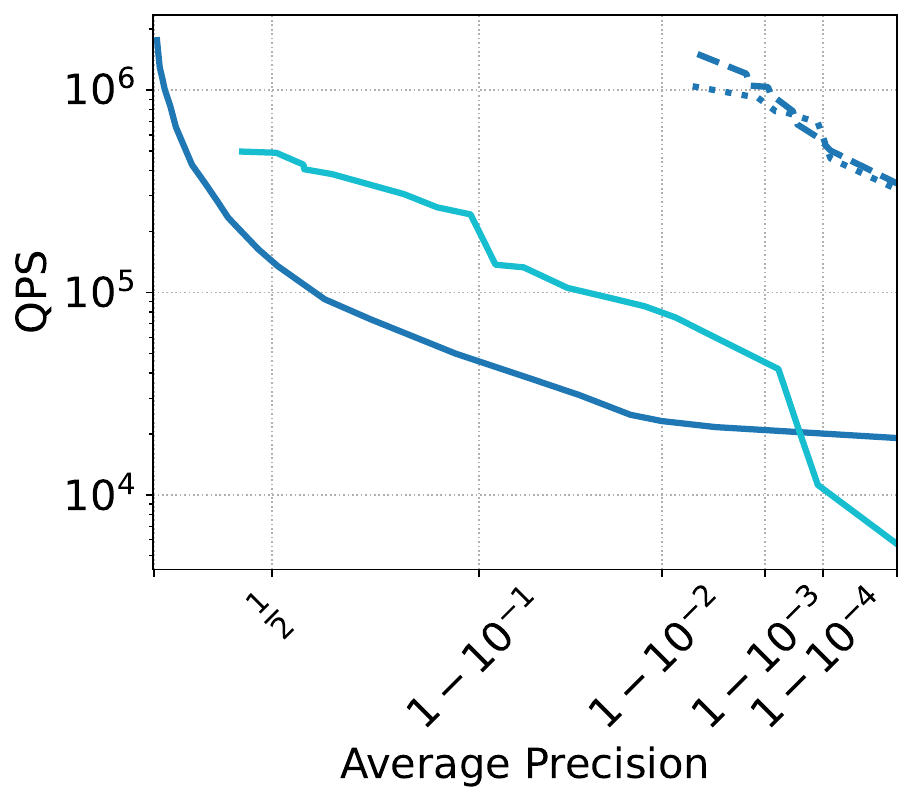}
		\caption{ BIGANN-1M}\label{fig:bigannap}
	\end{subfigure} 
	\begin{subfigure}[t]{.32\textwidth}
		\includegraphics[scale=.32]{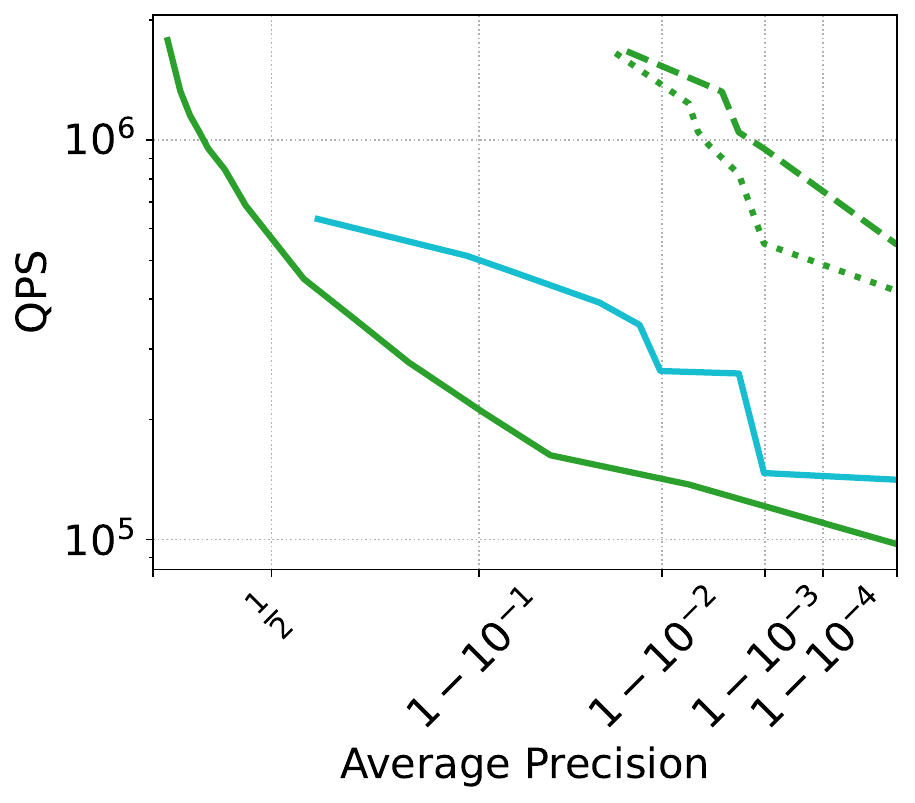}
		\caption{DEEP-1M}\label{fig:deepap}
	\end{subfigure}
	\begin{subfigure}[t]{.32\textwidth}
		\includegraphics[scale=.32]{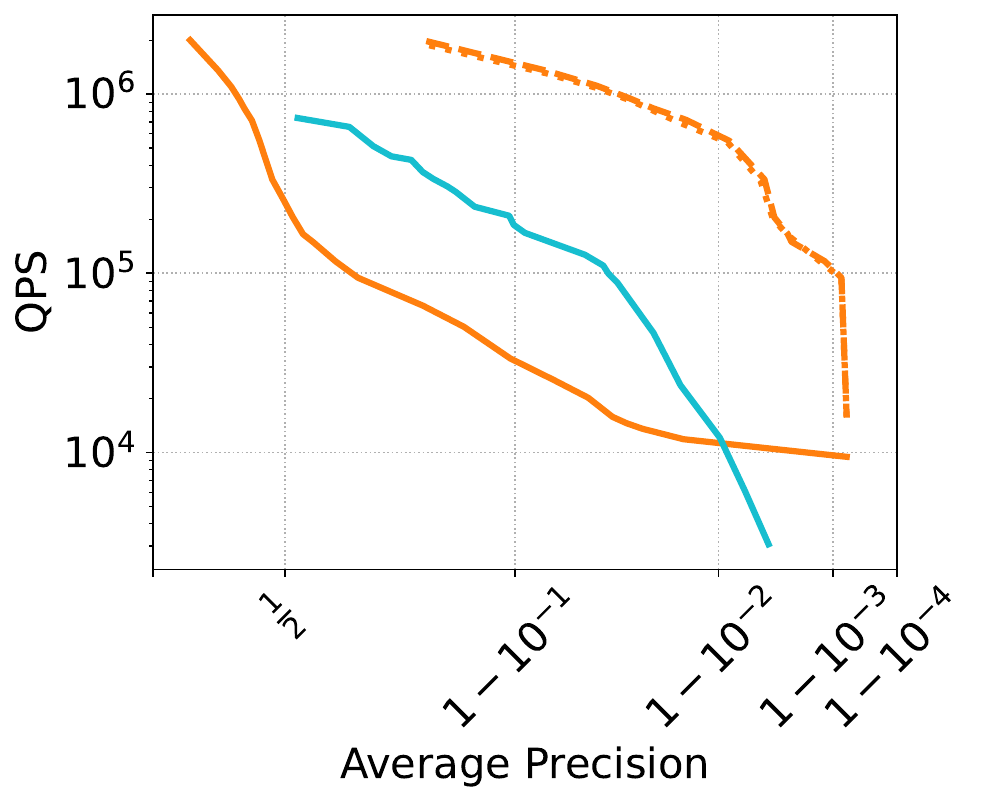}
		\caption{ MSTuring-1M}\label{fig:msturingap}
	\end{subfigure} \\
	\begin{subfigure}[t]{.32\textwidth}
		\includegraphics[scale=.32]{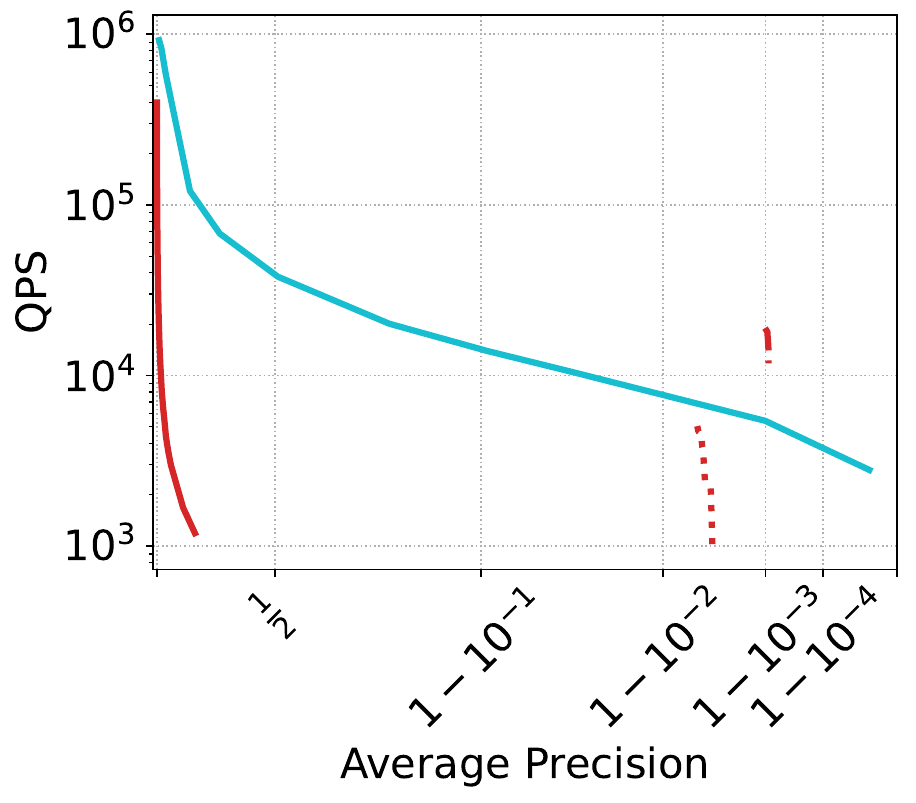}
		\caption{GIST-1M}\label{fig:gistap}
	\end{subfigure} 
	\begin{subfigure}[t]{.32\textwidth}
		\includegraphics[scale=.32]{new_figures/qps_recall/OpenAI-1M.pdf}
		\caption{ OpenAI-1M}\label{fig:openaiap}
	\end{subfigure}
	\begin{subfigure}[t]{.32\textwidth}
		\includegraphics[scale=.32]{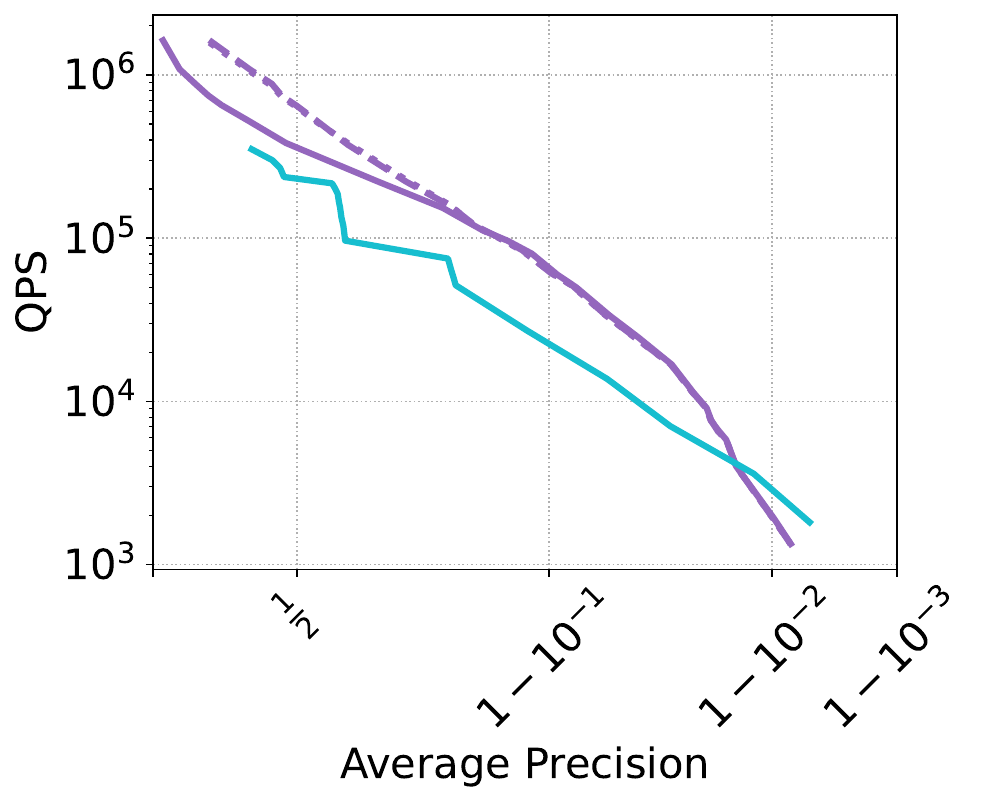}
		\caption{SSNPP-1M}\label{fig:ssnppap}
	\end{subfigure} 
	\begin{subfigure}[t]{.32\textwidth}
		\includegraphics[scale=.32]{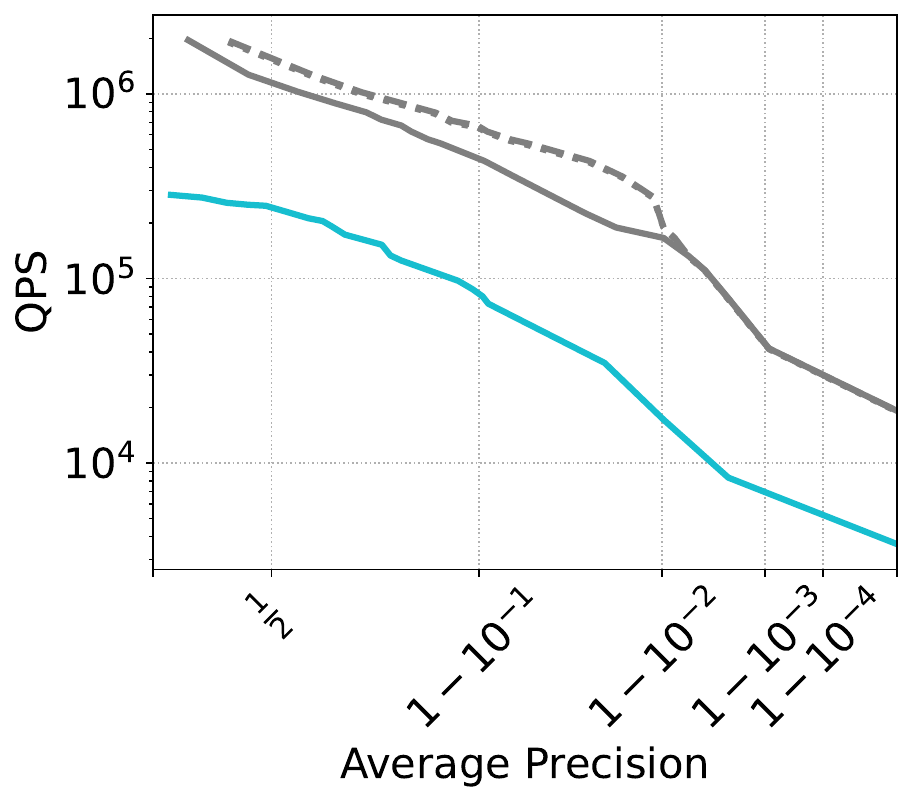}
		\caption{ Text2Image-1M}\label{fig:t2iap}
	\end{subfigure} 
	\begin{subfigure}[t]{.32\textwidth}
		\includegraphics[scale=.32]{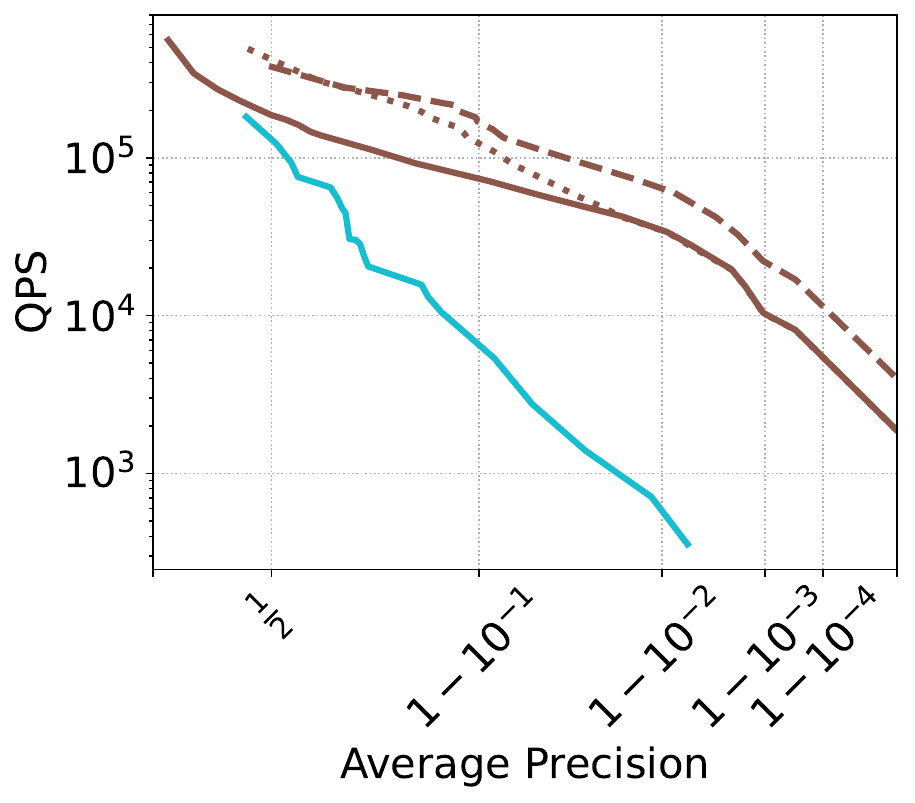}
		\caption{Wikipedia-1M}\label{fig:wikiap}
	\end{subfigure}
	\begin{subfigure}[t]{.32\textwidth}
		\includegraphics[scale=.32]{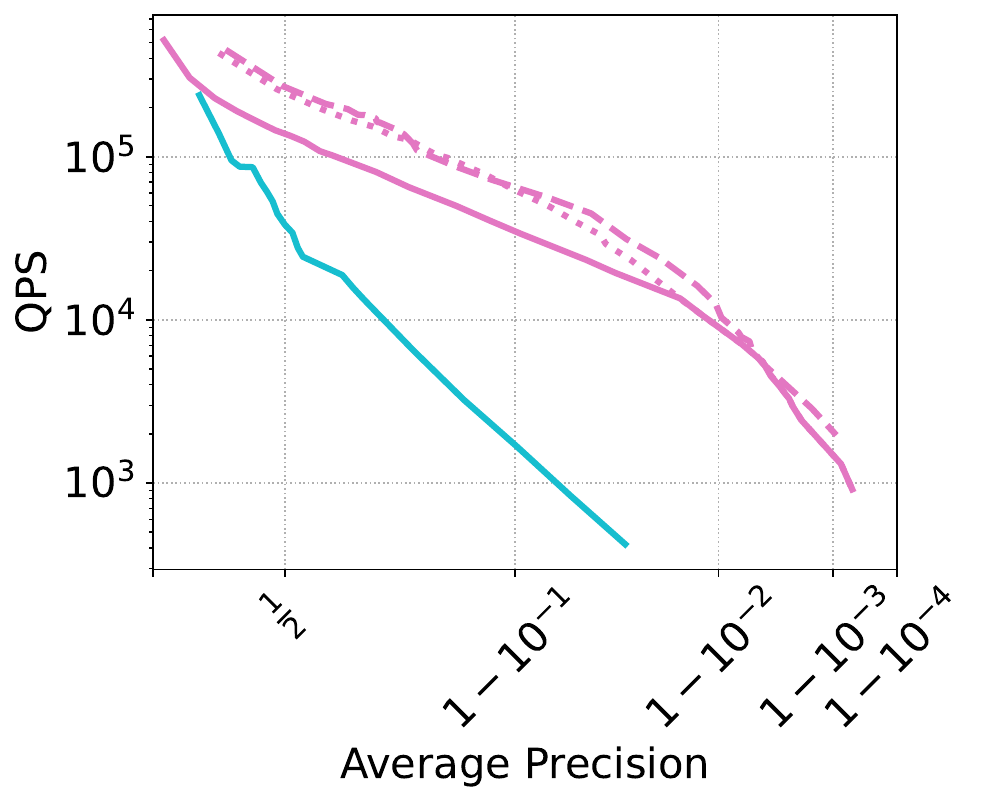}
		\caption{ MSMARCOWebSearch-1M}\label{fig:msmarcoap}
	\end{subfigure}
	\caption{ Average precision vs QPS for eight datasets and four range search algorithms. For GIST-1M, the lines for doubling search and greedy search are very short due to even the smallest of initial beam sizes producing recall in the .99-.999 range.}
	\label{fig:apgraphseight}
\end{figure*}

\paragraph{Competitors} To evaluate our algorithms against another standard baseline, we also use the range search algorithms from the well-known FAISS-IVF library~\cite{douze2024faiss}. At a high level, FAISS-IVF uses an inverted file index produced via clustering, with the option to construct an HNSW index over the cluster centroids to help select clusters more efficiently at query time. A query probes a selected number of cells and returns the candidates within the radius. The parameters chosen for FAISS, along with information on how we configured them, are shown in the supplemental material. To produce a curve of QPS versus average precision for FAISS, we varied both the number of cells probed and the $ef\_search$ of the centroid graph and report the Pareto frontier.

\begin{figure*}
	\centering
	\includegraphics[scale=.35]{figures/legends/threealg_legend.pdf} \\
	\includegraphics[scale=.35]{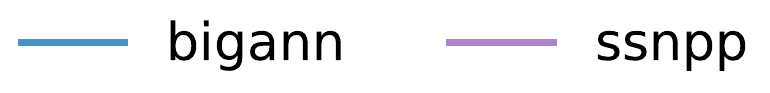} 
	\includegraphics[scale=.32]{figures/legends/FAISS_all_legend.pdf} \\
	\begin{subfigure}{\textwidth}
		\includegraphics[scale=.32]{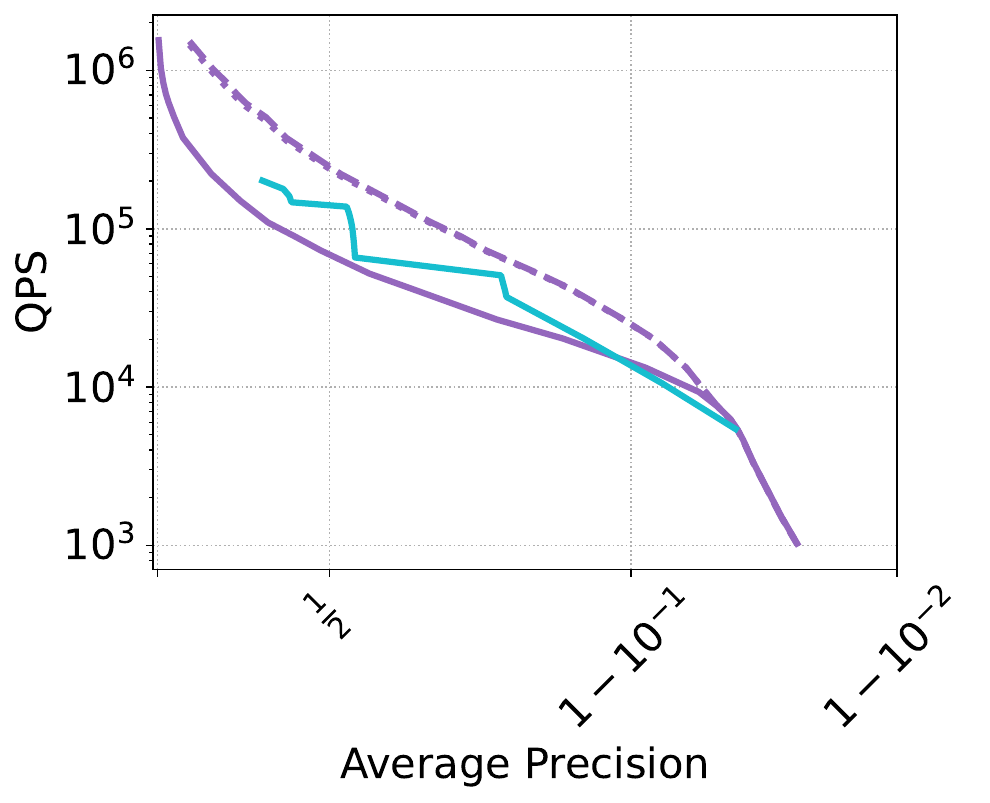} \hfil 
		\includegraphics[scale=.32]{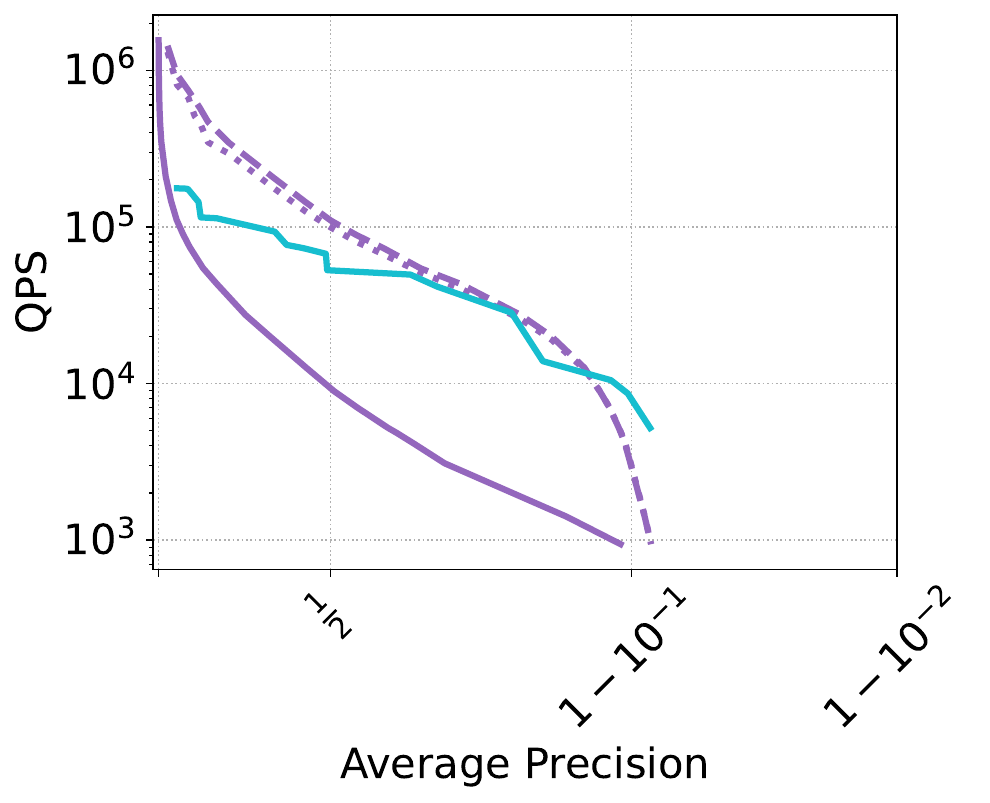} \hfil 
		\includegraphics[scale=.32]{new_figures/qps_recall/ssnpp-1B.pdf} 
		\caption{SSNPP}
	\end{subfigure} \hfil 
	\begin{subfigure}{\textwidth}
		\includegraphics[scale=.32]{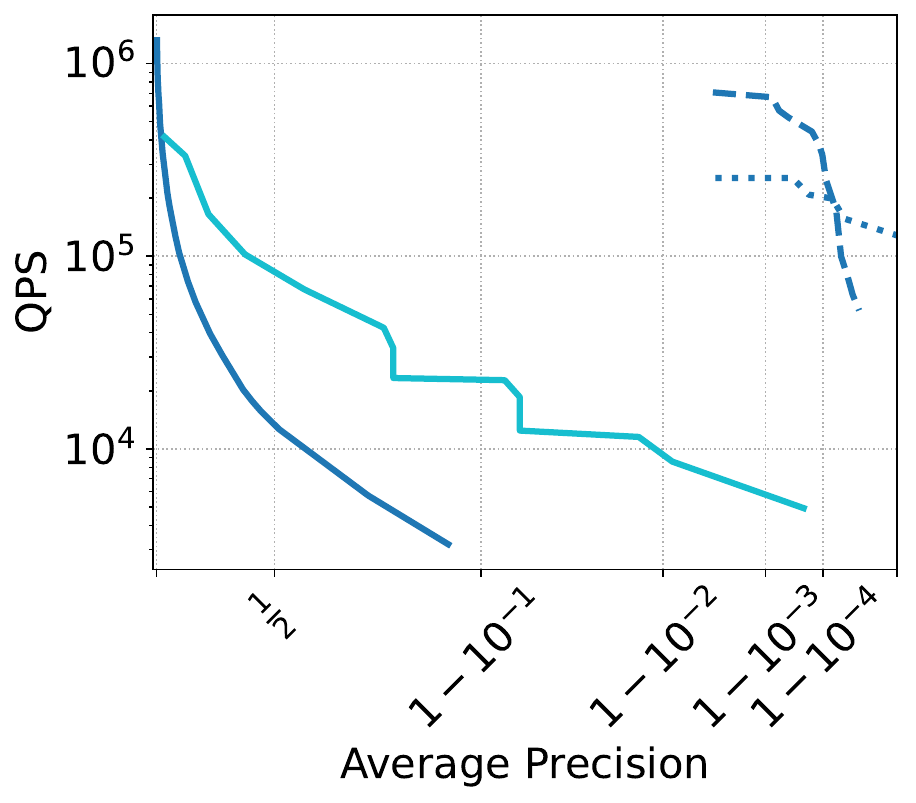}\hfil 
		\includegraphics[scale=.32]{new_figures/qps_recall/bigann-100M.pdf}\hfil 
		\includegraphics[scale=.32]{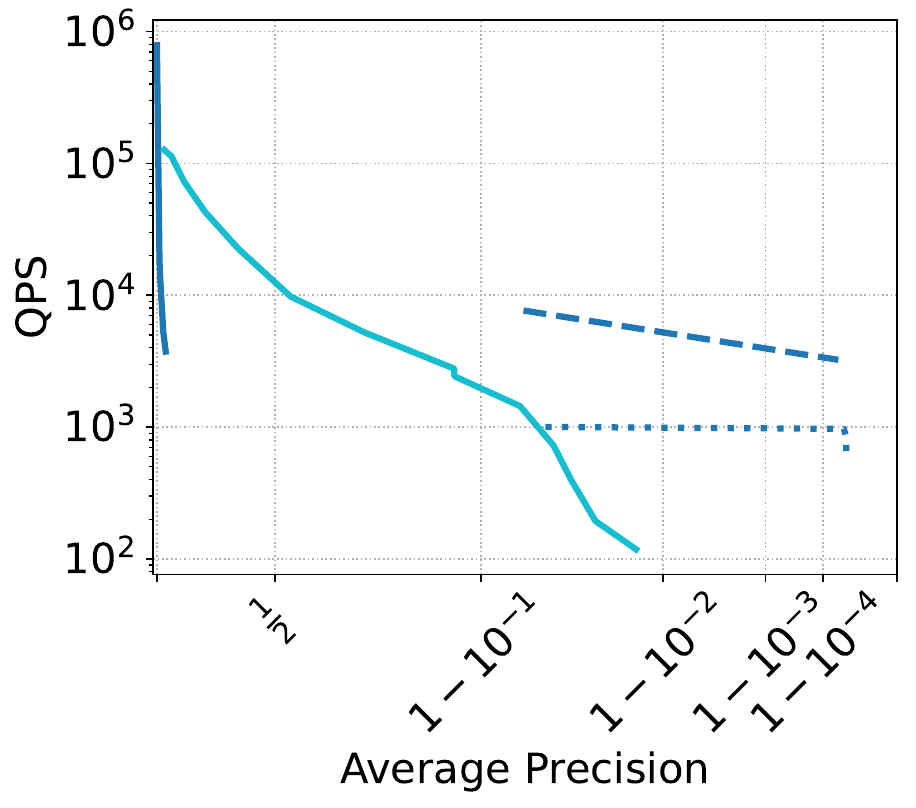} 
		\caption{BIGANN}
	\end{subfigure} 
	\caption{Average precision vs QPS for the SSNPP and BIGANN datasets. From left to right, slices of size 10 million, 100 million and 1 billion.}
	\label{fig:sizescaling}
\end{figure*}

\paragraph{QPS and Precision Tradeoffs} Our experimental evaluation plots queries per second (QPS) versus precision for nine datasets (Figure~\ref{fig:apgraphseight}). Figure~\ref{fig:sizescaling} examines size scaling effects for SSNPP and BIGANN, from 10 million to 1 billion. While our algorithms improve upon the naive baseline across all datasets, the degree of improvement and algorithm rankings vary. DEEP, BIGANN, and MSTuring achieve near-100x throughput gains over FAISS-IVF and the beam search baseline, with doubling beam search similar than greedy search at 1 million scale but up to 10x slower on larger datasets. OpenAI sees a 10x improvement over beam search and FAISS, with both doubling and greedy search performing similarly. Wikipedia and MSMARCO-WebSearch improve moderately over the FAISS and beam search baselines, with greedy search slightly ahead of doubling. GIST stands out due to its unique frequency distribution, as the only dataset with hundreds of points exceeding 10,000 range results, making greedy search more effective. Text2Image shows modest gains below 90\% recall. SSNPP, analyzed across four scales, exhibits little improvement at 1 million but increasingly benefits from the specialized range search algorithms as dataset size grows.

\paragraph{Comparison with FAISS-IVF} Greedy search and doubling search are overall faster than FAISS-IVF, with occasional exceptions at the highest recall values. Interestingly, FAISS-IVF is faster than the beam search baseline about half the time. These results reflect the fact that the FAISS algorithm is more intrinsically suited to range search than the beam search baseline, since it considers every point in each cell as a potential range candidate. However, greedy search and doubling search with early stopping are overall more successful at precisely searching the correct number of points for each range query.

\paragraph{Scaling} Figure~\ref{fig:sizescaling} examines the performance of our algorithms on two datasets at three different scales: 10 million, 100 million, and 1 billion. Similarly to existing range search benchmarks, we use the same radius for each data scale, so each order of magnitude effectively represents a more dense space, and the average number of results per query point increases. Unsurprisingly, the advantage of greedy search and doubling search increase significantly over the beam search baseline, which is only capable of returning up to about 1000 points per query before becoming hopelessly inefficient. Interestingly, we also see that the greedy search seems to have the most significant advantage over the doubling beam search as the density of the dataset increases (this is also true of GIST-1M, which has some extraordinarily dense outliers). This is likely because in denser graphs, the greedy search is particularly adept at efficiently finding all points within the radius, which are likely to be in a dense connected cluster in the graph structure. In the supplemental material we investigate this hypothesis directly by fixing one dataset and measuring performance of both algorithms while varying the chosen radius. Greedy search had better performance for larger choices of radius (indicating more matches per point on average).

\begin{figure*}
	\includegraphics[scale=.18]{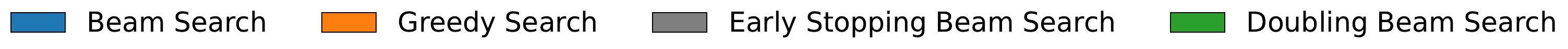} \\
		\begin{subfigure}{.32\textwidth}
		\includegraphics[scale=.2]{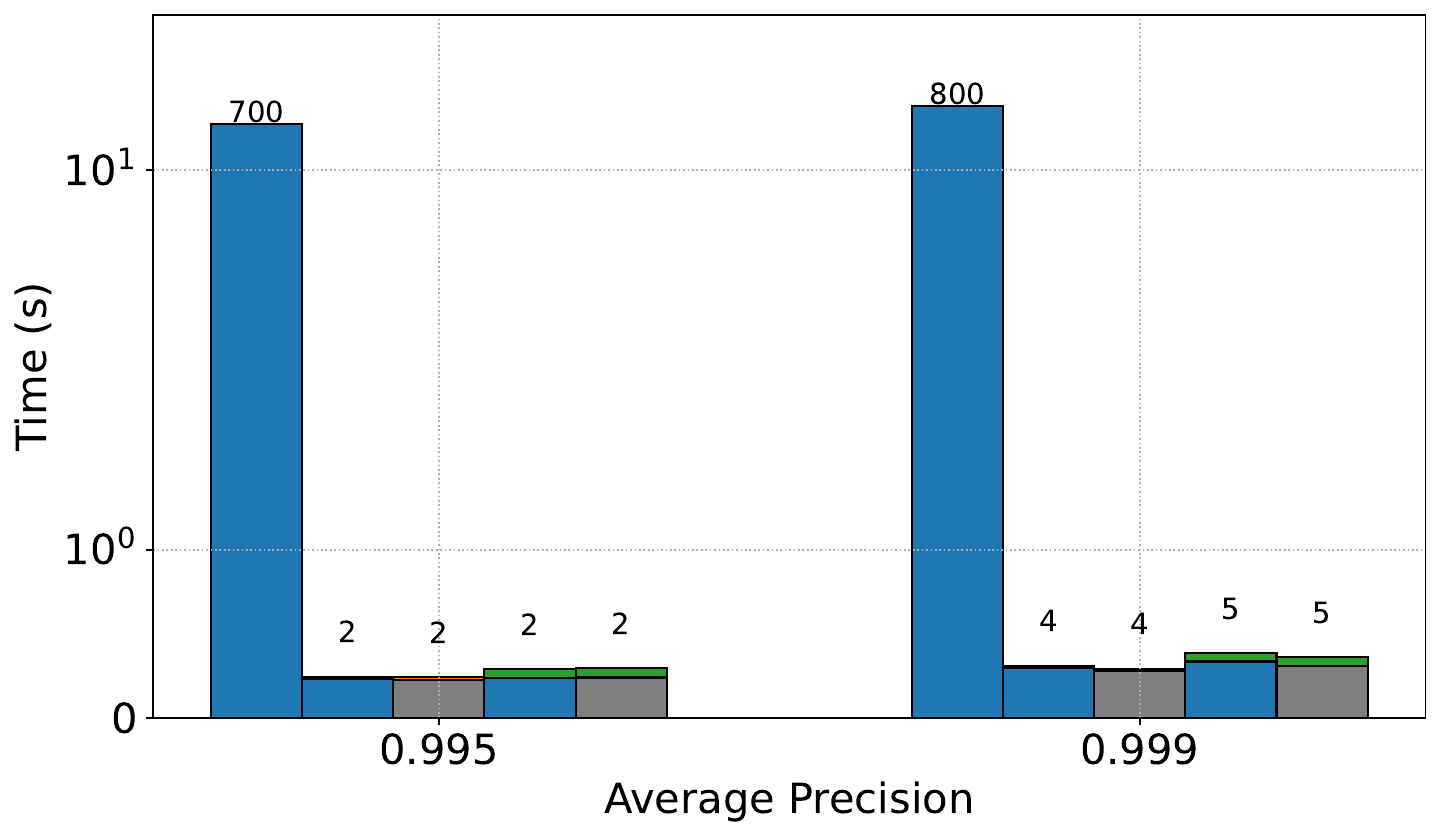}
		\caption{BIGANN-1M}\label{fig:bigann1mbar}
	\end{subfigure} \hfil
	\begin{subfigure}{.32\textwidth}
		\includegraphics[scale=.2]{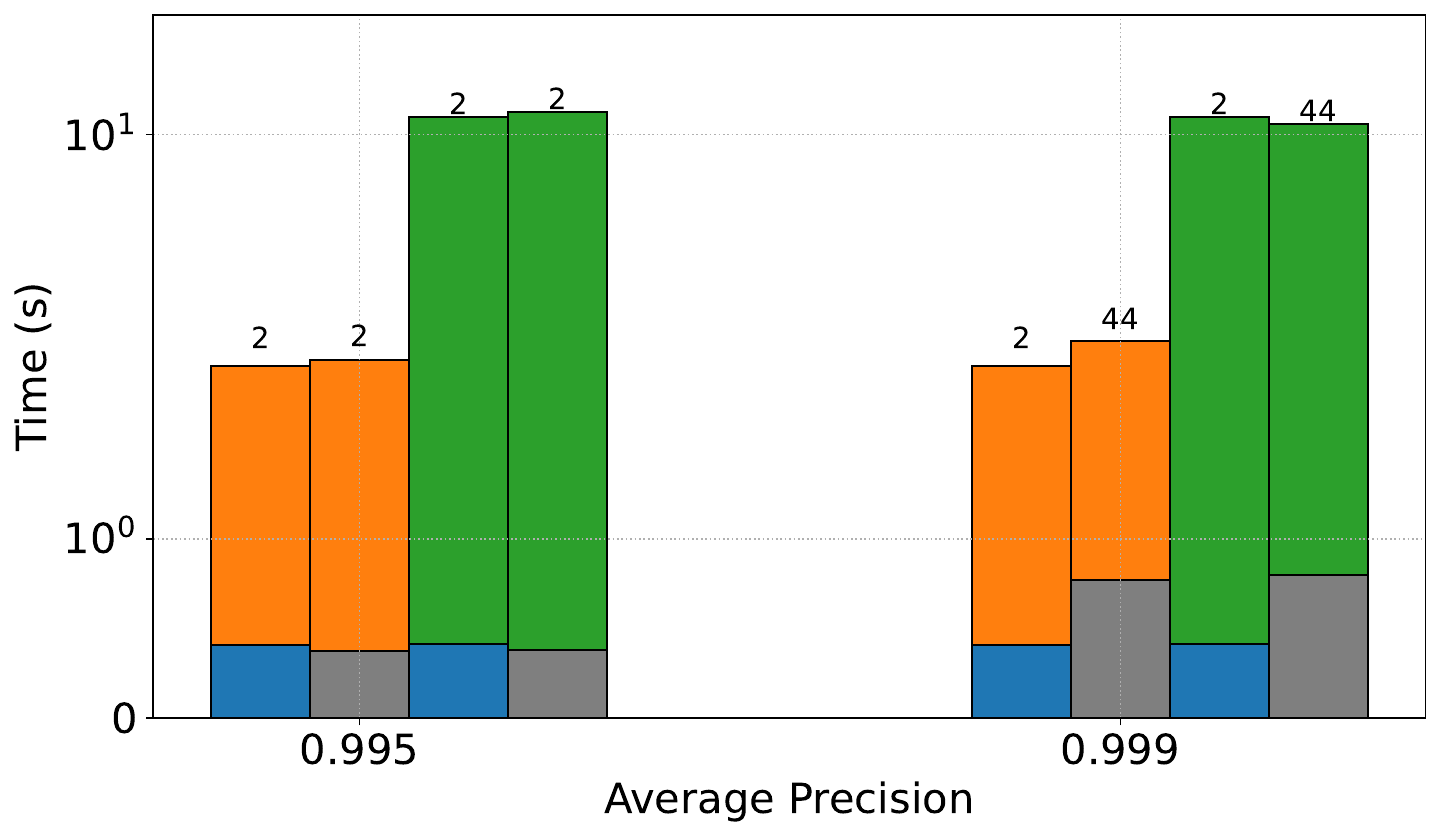}
		\caption{BIGANN-100M}\label{fig:bigann100mbar}
	\end{subfigure} \hfil
	\begin{subfigure}{.32\textwidth}
		\includegraphics[scale=.2]{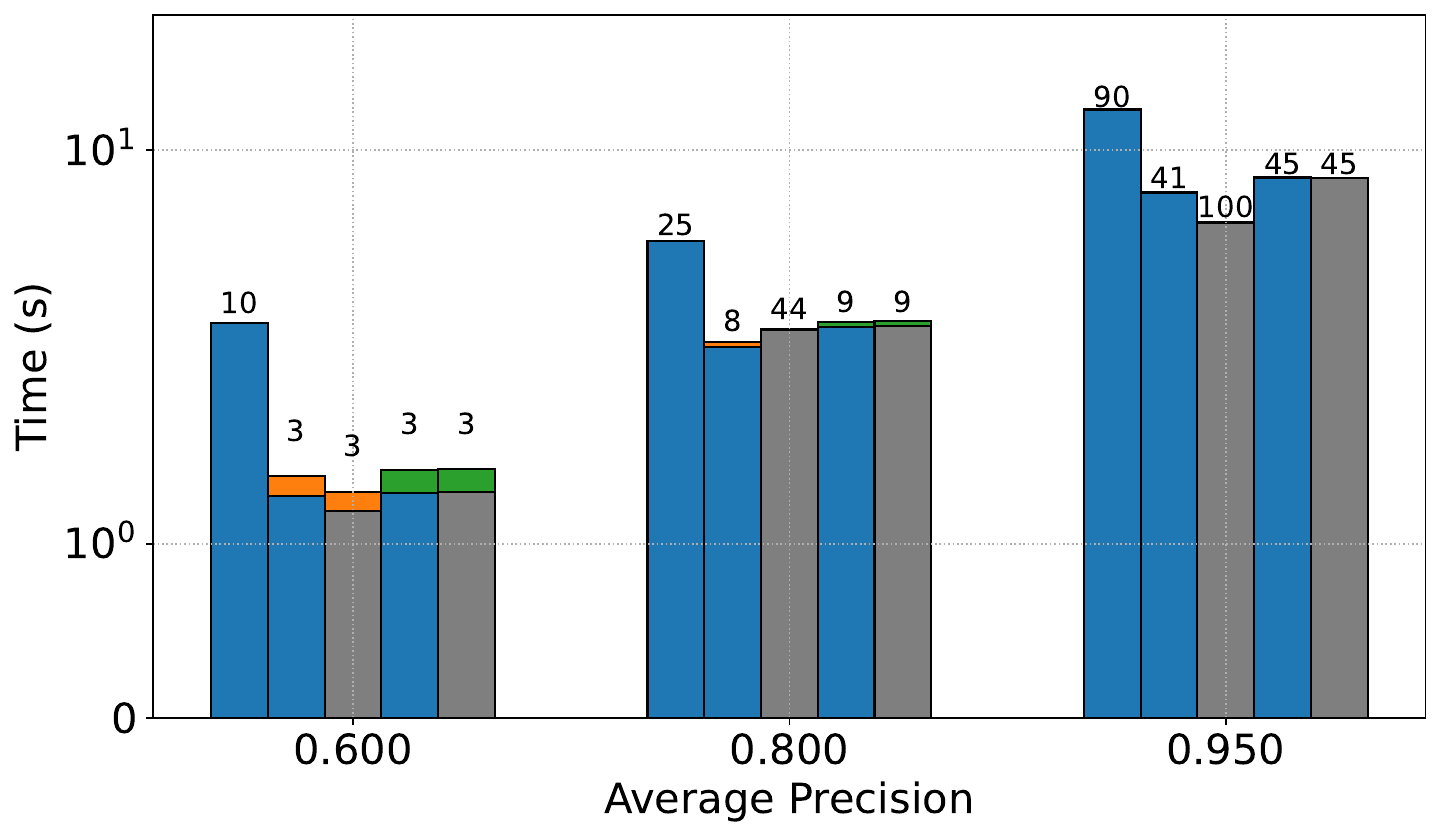}
		\caption{MSMARCOWebSearch-1M}\label{fig:msmarcobar}
	\end{subfigure}
	\caption{Figures breaking down the cost in seconds (single threaded time) of each type of search for three datasets for selected average precision. The label at the top of each column indicates the beam width of the initial search. Each collection from left to right shows: the beam search baseline (which is sometimes omitted when it cannot achieve the desired average precision), beam search followed by greedy search, beam search with early stopping followed by greedy search, beam search followed by doubling beam search, and beam search with early stopping followed by doubling beam search. Note that in some cases, the time spent on the second phase of search is so small that it is not visible.}\label{fig:barchart}
\end{figure*}

\paragraph{Analyzing Time Spent on Algorithm Components} To aid in our analysis, we investigate the time spent on different phases of each algorithm in Figure~\ref{fig:barchart}. Each algorithm consists of an initial beam search phase---either with or without the early stopping criteria in Algorithm~\ref{alg:earlystopping}---and then a following phase of either greedy search or successive calls to doubling beam search. We show this visual breakdown for three datasets at selected fixed average precision: BIGANN-1M, BIGANN-100M, and MSMarcoWebSearch-1M. The time breakdowns provide significant insight into the results in Figures~\ref{fig:apgraphseight} and~\ref{fig:sizescaling}. First, note that for all datasets, the initial beam search phase is a sizable portion of the overall cost of computation. This is because the initial beam search takes place on all points, but further searches only take place on points with one or more valid range result (since the distances used are exact, a point with zero results will never pass on to the second round of computation).

\paragraph{Greedy Search vs Doubling Search} Next, we examine the differences between doubling and greedy search. Doubling search spends more time in the second phase but achieves higher recall from a smaller starting beam. Dataset characteristics determine the best tradeoff between QPS and average precision. Greedy search requires a larger initial beam but has a cheaper second round since it only visits points within the radius. In contrast, doubling search starts with a smaller beam but spends more time in the second phase, performing additional distance comparisons and incurring overhead from repeated beam search calls. Experimental results suggest doubling search dominates greedy search when (a) a small initial beam is enough for doubling search to achieve high recall, and (b) few rounds of doubling are needed. These conditions hold for BIGANN-1M (and DEEP-1M), but GIST-1M requires excessive doubling rounds, and MSMARCOWebSearch-1M demands a large initial beam. Scaling studies in Figure~\ref{fig:sizescaling} show greedy search overtakes doubling search as match density increases, likely due to the growing number of doubling rounds required.

\begin{figure*}
	\includegraphics[scale=.35]{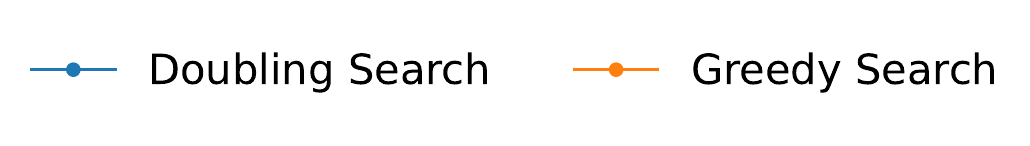} \\
	\includegraphics[scale=.22]{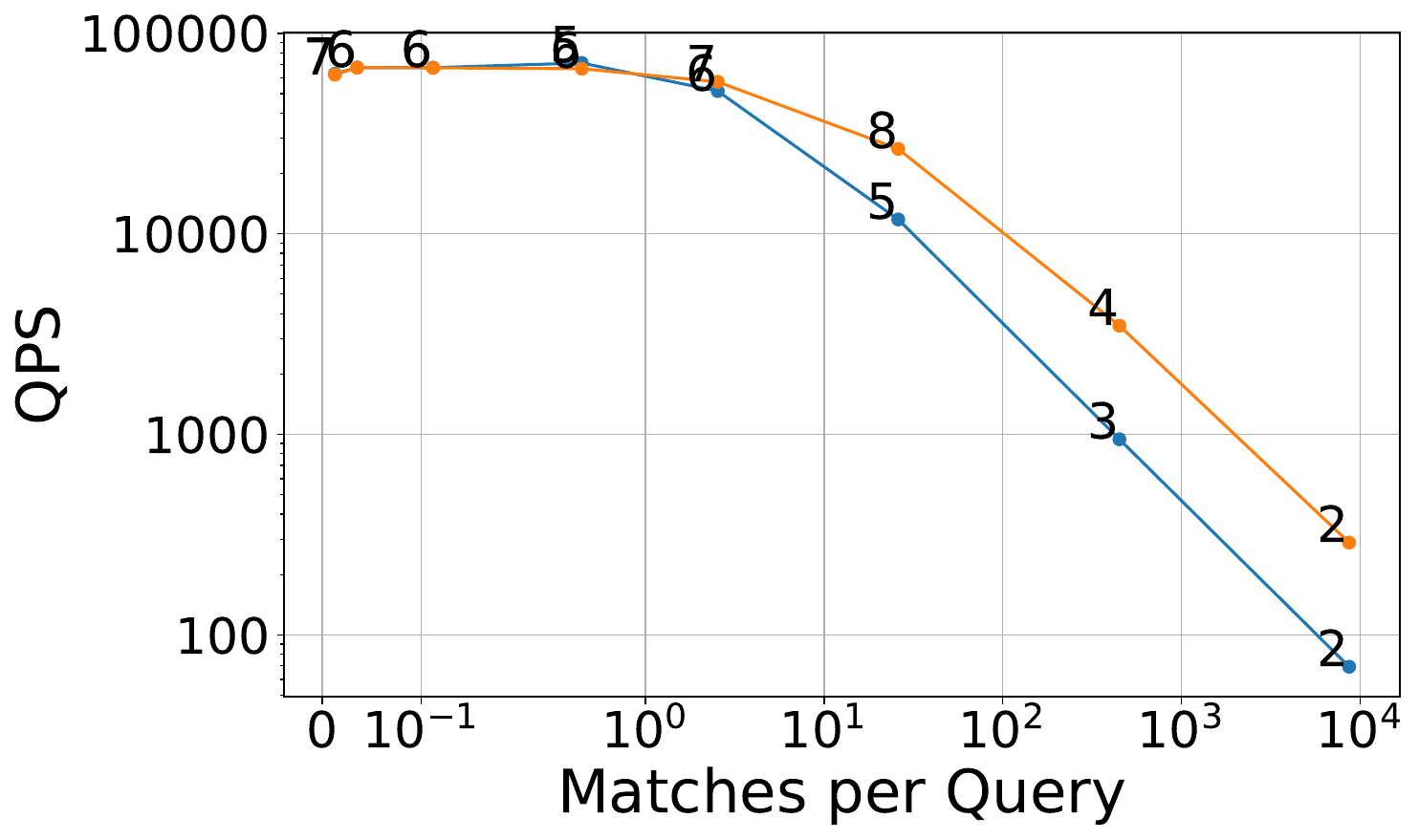} \hfil
	\includegraphics[scale=.22]{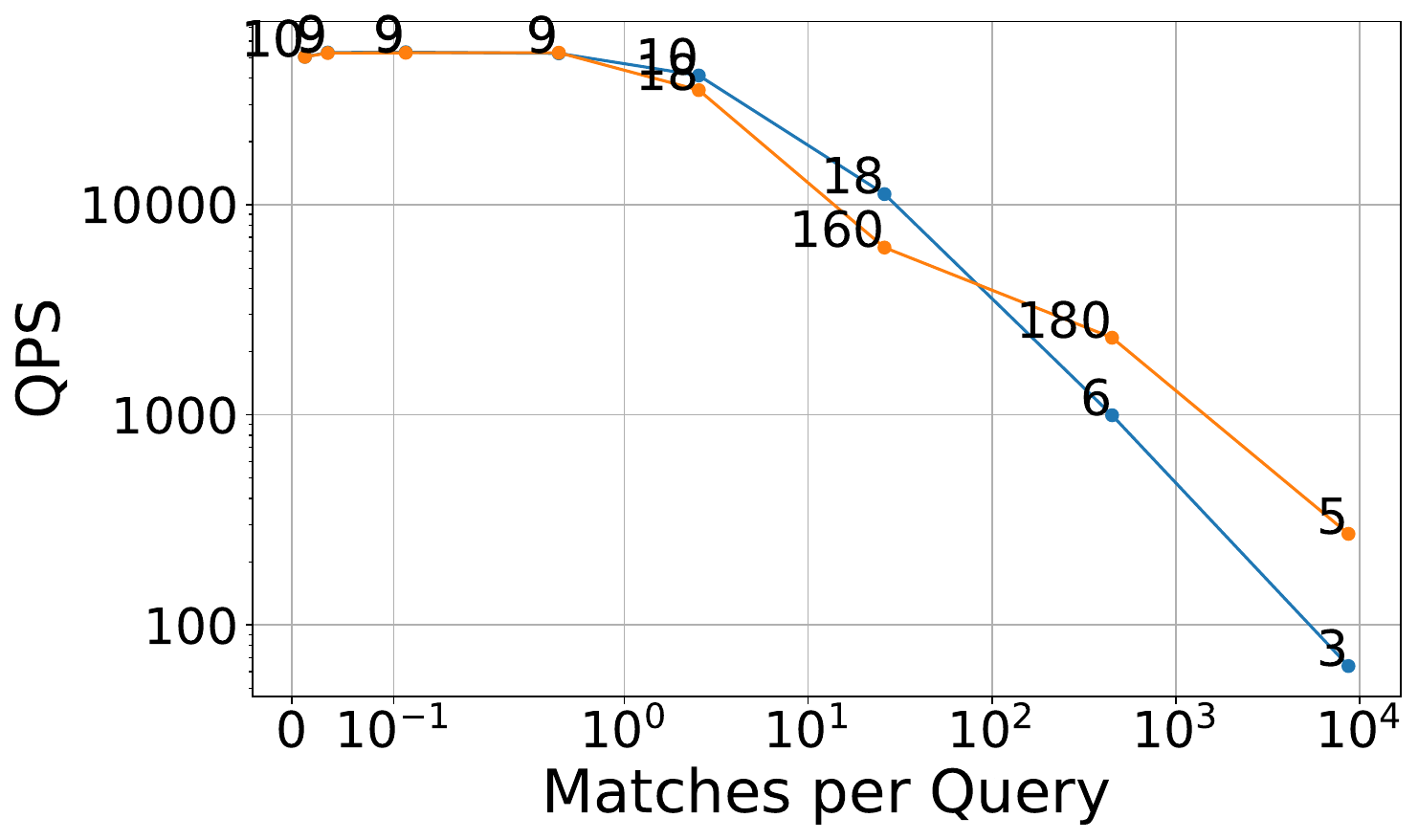} \hfil
	\includegraphics[scale=.22]{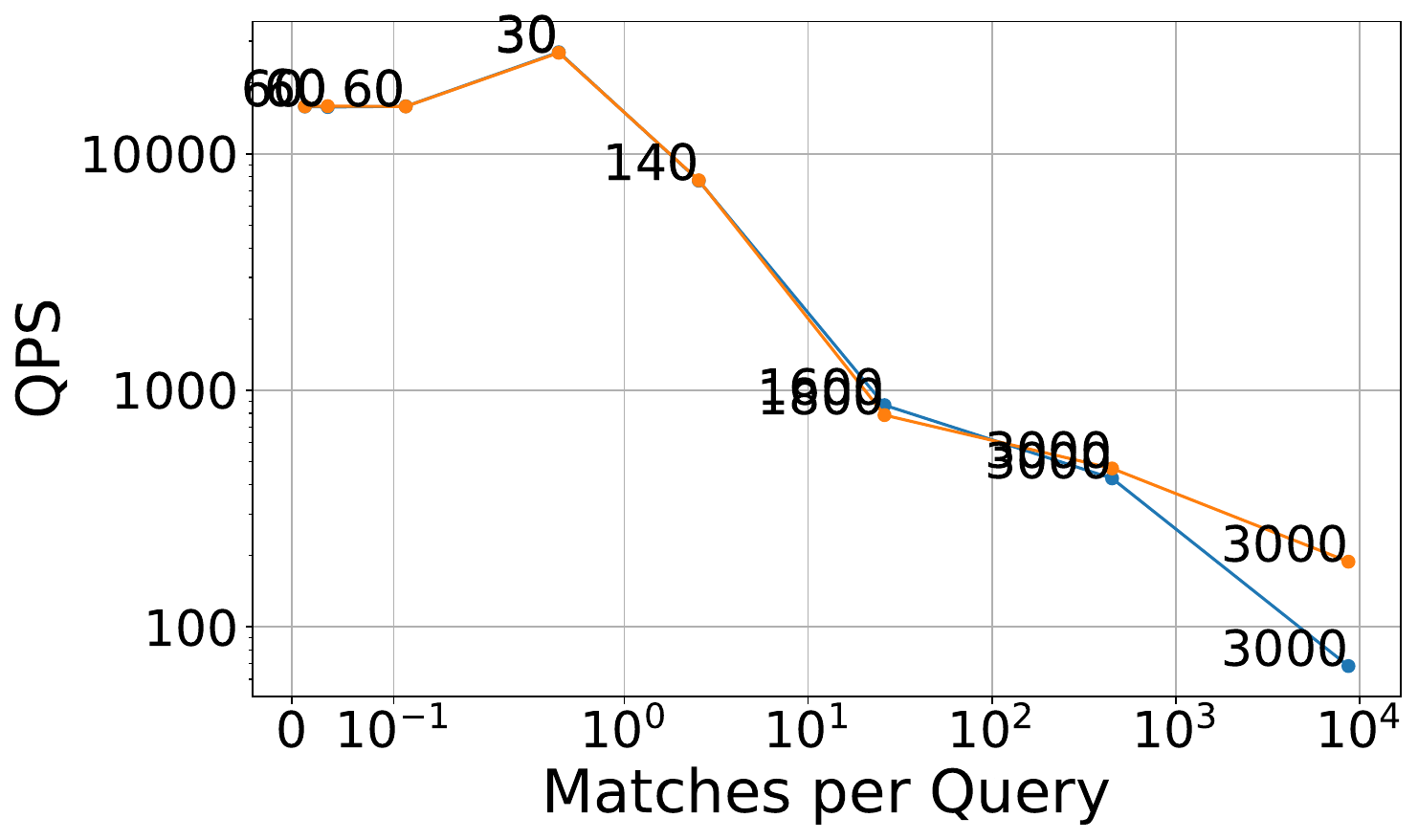}
	\caption{Figures showing the effect of varying the search radius (expressed here as average number of matches per query) on the performance of greedy search and doubling search. Each figure shows the Wikipedia-10M dataset for selected average precision (from left to right: .9, .95, .99). Each data point is annotated with the beam used for the initial beam search. Early stopping is not used for either search type.}\label{fig:qpsmatches}
\end{figure*}

\paragraph{Effects of Increasing Radius} To better isolate the reason for greedy search's superior performance, in Figure~\ref{fig:qpsmatches}, we measure the effect of varying the radius for a fixed dataset for three selected values of average precision, annotating each data point with the beam used for the initial beam search. To aid the reader's understanding, instead of marking the x-axis with the raw value of the radius used, we instead measure the average number of matches per query induced by the chosen radius. We see that in most cases greedy search outperforms or performs equally to doubling search, with the effect becoming more pronounced as the number of matches per point increases. This confirms the observations in the size scaling study.

\paragraph{Effects of Early Stopping} Next, we examine the impact of early stopping. It significantly reduces search time for a fixed beam; for example, in Figure~\ref{fig:bigann100mbar}, early stopping with beam 44 takes only about twice the time as a search with beam 2 and no early stopping. However, as Figure~\ref{fig:msmarcobar} shows, early stopping can reduce accuracy by prematurely terminating points with valid range results. Appendix~\ref{apdx} details its effects on selected datasets. Overall, early stopping is beneficial when there is clear separation between points with no results and those with one or more results.

\paragraph{Comparison with Top-$k$ Search} We add a final note on comparison with top-$k$ search. To investigate this, we ran top-10 searches on OpenAI-1M and MSTuring-1M, and measured the QPS at 90\% recall. For OpenAI-1M, the QPS at 90\% recall was around 5000, compared to around 10000 QPS for range search at 90\% precision. For MSTuring-1M, the QPS at 90\% recall for top-10 search was about 30000, compared to over 100000 for range search using our algorithms. This suggests range search may actually be an easier problem than top-$k$ search.

\section{Conclusion}

We present a set of algorithms for range retrieval on high-dimensional datasets using graph-based indices. To improve the performance of queries with no range results, we introduce an early stopping mechanism capable of significantly decreasing the time taken to terminate searches on such queries. We also introduce two extensions of beam search, greedy search and doubling search, for efficient range queries with hundreds to tens of thousands of results. Combined, our techniques achieved up to 100x speedup over a naive baseline, with 5-10x speedup in almost all cases. We introduce a benchmark of nine datasets, including datasets with up to 100 million points, and demonstrate strong performance and scalability for our algorithms on these datasets. 

Our results yield significant improvement over the naive baseline, and established a wide-ranging benchmark for performance. As we tackle a relatively unexplored problem, many areas for further improvement are still available. One natural extension would be to explore more sophisticated metrics for our early stopping algorithm, perhaps using machine learning methods, as better early termination of queries with few results has potential for significant impact on overall search times. Another would be to explore extensions of the range retrieval benchmark itself; for example, Szilvasy et al.~\cite{szilvasy2024vector} argue that the range search benchmark should differentiate between results which are very close to the query point and points which are close to the radius boundary, giving the former more weight. In preliminary experiments, we attempted to use a simple weighting mechanism based on a point's distance to the query point when calculating precision, but we were unable to find a weighting that changed the precision values significantly. However, future work identifying such mechanisms would be very practical and promising.

Finally, there is a notable lack of benchmarks designed for range searching. Contributions of new datasets with natural range search benchmarks would significantly further this area of exploration.

{
\begin{acks}                            
	
	This work was supported in part by the National Science Foundation
	grants CCF-2119352 and CCF-1919223. 
\end{acks}
}

\clearpage
\bibliography{../bibliography/strings,../bibliography/main}


\begin{thebibliography}{35}


\ifx \showCODEN    \undefined \def \showCODEN     #1{\unskip}     \fi
\ifx \showDOI      \undefined \def \showDOI       #1{#1}\fi
\ifx \showISBNx    \undefined \def \showISBNx     #1{\unskip}     \fi
\ifx \showISBNxiii \undefined \def \showISBNxiii  #1{\unskip}     \fi
\ifx \showISSN     \undefined \def \showISSN      #1{\unskip}     \fi
\ifx \showLCCN     \undefined \def \showLCCN      #1{\unskip}     \fi
\ifx \shownote     \undefined \def \shownote      #1{#1}          \fi
\ifx \showarticletitle \undefined \def \showarticletitle #1{#1}   \fi
\ifx \showURL      \undefined \def \showURL       {\relax}        \fi
\providecommand\bibfield[2]{#2}
\providecommand\bibinfo[2]{#2}
\providecommand\natexlab[1]{#1}
\providecommand\showeprint[2][]{arXiv:#2}

\bibitem[cha({[n.\,d.]})]%
        {chatzakis2025darth}
 \bibinfo{year}{[n.\,d.]}\natexlab{}.
\newblock  (\bibinfo{year}{[n.\,d.]}).
\newblock


\bibitem[par({[n.\,d.]})]%
        {parlay}
 \bibinfo{year}{[n.\,d.]}\natexlab{}.
\newblock \showarticletitle{{ParlayLib}: A Toolkit for Programming Parallel
  Algorithms on Shared-Memory Multicore Machines}.
\newblock
  \bibinfo{journal}{\emph{\url{https://cmuparlay.github.io/parlaylib/}}}
  (\bibinfo{year}{[n.\,d.]}).
\newblock


\bibitem[Amsaleg and Jegou(2010)]%
        {siftgist}
\bibfield{author}{\bibinfo{person}{Laurent Amsaleg} {and}
  \bibinfo{person}{Herve Jegou}.} \bibinfo{year}{2010}\natexlab{}.
\newblock \bibinfo{title}{Datasets for approximate nearest neighbor search}.
\newblock \bibinfo{howpublished}{\url{http://corpus-texmex.irisa.fr/}}.
\newblock


\bibitem[Aum{\"u}ller et~al\mbox{.}(2021)]%
        {annbench}
\bibfield{author}{\bibinfo{person}{Martin Aum{\"u}ller}, \bibinfo{person}{Erik
  Bernhardsson}, {and} \bibinfo{person}{Alexander Faithfull}.}
  \bibinfo{year}{2021}\natexlab{}.
\newblock \bibinfo{title}{{ANN} Benchmarks}.
\newblock \bibinfo{howpublished}{\url{http://ann-benchmarks.com/}}.
\newblock


\bibitem[Baranchuk and Babenko(2021)]%
        {Deep1B}
\bibfield{author}{\bibinfo{person}{Dmitry Baranchuk} {and}
  \bibinfo{person}{Artem Babenko}.} \bibinfo{year}{2021}\natexlab{}.
\newblock \bibinfo{title}{Deep1B Dataset}.
\newblock
  \bibinfo{howpublished}{\url{https://research.yandex.com/blog/benchmarks-for-billion-scale-similarity-search}}.
\newblock


\bibitem[Chazelle et~al\mbox{.}(2008)]%
        {chazelle2008approximate}
\bibfield{author}{\bibinfo{person}{Bernard Chazelle}, \bibinfo{person}{Ding
  Liu}, {and} \bibinfo{person}{Avner Magen}.} \bibinfo{year}{2008}\natexlab{}.
\newblock \showarticletitle{Approximate range searching in higher dimension}.
\newblock \bibinfo{journal}{\emph{Comput. Geom.}} \bibinfo{volume}{39},
  \bibinfo{number}{1} (\bibinfo{year}{2008}), \bibinfo{pages}{24--29}.
\newblock
\urldef\tempurl%
\url{https://doi.org/10.1016/j.comgeo.2007.05.008}
\showURL{%
\tempurl}


\bibitem[Chen et~al\mbox{.}(2024)]%
        {msmarco}
\bibfield{author}{\bibinfo{person}{Qi Chen}, \bibinfo{person}{Xiubo Geng},
  \bibinfo{person}{Corby Rosset}, \bibinfo{person}{Carolyn Buractaon},
  \bibinfo{person}{Jingwen Lu}, \bibinfo{person}{Tao Shen},
  \bibinfo{person}{Kun Zhou}, \bibinfo{person}{Chenyan Xiong},
  \bibinfo{person}{Yeyun Gong}, \bibinfo{person}{Paul Bennett},
  \bibinfo{person}{Nick Craswell}, \bibinfo{person}{Xing Xie},
  \bibinfo{person}{Fan Yang}, \bibinfo{person}{Bryan Tower},
  \bibinfo{person}{Nikhil Rao}, \bibinfo{person}{Anlei Dong},
  \bibinfo{person}{Wenqi Jiang}, \bibinfo{person}{Zheng Liu},
  \bibinfo{person}{Mingqin Li}, \bibinfo{person}{Chuanjie Liu},
  \bibinfo{person}{Zengzhong Li}, \bibinfo{person}{Rangan Majumder},
  \bibinfo{person}{Jennifer Neville}, \bibinfo{person}{Andy Oakley},
  \bibinfo{person}{Knut~Magne Risvik}, \bibinfo{person}{Harsha~Vardhan
  Simhadri}, \bibinfo{person}{Manik Varma}, \bibinfo{person}{Yujing Wang},
  \bibinfo{person}{Linjun Yang}, \bibinfo{person}{Mao Yang}, {and}
  \bibinfo{person}{Ce Zhang}.} \bibinfo{year}{2024}\natexlab{}.
\newblock \showarticletitle{MS MARCO Web Search: A Large-scale Information-rich
  Web Dataset with Millions of Real Click Labels}. In
  \bibinfo{booktitle}{\emph{Companion Proceedings of the ACM on Web Conference
  2024}} \emph{(\bibinfo{series}{WWW ’24})}.
\newblock


\bibitem[Douze(2022)]%
        {FAISS_Wiki}
\bibfield{author}{\bibinfo{person}{Matthijs Douze}.}
  \bibinfo{year}{2022}\natexlab{}.
\newblock \bibinfo{title}{Faiss Wiki}.
\newblock \bibinfo{howpublished}{Webpage}.
\newblock
\urldef\tempurl%
\url{https://github.com/facebookresearch/faiss/wiki}
\showURL{%
Retrieved April 29, 2025 from \tempurl}


\bibitem[Douze et~al\mbox{.}(2024)]%
        {douze2024faiss}
\bibfield{author}{\bibinfo{person}{Matthijs Douze}, \bibinfo{person}{Alexandr
  Guzhva}, \bibinfo{person}{Chengqi Deng}, \bibinfo{person}{Jeff Johnson},
  \bibinfo{person}{Gergely Szilvasy}, \bibinfo{person}{Pierre{-}Emmanuel
  Mazar{\'{e}}}, \bibinfo{person}{Maria Lomeli}, \bibinfo{person}{Lucas
  Hosseini}, {and} \bibinfo{person}{Herv{\'{e}} J{\'{e}}gou}.}
  \bibinfo{year}{2024}\natexlab{}.
\newblock \showarticletitle{The Faiss library}.
\newblock \bibinfo{journal}{\emph{CoRR}} (\bibinfo{year}{2024}).
\newblock
\urldef\tempurl%
\url{https://doi.org/10.48550/arXiv.2401.08281}
\showURL{%
\tempurl}


\bibitem[Douze et~al\mbox{.}(2021)]%
        {douze2021image}
\bibfield{author}{\bibinfo{person}{Matthijs Douze}, \bibinfo{person}{Giorgos
  Tolias}, \bibinfo{person}{Ed Pizzi}, \bibinfo{person}{Zo{\"{e}} Papakipos},
  \bibinfo{person}{Lowik Chanussot}, \bibinfo{person}{Filip Radenovic},
  \bibinfo{person}{Tom{\'{a}}s Jen{\'{\i}}cek}, \bibinfo{person}{Maxim
  Maximov}, \bibinfo{person}{Laura Leal{-}Taix{\'{e}}}, \bibinfo{person}{Ismail
  Elezi}, \bibinfo{person}{Ondrej Chum}, {and} \bibinfo{person}{Cristian
  Canton{-}Ferrer}.} \bibinfo{year}{2021}\natexlab{}.
\newblock \showarticletitle{The 2021 Image Similarity Dataset and Challenge}.
\newblock \bibinfo{journal}{\emph{CoRR}} (\bibinfo{year}{2021}).
\newblock
\urldef\tempurl%
\url{https://arxiv.org/abs/2106.09672}
\showURL{%
\tempurl}


\bibitem[Engels et~al\mbox{.}(2024)]%
        {engels2024approximate}
\bibfield{author}{\bibinfo{person}{Joshua Engels}, \bibinfo{person}{Benjamin
  Landrum}, \bibinfo{person}{Shangdi Yu}, \bibinfo{person}{Laxman Dhulipala},
  {and} \bibinfo{person}{Julian Shun}.} \bibinfo{year}{2024}\natexlab{}.
\newblock \showarticletitle{Approximate Nearest Neighbor Search with Window
  Filters}. In \bibinfo{booktitle}{\emph{Forty-first International Conference
  on Machine Learning, {ICML} 2024, Vienna, Austria, July 21-27, 2024}}.
  \bibinfo{publisher}{OpenReview.net}.
\newblock
\urldef\tempurl%
\url{https://openreview.net/forum?id=8t8zBaGFar}
\showURL{%
\tempurl}


\bibitem[Fu et~al\mbox{.}(2019)]%
        {fu2019nsg}
\bibfield{author}{\bibinfo{person}{Cong Fu}, \bibinfo{person}{Chao Xiang},
  \bibinfo{person}{Changxu Wang}, {and} \bibinfo{person}{Deng Cai}.}
  \bibinfo{year}{2019}\natexlab{}.
\newblock \showarticletitle{Fast Approximate Nearest Neighbor Search With The
  Navigating Spreading-out Graph}.
\newblock \bibinfo{journal}{\emph{Proc. {VLDB} Endow.}} \bibinfo{volume}{12},
  \bibinfo{number}{5} (\bibinfo{year}{2019}), \bibinfo{pages}{461--474}.
\newblock
\urldef\tempurl%
\url{http://www.vldb.org/pvldb/vol12/p461-fu.pdf}
\showURL{%
\tempurl}


\bibitem[Greene et~al\mbox{.}(2022)]%
        {openai}
\bibfield{author}{\bibinfo{person}{Ryan Greene}, \bibinfo{person}{Ted Sanders},
  \bibinfo{person}{Lilian Weng}, {and} \bibinfo{person}{Arvind Neelakantan}.}
  \bibinfo{year}{2022}\natexlab{}.
\newblock \bibinfo{title}{New and improved embedding model}.
\newblock
  \bibinfo{howpublished}{\url{https://openai.com/index/new-and-improved-embedding-model/}}.
\newblock


\bibitem[Halcrow et~al\mbox{.}(2020)]%
        {Grale20}
\bibfield{author}{\bibinfo{person}{Jonathan Halcrow},
  \bibinfo{person}{Alexandru Mosoi}, \bibinfo{person}{Sam Ruth}, {and}
  \bibinfo{person}{Bryan Perozzi}.} \bibinfo{year}{2020}\natexlab{}.
\newblock \showarticletitle{Grale: Designing Networks for Graph Learning}. In
  \bibinfo{booktitle}{\emph{{SIGKDD} Conference on Knowledge Discovery and Data
  Mining (KDD)}}. \bibinfo{pages}{2523–2532}.
\newblock
\showISBNx{9781450379984}
\urldef\tempurl%
\url{https://doi.org/10.1145/3394486.3403302}
\showDOI{\tempurl}


\bibitem[Indyk and Motwani(1998)]%
        {indyk1998towards}
\bibfield{author}{\bibinfo{person}{Piotr Indyk} {and} \bibinfo{person}{Rajeev
  Motwani}.} \bibinfo{year}{1998}\natexlab{}.
\newblock \showarticletitle{Approximate Nearest Neighbors: Towards Removing the
  Curse of Dimensionality}. In \bibinfo{booktitle}{\emph{Proceedings of the
  Thirtieth Annual {ACM} Symposium on the Theory of Computing, Dallas, Texas,
  USA, May 23-26, 1998}}. \bibinfo{publisher}{{ACM}},
  \bibinfo{pages}{604--613}.
\newblock
\urldef\tempurl%
\url{https://doi.org/10.1145/276698.276876}
\showURL{%
\tempurl}


\bibitem[Jegou et~al\mbox{.}(2010)]%
        {jegou2010product}
\bibfield{author}{\bibinfo{person}{Herve Jegou}, \bibinfo{person}{Matthijs
  Douze}, {and} \bibinfo{person}{Cordelia Schmid}.}
  \bibinfo{year}{2010}\natexlab{}.
\newblock \showarticletitle{Product quantization for nearest neighbor search}.
\newblock \bibinfo{journal}{\emph{IEEE Transactions on Pattern Analysis and
  Machine Intelligence (TPAMI)}} \bibinfo{volume}{33}, \bibinfo{number}{1}
  (\bibinfo{year}{2010}).
\newblock


\bibitem[Li et~al\mbox{.}(2020b)]%
        {li2020improving}
\bibfield{author}{\bibinfo{person}{Conglong Li}, \bibinfo{person}{Minjia
  Zhang}, \bibinfo{person}{David~G. Andersen}, {and} \bibinfo{person}{Yuxiong
  He}.} \bibinfo{year}{2020}\natexlab{b}.
\newblock \showarticletitle{Improving Approximate Nearest Neighbor Search
  through Learned Adaptive Early Termination}. In
  \bibinfo{booktitle}{\emph{Proceedings of the 2020 International Conference on
  Management of Data, {SIGMOD} Conference 2020, online conference [Portland,
  OR, USA], June 14-19, 2020}}. \bibinfo{publisher}{{ACM}},
  \bibinfo{pages}{2539--2554}.
\newblock
\urldef\tempurl%
\url{https://doi.org/10.1145/3318464.3380600}
\showURL{%
\tempurl}


\bibitem[Li et~al\mbox{.}(2020a)]%
        {li2020density}
\bibfield{author}{\bibinfo{person}{Hao Li}, \bibinfo{person}{Xiaojie Liu},
  \bibinfo{person}{Tao Li}, {and} \bibinfo{person}{Rundong Gan}.}
  \bibinfo{year}{2020}\natexlab{a}.
\newblock \showarticletitle{A novel density-based clustering algorithm using
  nearest neighbor graph}.
\newblock \bibinfo{journal}{\emph{Pattern Recognit.}}  \bibinfo{volume}{102}
  (\bibinfo{year}{2020}), \bibinfo{pages}{107206}.
\newblock
\urldef\tempurl%
\url{https://doi.org/10.1016/j.patcog.2020.107206}
\showURL{%
\tempurl}


\bibitem[Malkov and Yashunin(2018)]%
        {malkov2018efficient}
\bibfield{author}{\bibinfo{person}{Yury Malkov} {and} \bibinfo{person}{Dmitry
  Yashunin}.} \bibinfo{year}{2018}\natexlab{}.
\newblock \showarticletitle{Efficient and robust approximate nearest neighbor
  search using hierarchical navigable small world graphs}.
\newblock \bibinfo{journal}{\emph{IEEE Transactions on Pattern Analysis and
  Machine Intelligence}} \bibinfo{volume}{42}, \bibinfo{number}{4}
  (\bibinfo{year}{2018}).
\newblock


\bibitem[Manohar et~al\mbox{.}(2024)]%
        {manohar2024parlayann}
\bibfield{author}{\bibinfo{person}{Magdalen~Dobson Manohar},
  \bibinfo{person}{Zheqi Shen}, \bibinfo{person}{Guy~E. Blelloch},
  \bibinfo{person}{Laxman Dhulipala}, \bibinfo{person}{Yan Gu},
  \bibinfo{person}{Harsha~Vardhan Simhadri}, {and} \bibinfo{person}{Yihan
  Sun}.} \bibinfo{year}{2024}\natexlab{}.
\newblock \showarticletitle{ParlayANN: Scalable and Deterministic Parallel
  Graph-Based Approximate Nearest Neighbor Search Algorithms}. In
  \bibinfo{booktitle}{\emph{Proceedings of the 29th {ACM} {SIGPLAN} Annual
  Symposium on Principles and Practice of Parallel Programming, PPoPP 2024,
  Edinburgh, United Kingdom, March 2-6, 2024}}. \bibinfo{publisher}{{ACM}},
  \bibinfo{pages}{270--285}.
\newblock
\urldef\tempurl%
\url{https://doi.org/10.1145/3627535.3638475}
\showURL{%
\tempurl}


\bibitem[Manohar et~al\mbox{.}(2025)]%
        {parlayann}
\bibfield{author}{\bibinfo{person}{Magdalen~Dobson Manohar},
  \bibinfo{person}{Zheqi Shen}, \bibinfo{person}{Guy~E. Blelloch},
  \bibinfo{person}{Laxman Dhulipala}, \bibinfo{person}{Yan Gu},
  \bibinfo{person}{Harsha~Vardhan Simhadri}, {and} \bibinfo{person}{Yihan
  Sun}.} \bibinfo{year}{2025}\natexlab{}.
\newblock \bibinfo{title}{ParlayANN}.
\newblock \bibinfo{howpublished}{Webpage}.
\newblock
\urldef\tempurl%
\url{https://github.com/cmuparlay/ParlayANN}
\showURL{%
Retrieved February 7, 2025 from \tempurl}


\bibitem[MetaAI(2020)]%
        {simsearchnet}
\bibfield{author}{\bibinfo{person}{MetaAI}.} \bibinfo{year}{2020}\natexlab{}.
\newblock \bibinfo{title}{Using AI to detect COVID-19 misinformation and
  exploitative content}.
\newblock \bibinfo{howpublished}{Webpage}.
\newblock
\urldef\tempurl%
\url{https://ai.facebook.com/blog/using-ai-to-detect-covid-19-misinformation-and-exploitative-content/}
\showURL{%
Retrieved January 26, 2022 from \tempurl}


\bibitem[Reimers(2022)]%
        {wikipedia}
\bibfield{author}{\bibinfo{person}{Nils Reimers}.}
  \bibinfo{year}{2022}\natexlab{}.
\newblock \bibinfo{title}{Datasets: Cohere/wikipedia-22-12-en-embeddings}.
\newblock
  \bibinfo{howpublished}{\url{https://huggingface.co/datasets/Cohere/wikipedia-22-12-en-embeddings}}.
\newblock


\bibitem[Schroff et~al\mbox{.}(2015)]%
        {schroff2015facenet}
\bibfield{author}{\bibinfo{person}{Florian Schroff}, \bibinfo{person}{Dmitry
  Kalenichenko}, {and} \bibinfo{person}{James Philbin}.}
  \bibinfo{year}{2015}\natexlab{}.
\newblock \showarticletitle{FaceNet: {A} unified embedding for face recognition
  and clustering}. In \bibinfo{booktitle}{\emph{{IEEE} Conference on Computer
  Vision and Pattern Recognition, {CVPR} 2015, Boston, MA, USA, June 7-12,
  2015}}. \bibinfo{publisher}{{IEEE} Computer Society},
  \bibinfo{pages}{815--823}.
\newblock
\urldef\tempurl%
\url{https://doi.org/10.1109/CVPR.2015.7298682}
\showURL{%
\tempurl}


\bibitem[Simhadri et~al\mbox{.}(2021b)]%
        {bigann}
\bibfield{author}{\bibinfo{person}{Harsha Simhadri}, \bibinfo{person}{George
  Williams}, \bibinfo{person}{Martin Aumüller}, \bibinfo{person}{Artem
  Babenko}, \bibinfo{person}{Dmitry Baranchuk}, \bibinfo{person}{Qi Chen},
  \bibinfo{person}{Matthijs Douze}, \bibinfo{person}{Lucas Hosseini},
  \bibinfo{person}{Ravishankar Krishnaswamy}, \bibinfo{person}{Gopal
  Srinivasa}, \bibinfo{person}{Suhas~Jayaram Subramanya}, {and}
  \bibinfo{person}{Jingdong Wang}.} \bibinfo{year}{2021}\natexlab{b}.
\newblock \bibinfo{title}{{BigANN} Benchmarks: Billion-Scale Approximate
  Nearest Neighbor Search Challenge}.
\newblock \bibinfo{howpublished}{\url{https://big-ann-benchmarks.com/}}.
\newblock


\bibitem[Simhadri et~al\mbox{.}(2021a)]%
        {simhadri2021results}
\bibfield{author}{\bibinfo{person}{Harsha~Vardhan Simhadri},
  \bibinfo{person}{George Williams}, \bibinfo{person}{Martin Aum{\"{u}}ller},
  \bibinfo{person}{Matthijs Douze}, \bibinfo{person}{Artem Babenko},
  \bibinfo{person}{Dmitry Baranchuk}, \bibinfo{person}{Qi Chen},
  \bibinfo{person}{Lucas Hosseini}, \bibinfo{person}{Ravishankar Krishnaswamy},
  \bibinfo{person}{Gopal Srinivasa}, \bibinfo{person}{Suhas~Jayaram
  Subramanya}, {and} \bibinfo{person}{Jingdong Wang}.}
  \bibinfo{year}{2021}\natexlab{a}.
\newblock \showarticletitle{Results of the NeurIPS'21 Challenge on
  Billion-Scale Approximate Nearest Neighbor Search}. In
  \bibinfo{booktitle}{\emph{NeurIPS 2021 Competitions and Demonstrations Track,
  6-14 December 2021, Online}} \emph{(\bibinfo{series}{Proceedings of Machine
  Learning Research}, Vol.~\bibinfo{volume}{176})}.
  \bibinfo{publisher}{{PMLR}}, \bibinfo{pages}{177--189}.
\newblock
\urldef\tempurl%
\url{https://proceedings.mlr.press/v176/simhadri22a.html}
\showURL{%
\tempurl}


\bibitem[Subramanya et~al\mbox{.}(2019)]%
        {subramanya2019diskann}
\bibfield{author}{\bibinfo{person}{Suhas~Jayaram Subramanya},
  \bibinfo{person}{Devvrit}, \bibinfo{person}{Harsha~Vardhan Simhadri},
  \bibinfo{person}{Ravishankar Krishnaswamy}, {and} \bibinfo{person}{Rohan
  Kadekodi}.} \bibinfo{year}{2019}\natexlab{}.
\newblock \showarticletitle{Rand-NSG: Fast Accurate Billion-point Nearest
  Neighbor Search on a Single Node}. In \bibinfo{booktitle}{\emph{Advances in
  Neural Information Processing Systems 32: Annual Conference on Neural
  Information Processing Systems 2019, NeurIPS 2019, December 8-14, 2019,
  Vancouver, BC, Canada}}. \bibinfo{pages}{13748--13758}.
\newblock
\urldef\tempurl%
\url{https://proceedings.neurips.cc/paper/2019/hash/09853c7fb1d3f8ee67a61b6bf4a7f8e6-Abstract.html}
\showURL{%
\tempurl}


\bibitem[Szilvasy et~al\mbox{.}(2024)]%
        {szilvasy2024vector}
\bibfield{author}{\bibinfo{person}{Gergely Szilvasy},
  \bibinfo{person}{Pierre{-}Emmanuel Mazar{\'{e}}}, {and}
  \bibinfo{person}{Matthijs Douze}.} \bibinfo{year}{2024}\natexlab{}.
\newblock \showarticletitle{Vector search with small radiuses}.
\newblock \bibinfo{journal}{\emph{CoRR}}  \bibinfo{volume}{abs/2403.10746}
  (\bibinfo{year}{2024}).
\newblock
\urldef\tempurl%
\url{https://doi.org/10.48550/arXiv.2403.10746}
\showURL{%
\tempurl}


\bibitem[Wang et~al\mbox{.}(2013)]%
        {wang2013pltree}
\bibfield{author}{\bibinfo{person}{Jie Wang}, \bibinfo{person}{Jian Lu},
  \bibinfo{person}{Zheng Fang}, \bibinfo{person}{Tingjian Ge}, {and}
  \bibinfo{person}{Cindy~X. Chen}.} \bibinfo{year}{2013}\natexlab{}.
\newblock \showarticletitle{PL-Tree: An Efficient Indexing Method for
  High-Dimensional Data}. In \bibinfo{booktitle}{\emph{Advances in Spatial and
  Temporal Databases - 13th International Symposium, {SSTD} 2013, Munich,
  Germany, August 21-23, 2013. Proceedings}} \emph{(\bibinfo{series}{Lecture
  Notes in Computer Science}, Vol.~\bibinfo{volume}{8098})}.
  \bibinfo{publisher}{Springer}, \bibinfo{pages}{183--200}.
\newblock
\urldef\tempurl%
\url{https://doi.org/10.1007/978-3-642-40235-7\_11}
\showURL{%
\tempurl}


\bibitem[Wang et~al\mbox{.}(2024)]%
        {wang2024starling}
\bibfield{author}{\bibinfo{person}{Mengzhao Wang}, \bibinfo{person}{Weizhi Xu},
  \bibinfo{person}{Xiaomeng Yi}, \bibinfo{person}{Songlin Wu},
  \bibinfo{person}{Zhangyang Peng}, \bibinfo{person}{Xiangyu Ke},
  \bibinfo{person}{Yunjun Gao}, \bibinfo{person}{Xiaoliang Xu},
  \bibinfo{person}{Rentong Guo}, {and} \bibinfo{person}{Charles Xie}.}
  \bibinfo{year}{2024}\natexlab{}.
\newblock \showarticletitle{Starling: An I/O-Efficient Disk-Resident Graph
  Index Framework for High-Dimensional Vector Similarity Search on Data
  Segment}.
\newblock \bibinfo{journal}{\emph{Proc. {ACM} Manag. Data}}
  \bibinfo{volume}{2}, \bibinfo{number}{1} (\bibinfo{year}{2024}),
  \bibinfo{pages}{V2mod014:1--V2mod014:27}.
\newblock
\urldef\tempurl%
\url{https://doi.org/10.1145/3639269}
\showURL{%
\tempurl}


\bibitem[Wang et~al\mbox{.}(2021)]%
        {wang2021comprehensive}
\bibfield{author}{\bibinfo{person}{Mengzhao Wang}, \bibinfo{person}{Xiaoliang
  Xu}, \bibinfo{person}{Qiang Yue}, {and} \bibinfo{person}{Yuxiang Wang}.}
  \bibinfo{year}{2021}\natexlab{}.
\newblock \showarticletitle{A Comprehensive Survey and Experimental Comparison
  of Graph-Based Approximate Nearest Neighbor Search}.
\newblock \bibinfo{journal}{\emph{Proc. {VLDB} Endow.}} \bibinfo{volume}{14},
  \bibinfo{number}{11} (\bibinfo{year}{2021}), \bibinfo{pages}{1964--1978}.
\newblock
\urldef\tempurl%
\url{http://www.vldb.org/pvldb/vol14/p1964-wang.pdf}
\showURL{%
\tempurl}


\bibitem[Xu et~al\mbox{.}(2021)]%
        {xu2021twostage}
\bibfield{author}{\bibinfo{person}{Xiaoliang Xu}, \bibinfo{person}{Mengzhao
  Wang}, \bibinfo{person}{Yuxiang Wang}, {and} \bibinfo{person}{Dingcheng Ma}.}
  \bibinfo{year}{2021}\natexlab{}.
\newblock \showarticletitle{Two-stage routing with optimized guided search and
  greedy algorithm on proximity graph}.
\newblock \bibinfo{journal}{\emph{Knowl. Based Syst.}}  \bibinfo{volume}{229}
  (\bibinfo{year}{2021}), \bibinfo{pages}{107305}.
\newblock
\urldef\tempurl%
\url{https://doi.org/10.1016/j.knosys.2021.107305}
\showURL{%
\tempurl}


\bibitem[Xu et~al\mbox{.}(2024)]%
        {xu2024irangegraph}
\bibfield{author}{\bibinfo{person}{Yuexuan Xu}, \bibinfo{person}{Jianyang Gao},
  \bibinfo{person}{Yutong Gou}, \bibinfo{person}{Cheng Long}, {and}
  \bibinfo{person}{Christian~S. Jensen}.} \bibinfo{year}{2024}\natexlab{}.
\newblock \showarticletitle{iRangeGraph: Improvising Range-dedicated Graphs for
  Range-filtering Nearest Neighbor Search}.
\newblock \bibinfo{journal}{\emph{Proc. {ACM} Manag. Data}}
  \bibinfo{volume}{2}, \bibinfo{number}{6} (\bibinfo{year}{2024}),
  \bibinfo{pages}{239:1--239:26}.
\newblock
\urldef\tempurl%
\url{https://doi.org/10.1145/3698814}
\showURL{%
\tempurl}


\bibitem[Zhang et~al\mbox{.}(2019)]%
        {msturing}
\bibfield{author}{\bibinfo{person}{Hongfei Zhang}, \bibinfo{person}{Xia Song},
  \bibinfo{person}{Chenyan Xiong}, \bibinfo{person}{Corby Rosset},
  \bibinfo{person}{Paul~N. Bennett}, \bibinfo{person}{Nick Craswell}, {and}
  \bibinfo{person}{Saurabh Tiwary}.} \bibinfo{year}{2019}\natexlab{}.
\newblock \showarticletitle{Generic Intent Representation in Web Search}. In
  \bibinfo{booktitle}{\emph{Proceedings of the 42nd International {ACM} {SIGIR}
  Conference on Research and Development in Information Retrieval, {SIGIR}
  2019, Paris, France, July 21-25, 2019}}. \bibinfo{publisher}{{ACM}},
  \bibinfo{pages}{65--74}.
\newblock
\urldef\tempurl%
\url{https://doi.org/10.1145/3331184.3331198}
\showURL{%
\tempurl}


\bibitem[Zuo et~al\mbox{.}(2024)]%
        {zuo2024serf}
\bibfield{author}{\bibinfo{person}{Chaoji Zuo}, \bibinfo{person}{Miao Qiao},
  \bibinfo{person}{Wenchao Zhou}, \bibinfo{person}{Feifei Li}, {and}
  \bibinfo{person}{Dong Deng}.} \bibinfo{year}{2024}\natexlab{}.
\newblock \showarticletitle{SeRF: Segment Graph for Range-Filtering Approximate
  Nearest Neighbor Search}.
\newblock \bibinfo{journal}{\emph{Proc. {ACM} Manag. Data}}
  \bibinfo{volume}{2}, \bibinfo{number}{1} (\bibinfo{year}{2024}),
  \bibinfo{pages}{69:1--69:26}.
\newblock
\urldef\tempurl%
\url{https://doi.org/10.1145/3639324}
\showURL{%
\tempurl}


\end{thebibliography}

\unless\ifx\fullversion\undefined

\clearpage
\appendix

\section{Supplementary Material}~\label{apdx}

\subsection{Additional Data on Early Stopping}

We provide data on early stopping metrics for four datasets in Figures~\ref{fig:bigann_earlystop}, \ref{fig:deep_earlystop}, \ref{fig:wikipedia_earlystop},  and~\ref{fig:ssnpp_earlystop}. In each figure, we histogram the values of each metric, first with all results included, and then with only points where the search has not already found a point inside the radius. The figure also shows vertical lines indicating a potential cutoff for Algorithm~\ref{alg:earlystopping}; points with more than one range result to the right of the line would be ``incorrectly" cut off from further search, while points with zero points to the right of the line would be ``incorrectly" allowed to proceed instead of terminating early. The goal is to identify a cutoff that yields a more favorable QPS/recall curve. 


\begin{figure*}
	\centering
	\includegraphics[scale=.35]{figures/dist_histograms/visited/bigann_test_legend.pdf} \\
	\begin{subfigure}{.24\textwidth}
		\includegraphics[scale=.24]{figures/dist_histograms/visited/Step20/bigann-1M.pdf}
	\end{subfigure}
	\begin{subfigure}{.24\textwidth}
		\includegraphics[scale=.24]{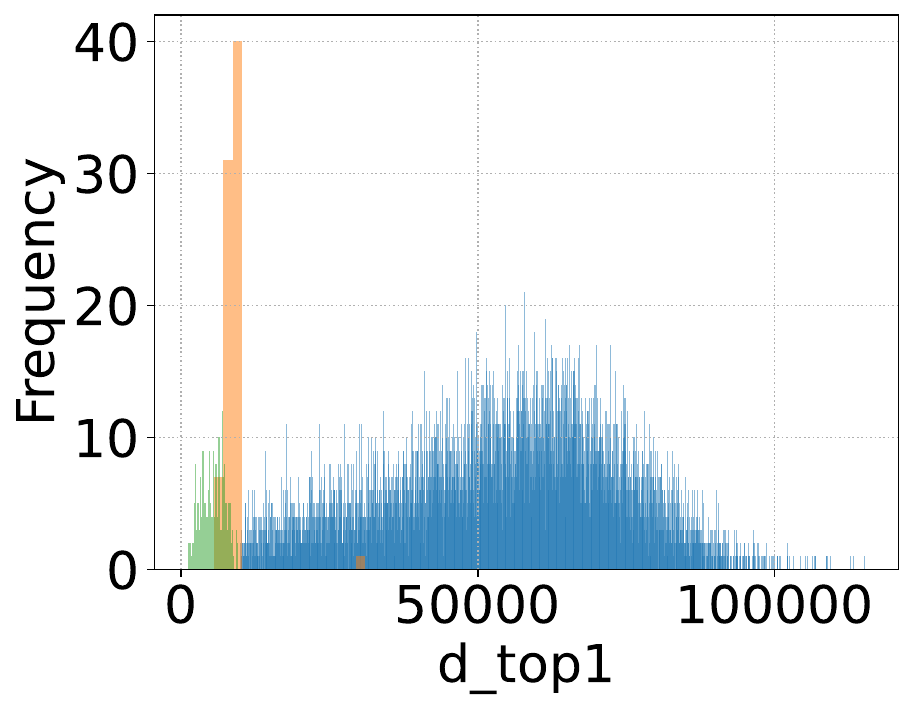}
	\end{subfigure}
	\begin{subfigure}{.24\textwidth}
		\includegraphics[scale=.24]{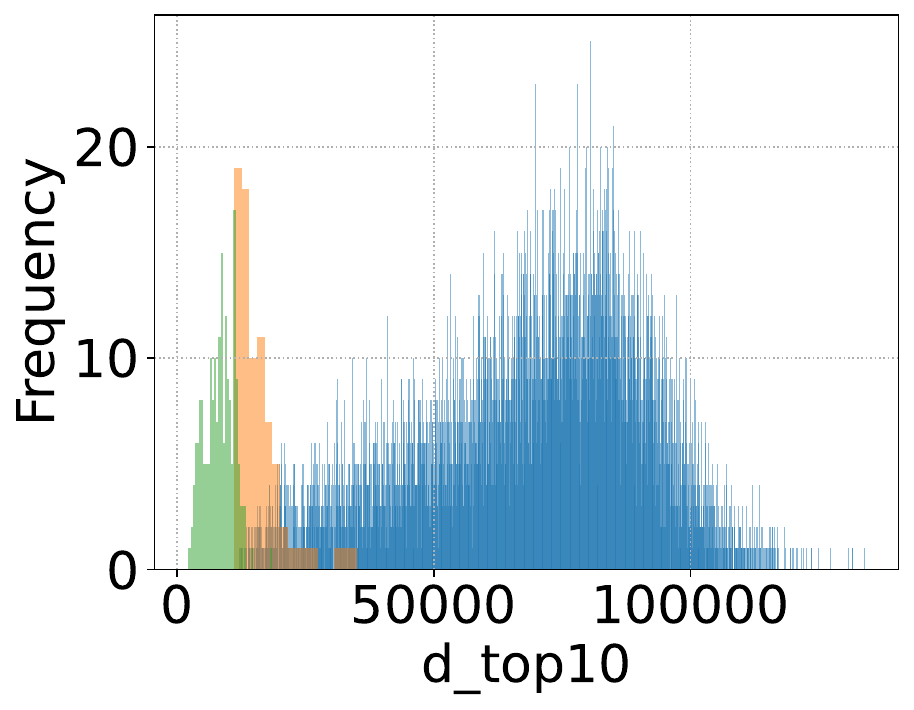}
	\end{subfigure}
	\begin{subfigure}{.24\textwidth}
		\includegraphics[scale=.24]{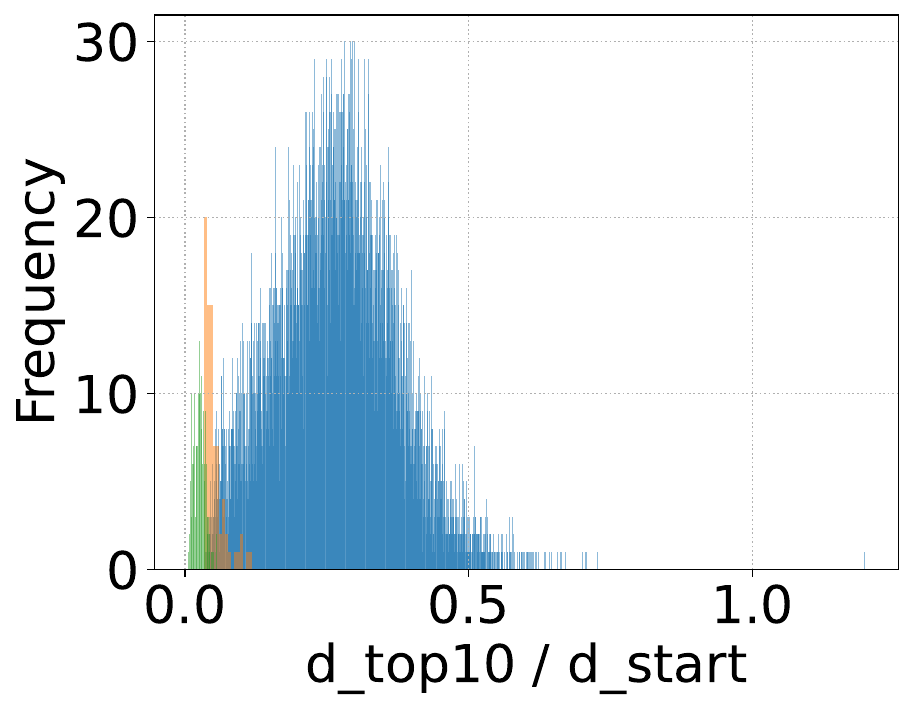}
	\end{subfigure}
	\begin{subfigure}{.24\textwidth}
		\includegraphics[scale=.24]{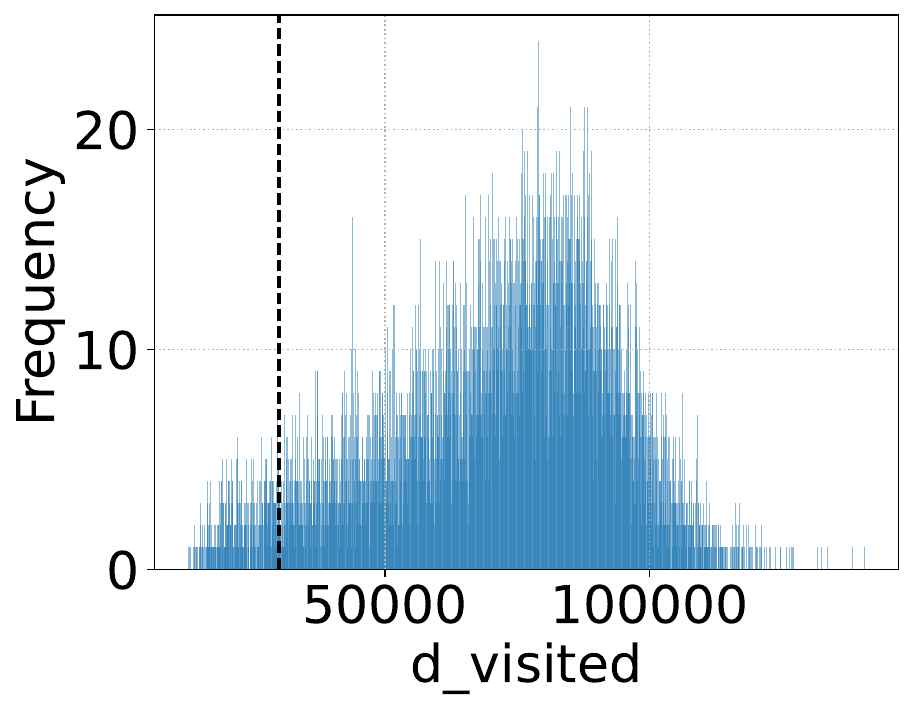}
	\end{subfigure}
	\begin{subfigure}{.24\textwidth}
		\includegraphics[scale=.24]{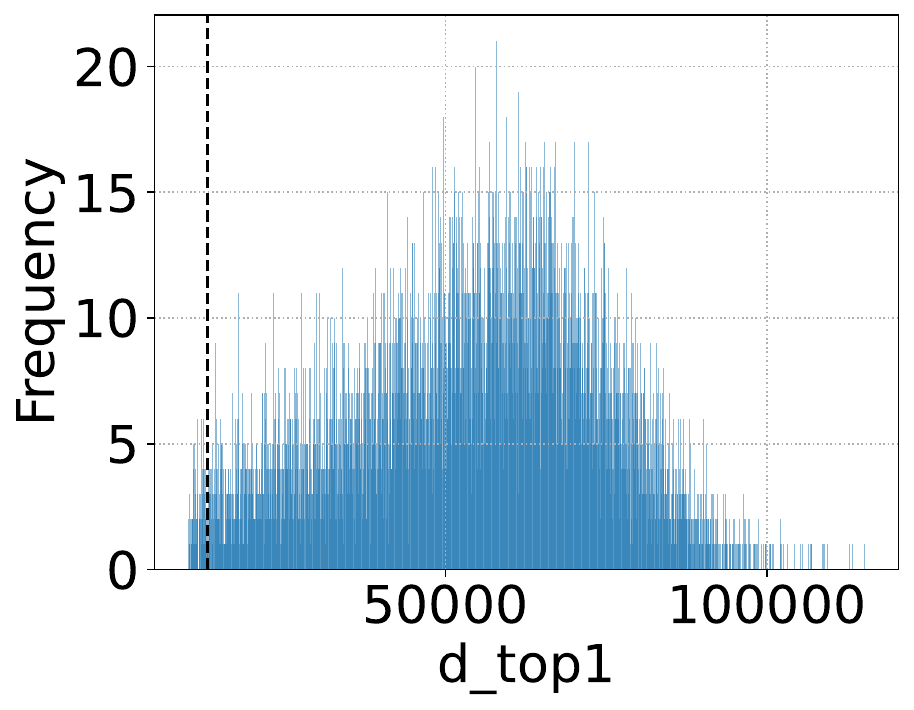}
	\end{subfigure}
	\begin{subfigure}{.24\textwidth}
		\includegraphics[scale=.24]{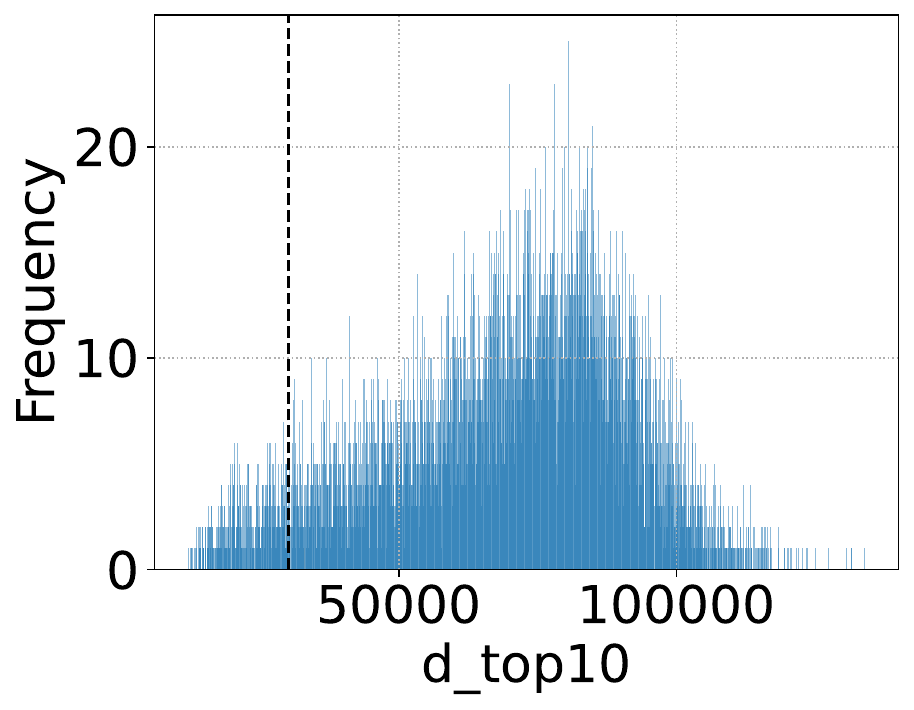}
	\end{subfigure}
	\begin{subfigure}{.24\textwidth}
		\includegraphics[scale=.24]{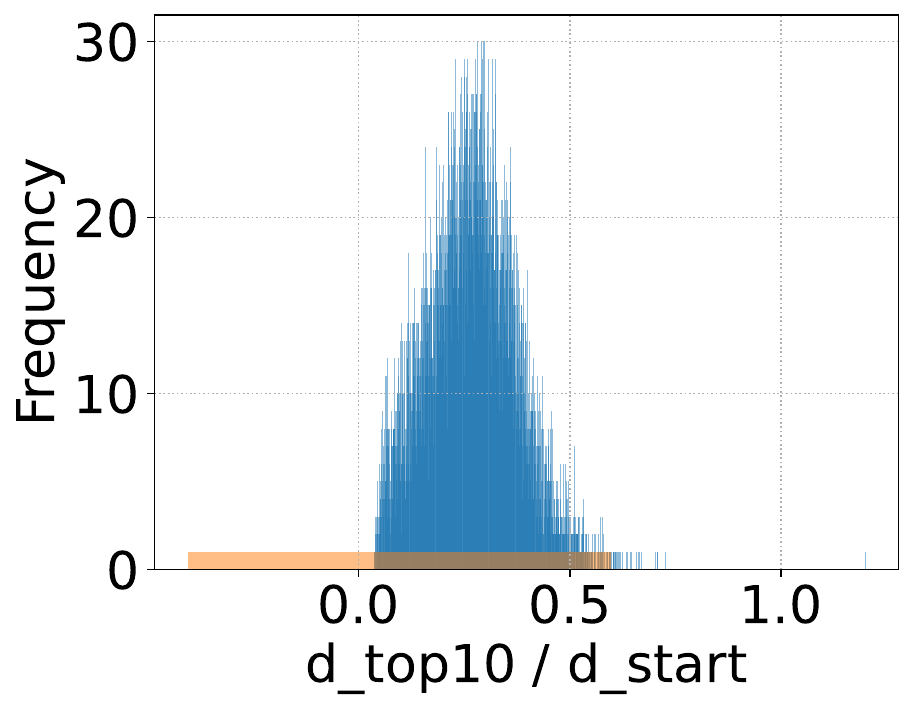}
	\end{subfigure}
	\caption{Early stopping metrics for BIGANN-1M, taken at step 20 of a beam search with beam 100. The top row shows all results, while the bottom row shows only results for beam searches that have not yet found a candidate within the radius.}
	\label{fig:bigann_earlystop}
\end{figure*}

\begin{figure*}
	\centering
	\includegraphics[scale=.35]{figures/dist_histograms/visited/bigann_test_legend.pdf} \\
	\begin{subfigure}{.24\textwidth}
		\includegraphics[scale=.24]{figures/dist_histograms/visited/Step20/deep-1M.pdf}
	\end{subfigure}
	\begin{subfigure}{.24\textwidth}
		\includegraphics[scale=.24]{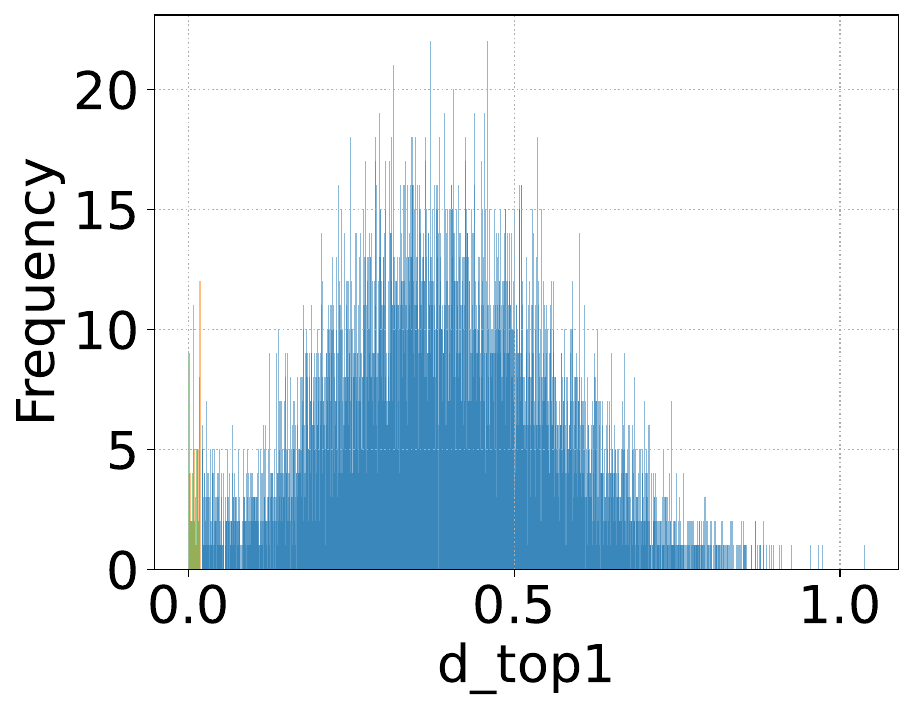}
	\end{subfigure}
	\begin{subfigure}{.24\textwidth}
		\includegraphics[scale=.24]{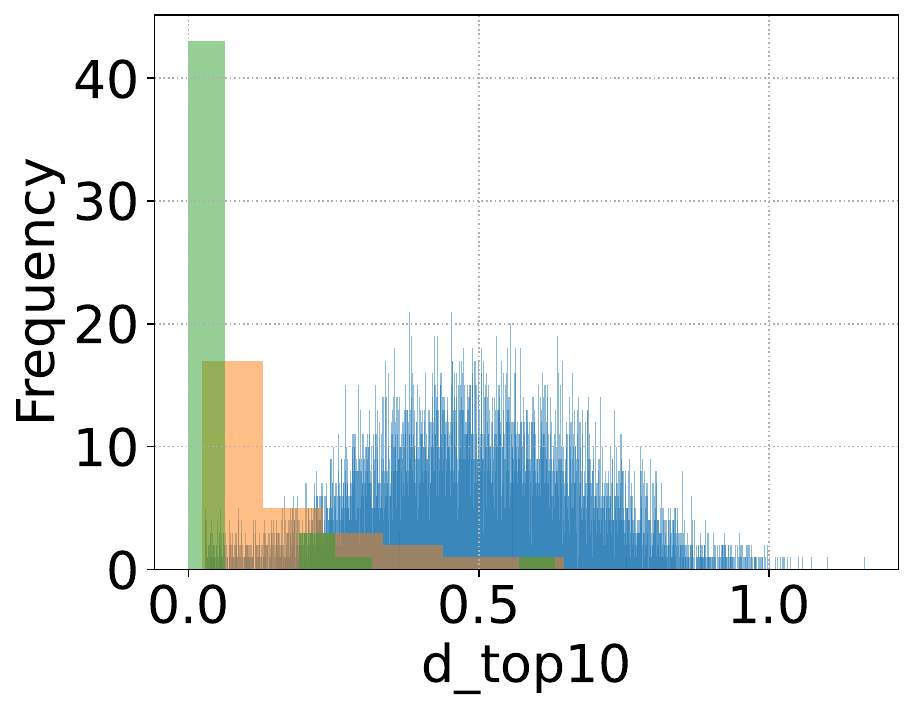}
	\end{subfigure}
	\begin{subfigure}{.24\textwidth}
		\includegraphics[scale=.24]{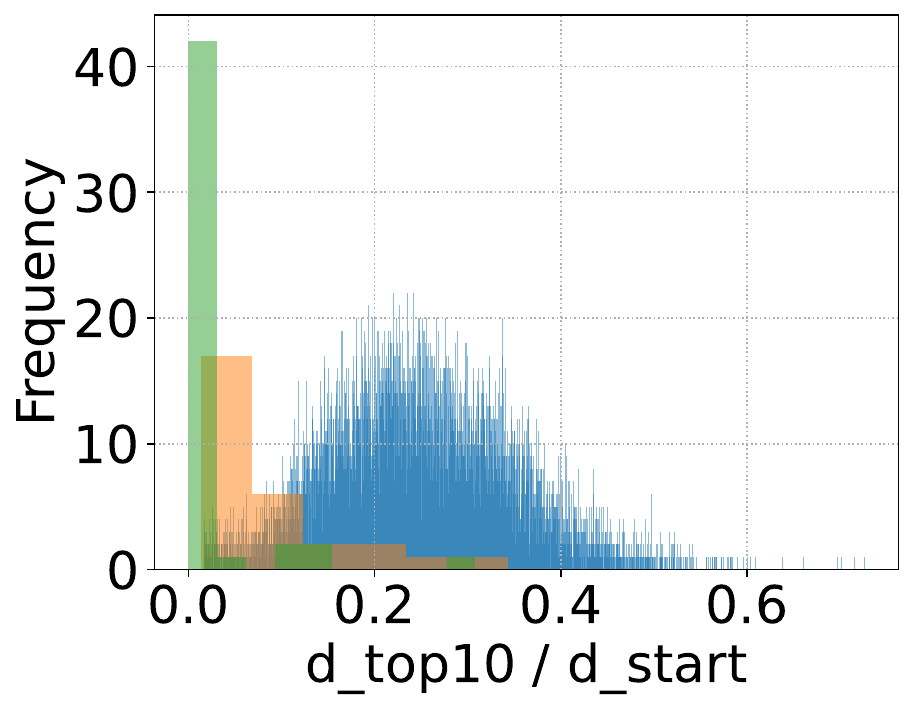}
	\end{subfigure}
	\begin{subfigure}{.24\textwidth}
		\includegraphics[scale=.24]{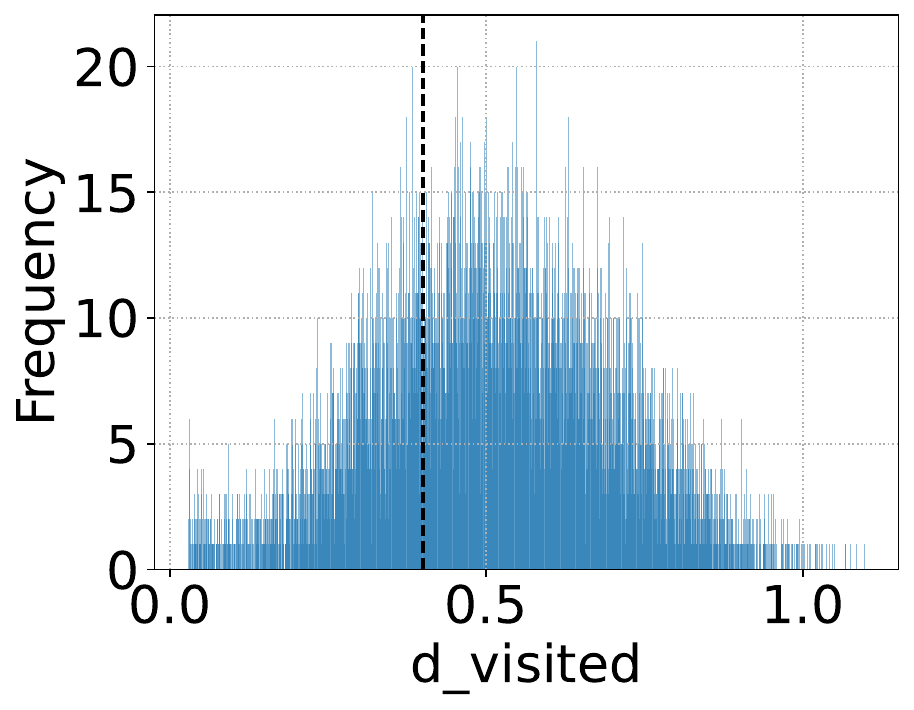}
	\end{subfigure}
	\begin{subfigure}{.24\textwidth}
		\includegraphics[scale=.24]{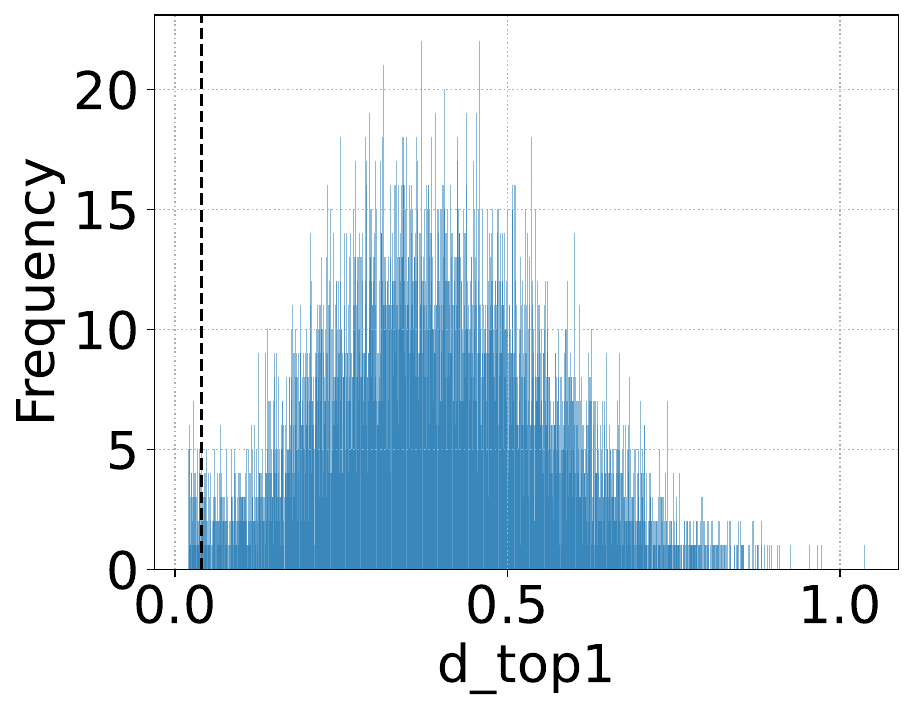}
	\end{subfigure}
	\begin{subfigure}{.24\textwidth}
		\includegraphics[scale=.24]{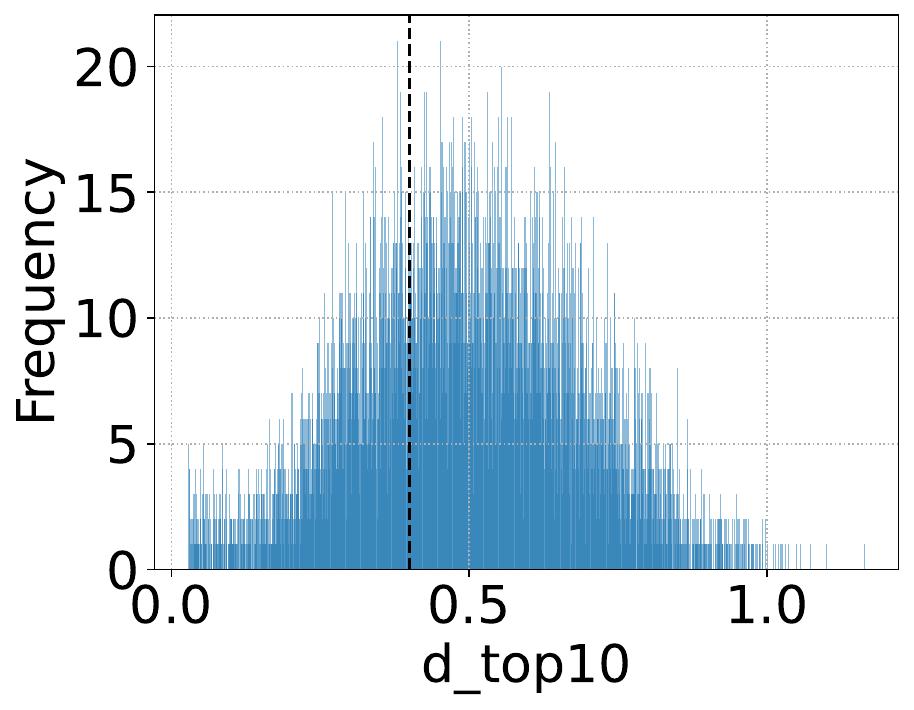}
	\end{subfigure}
	\begin{subfigure}{.24\textwidth}
		\includegraphics[scale=.24]{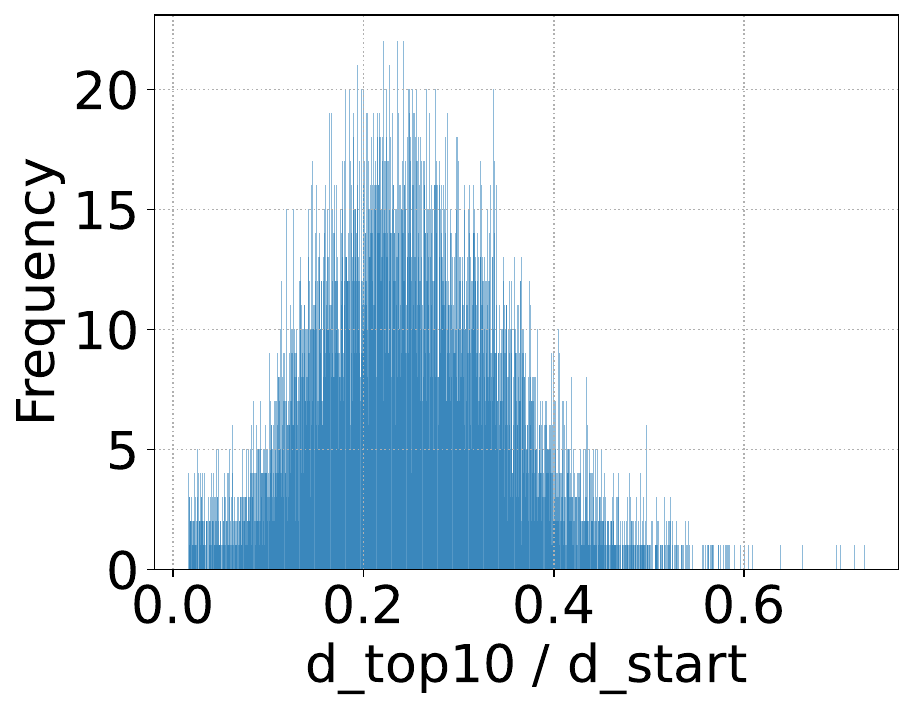}
	\end{subfigure}
	\caption{Early stopping metrics for DEEP-1M, taken at step 20 of a beam search with beam 100. The top row shows all results, while the bottom row shows only results for beam searches that have not yet found a candidate within the radius.}
	\label{fig:deep_earlystop}
\end{figure*}

\begin{figure*}
	\centering
	\includegraphics[scale=.35]{figures/dist_histograms/visited/bigann_test_legend.pdf} \\
	\begin{subfigure}{.24\textwidth}
		\includegraphics[scale=.24]{figures/dist_histograms/visited/Step20/wikipedia-1M.pdf}
	\end{subfigure}
	\begin{subfigure}{.24\textwidth}
		\includegraphics[scale=.24]{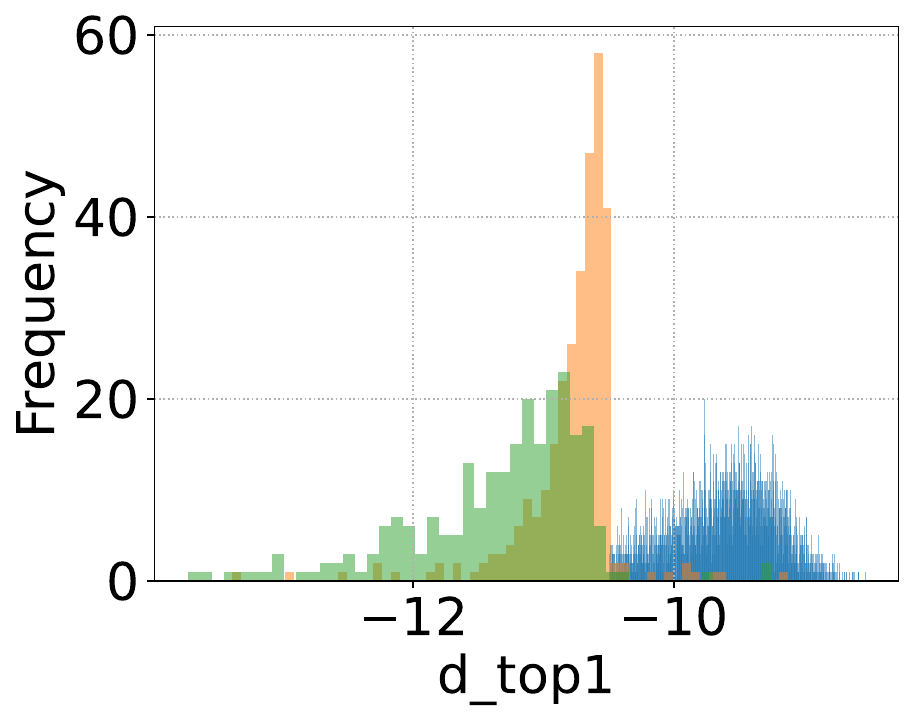}
	\end{subfigure}
	\begin{subfigure}{.24\textwidth}
		\includegraphics[scale=.24]{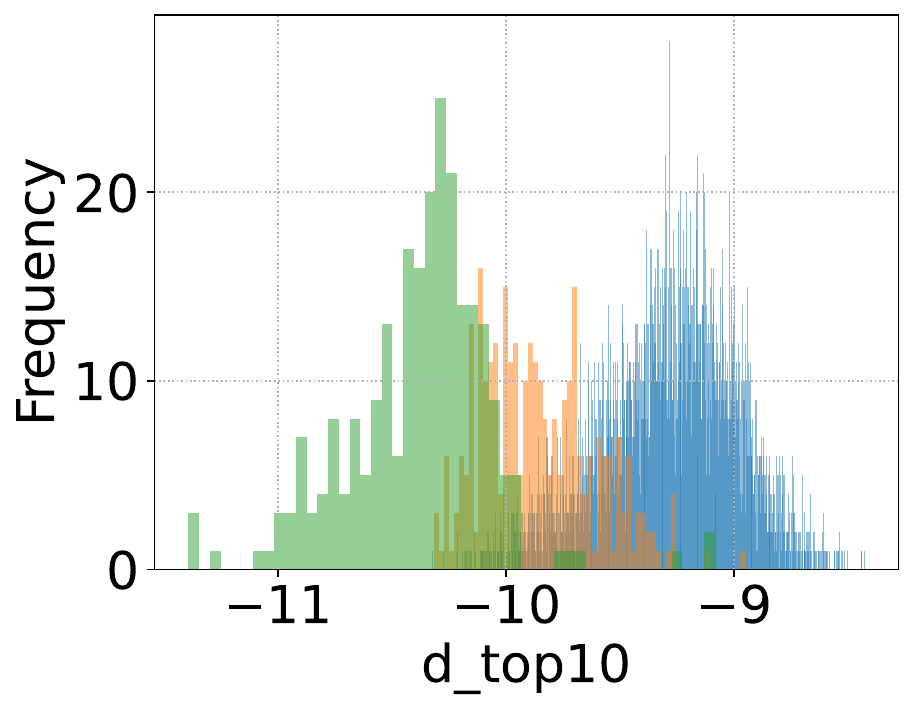}
	\end{subfigure}
	\begin{subfigure}{.24\textwidth}
		\includegraphics[scale=.24]{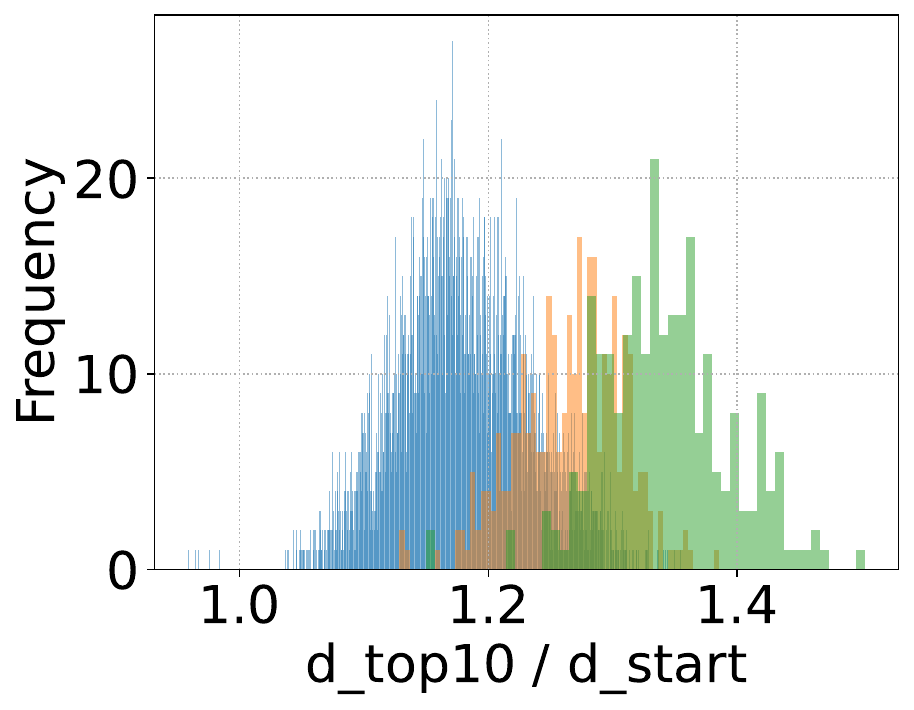}
	\end{subfigure}
	\begin{subfigure}{.24\textwidth}
		\includegraphics[scale=.24]{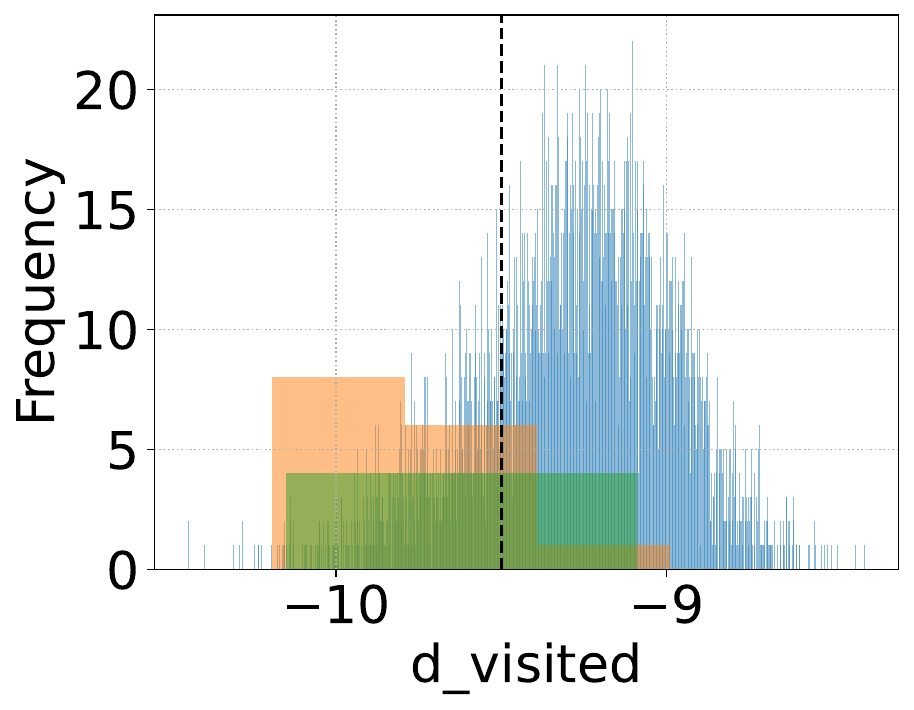}
	\end{subfigure}
	\begin{subfigure}{.24\textwidth}
		\includegraphics[scale=.24]{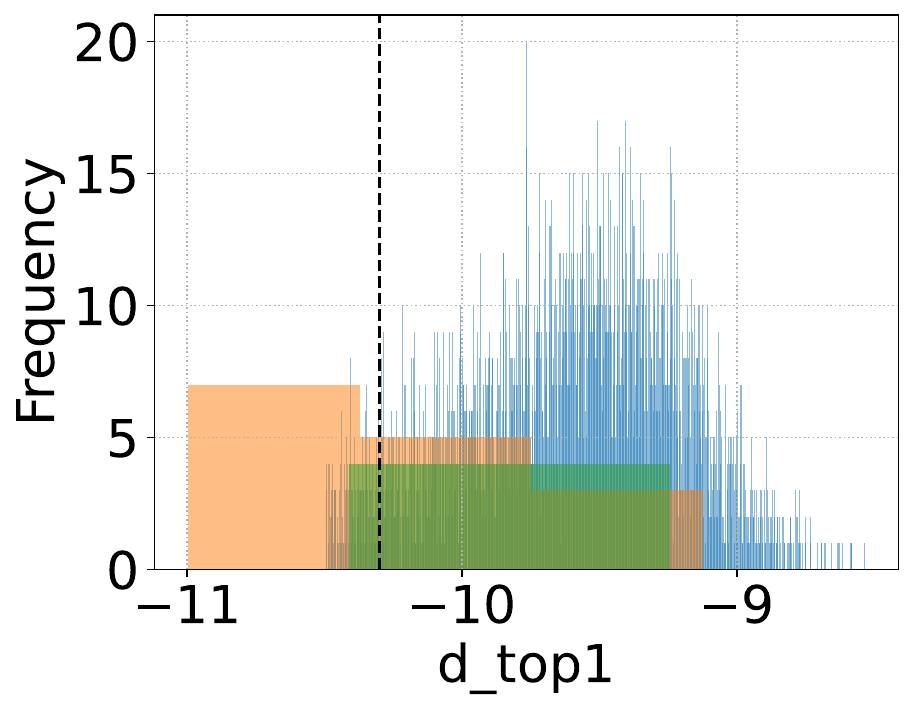}
	\end{subfigure}
	\begin{subfigure}{.24\textwidth}
		\includegraphics[scale=.24]{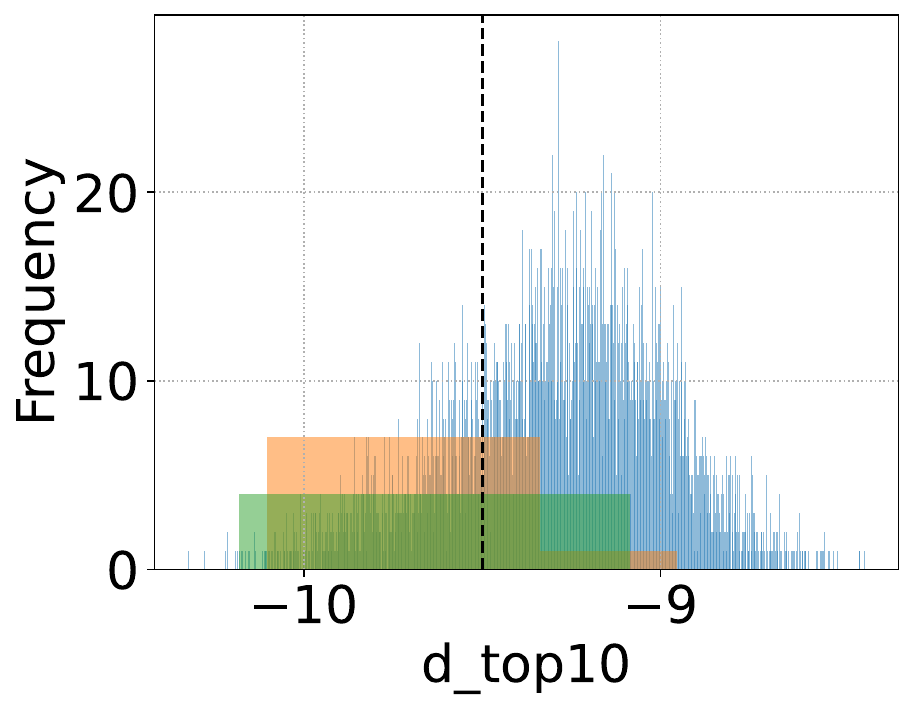}
	\end{subfigure}
	\begin{subfigure}{.24\textwidth}
		\includegraphics[scale=.24]{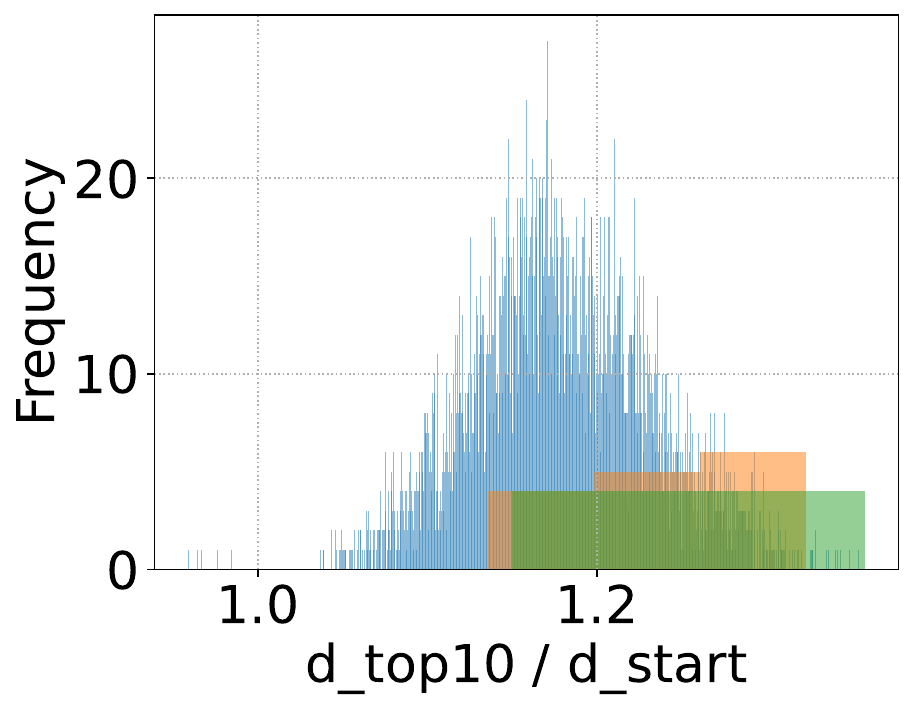}
	\end{subfigure}
	\caption{Early stopping metrics for Wikipedia-1M, taken at step 20 of a beam search with beam 100. The top row shows all results, while the bottom row shows only results for beam searches that have not yet found a candidate within the radius.}
	\label{fig:wikipedia_earlystop}
\end{figure*}

%
%
%
%
%
%

\begin{figure*}
	\centering
	\includegraphics[scale=.35]{figures/dist_histograms/visited/bigann_test_legend.pdf}  \\
	\begin{subfigure}{.24\textwidth}
		\includegraphics[scale=.24]{figures/dist_histograms/visited/Step20/ssnpp-1M.pdf}
	\end{subfigure}
	\begin{subfigure}{.24\textwidth}
		\includegraphics[scale=.24]{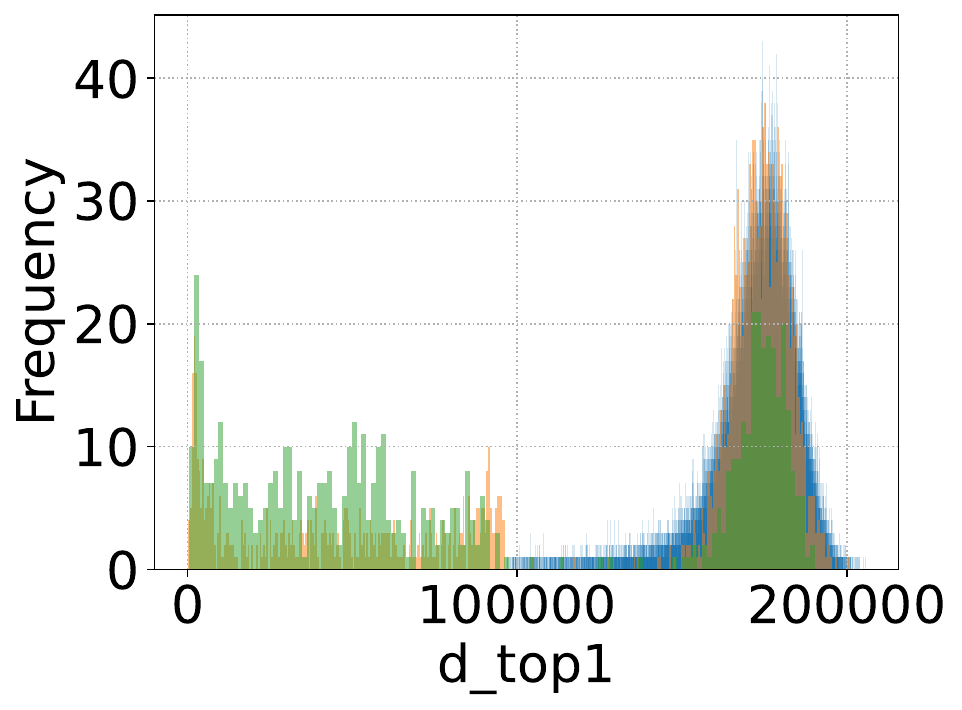}
	\end{subfigure}
	\begin{subfigure}{.24\textwidth}
		\includegraphics[scale=.24]{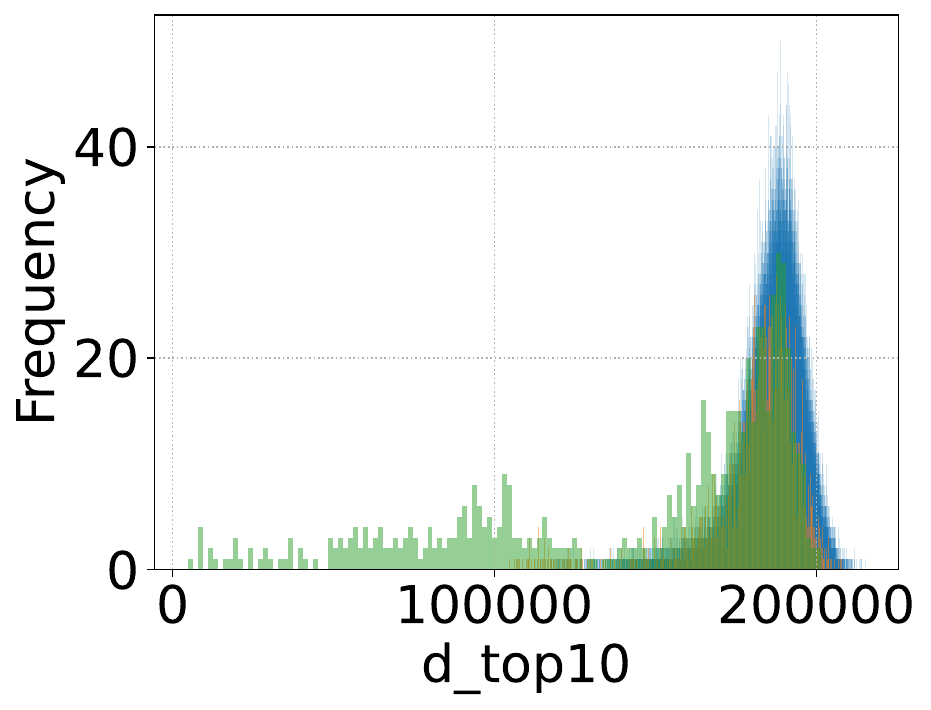}
	\end{subfigure}
	\begin{subfigure}{.24\textwidth}
		\includegraphics[scale=.24]{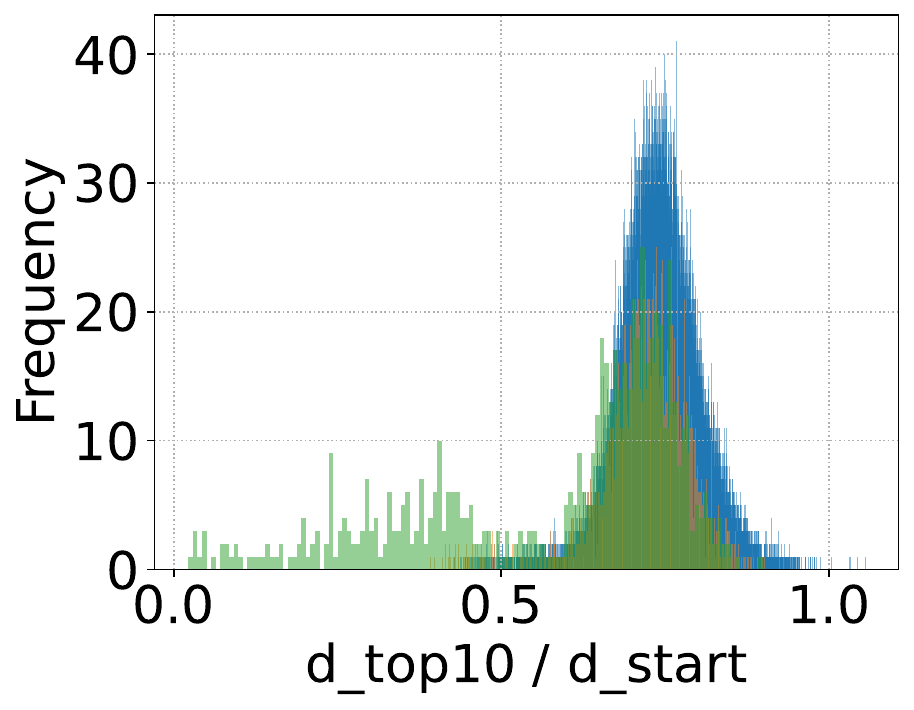}
	\end{subfigure}
	\begin{subfigure}{.24\textwidth}
		\includegraphics[scale=.24]{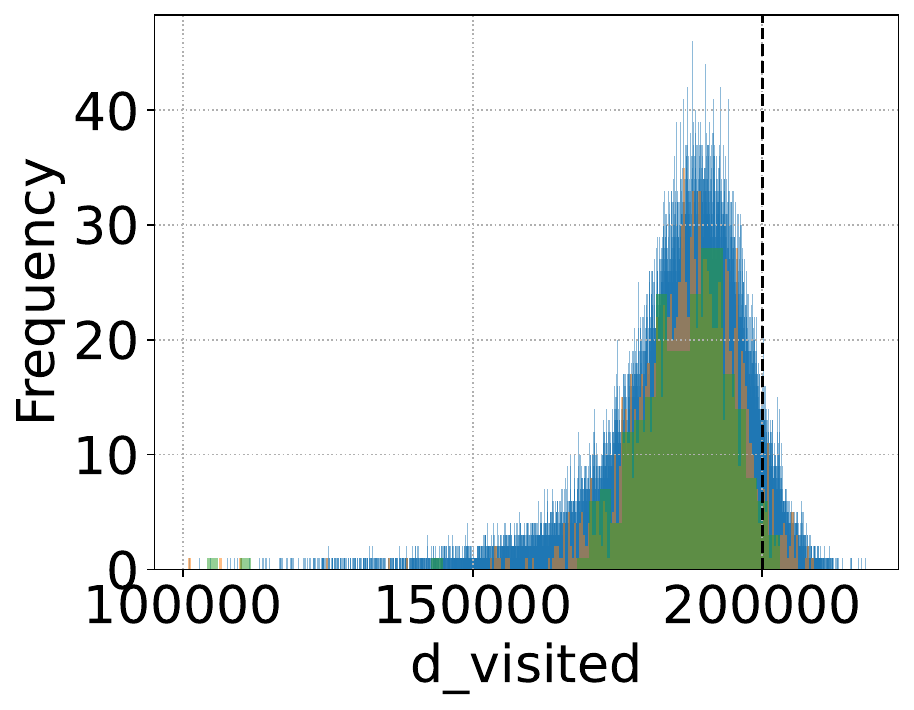}
	\end{subfigure}
	\begin{subfigure}{.24\textwidth}
		\includegraphics[scale=.24]{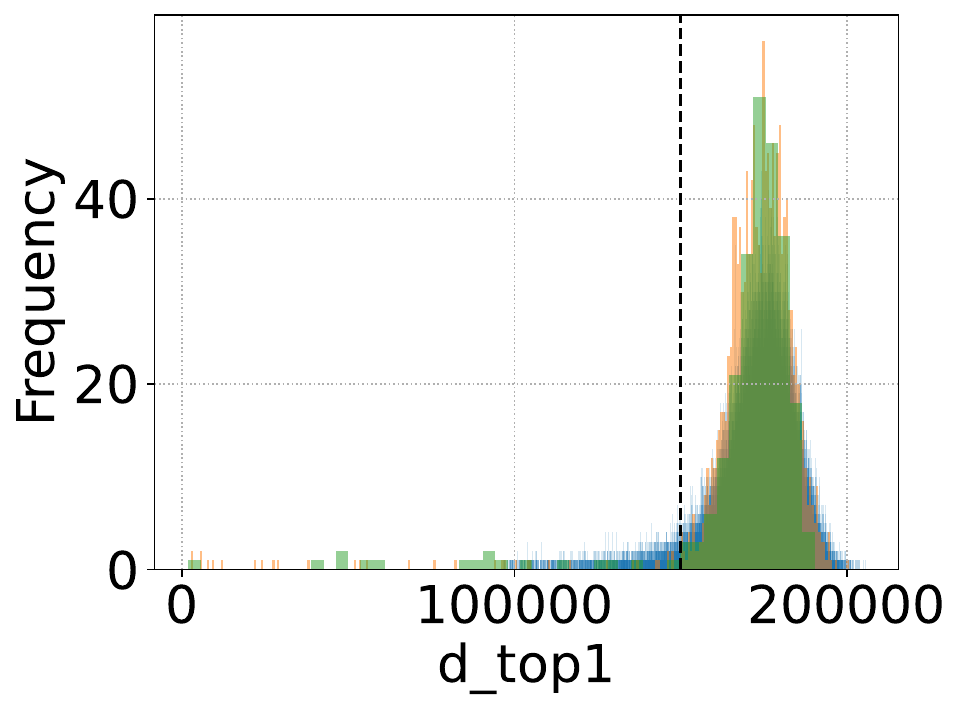}
	\end{subfigure}
	\begin{subfigure}{.24\textwidth}
		\includegraphics[scale=.24]{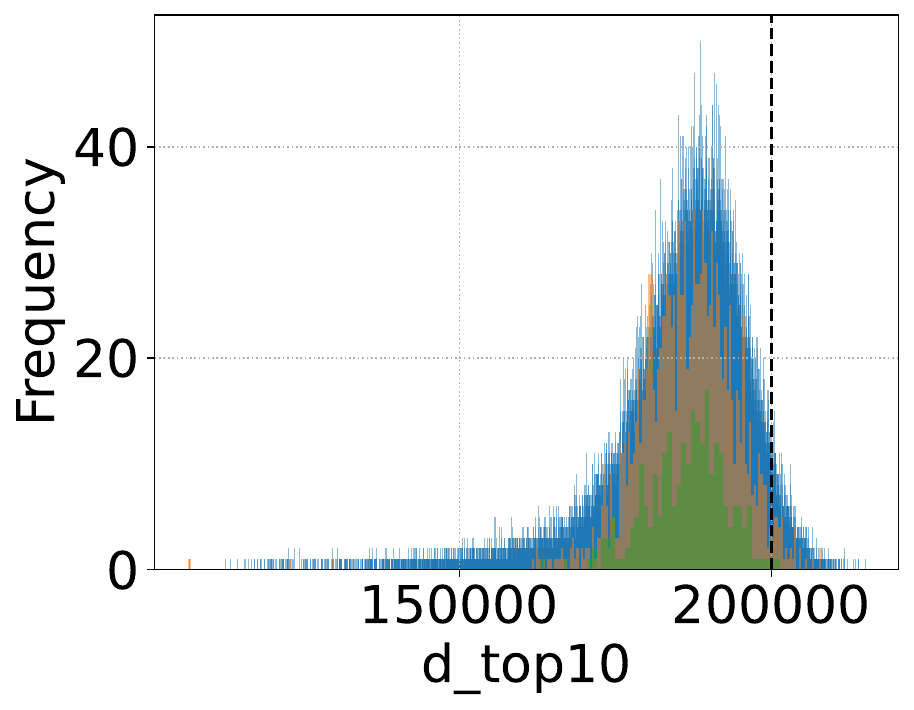}
	\end{subfigure}
	\begin{subfigure}{.24\textwidth}
		\includegraphics[scale=.24]{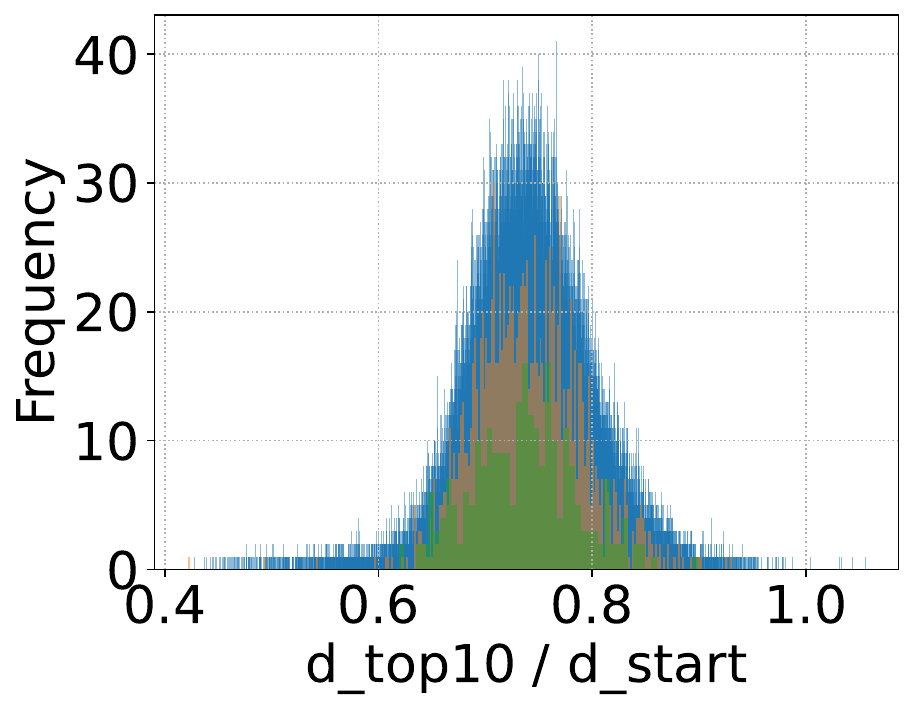}
	\end{subfigure}
	\caption{Early stopping metrics for SSNPP-1M, taken at step 20 of a beam search with beam 100. The top row shows all results, while the bottom row shows only results for beam searches that have not yet found a candidate within the radius.}
	\label{fig:ssnpp_earlystop}
\end{figure*}

In Figure~\ref{fig:ppgraphearlystoppinggreedy} and Figure~\ref{fig:ppgraphearlystoppingdoubling}, we examine the impact of early stopping on each dataset at the 1 million scale, measuring improvement over both greedy search and doubling search.

\begin{figure*}
	\centering
	\includegraphics[scale=.35]{figures/legends/nine_datasets_legend.pdf}\\
	\includegraphics[scale=.35]{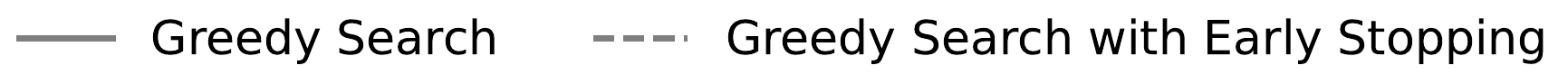} \\
	\begin{subfigure}{.3\textwidth}
		\includegraphics[scale=.3]{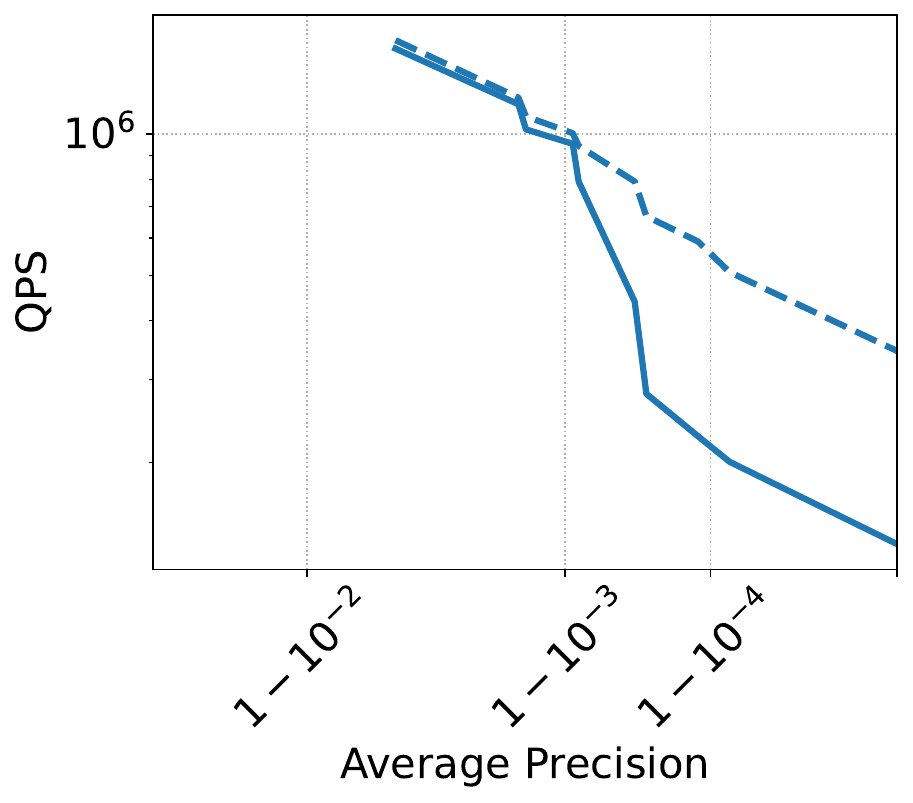}
		\caption{BIGANN-1M}\label{fig:bigannpp_greedy}
	\end{subfigure} 
	\begin{subfigure}{.3\textwidth}
		\includegraphics[scale=.3]{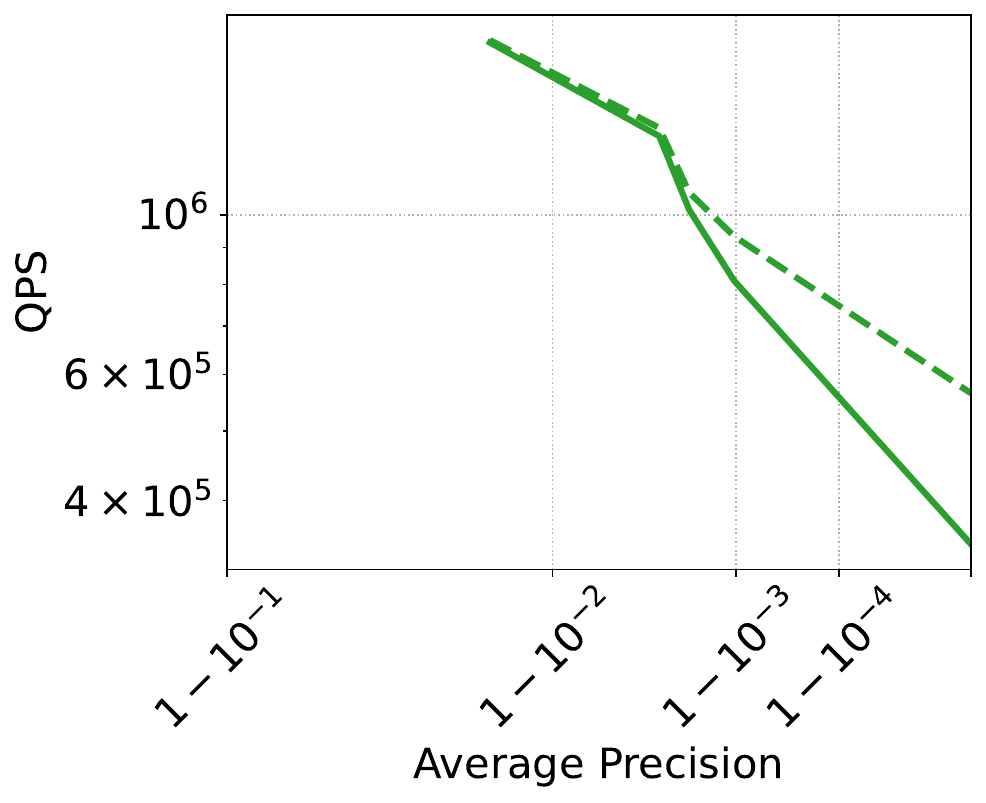}
		\caption{DEEP-1M}\label{fig:deeppp_greedy}
	\end{subfigure}
	\begin{subfigure}{.3\textwidth}
		\includegraphics[scale=.3]{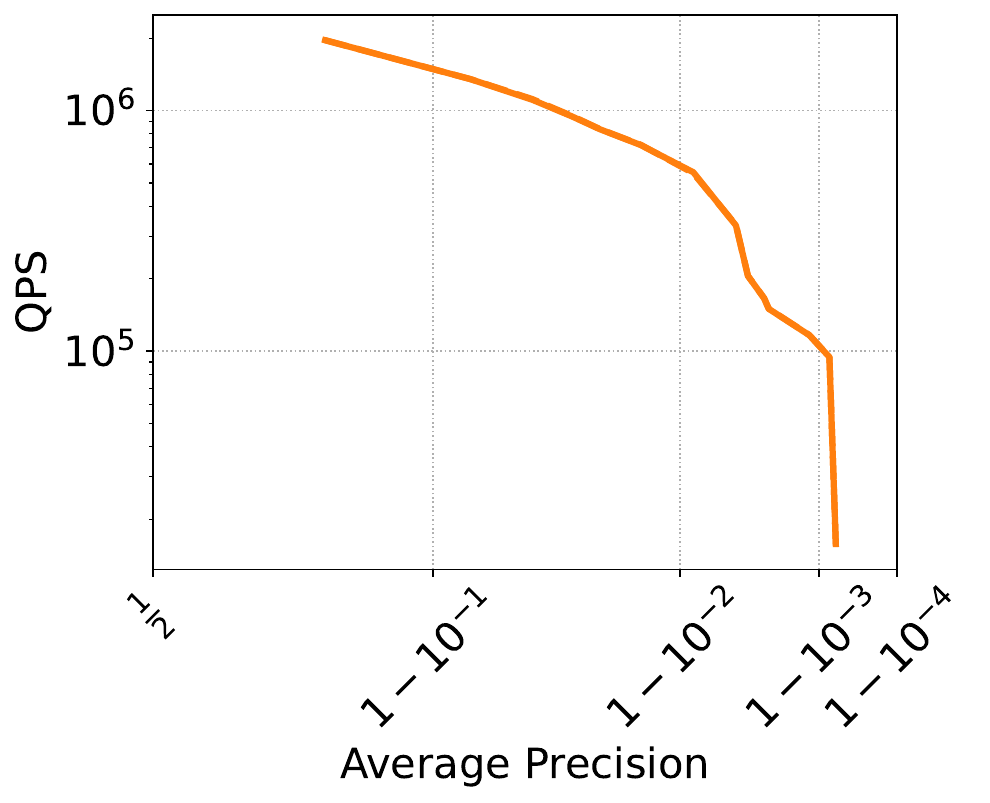}
		\caption{MSTuring-1M}\label{fig:msturingpp_greedy}
	\end{subfigure} \hfil
	\begin{subfigure}{.3\textwidth}
		\includegraphics[scale=.3]{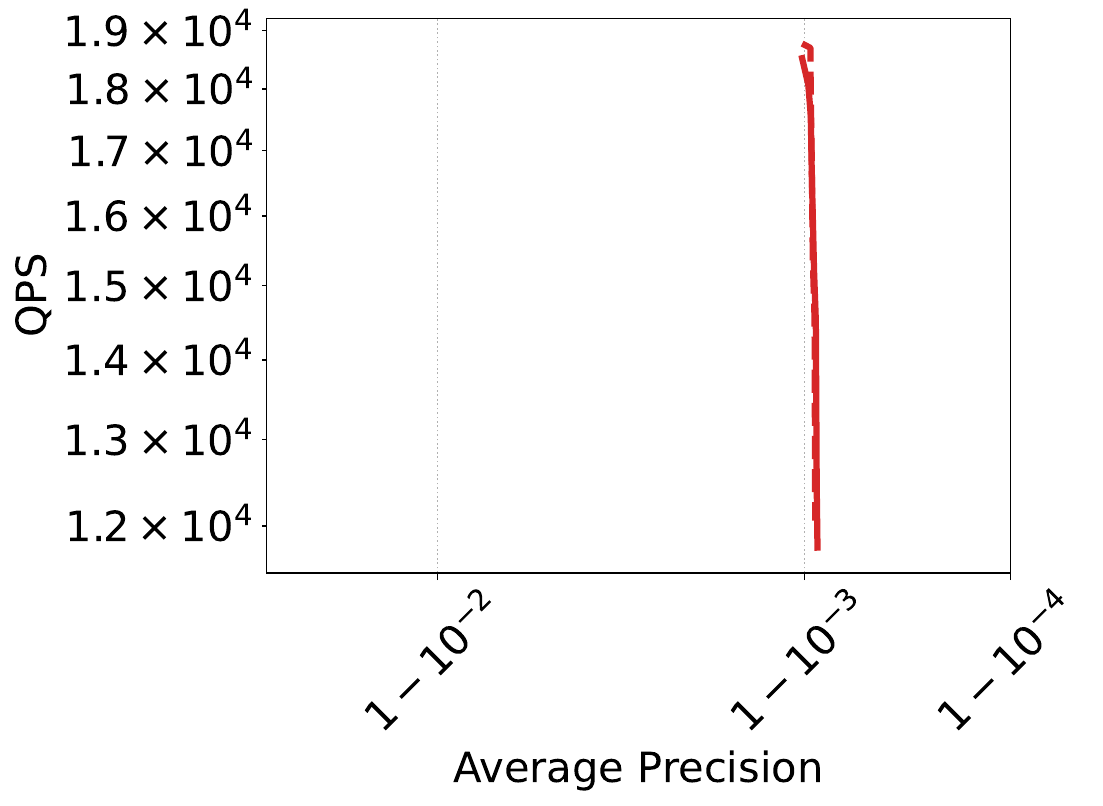}
		\caption{GIST-1M}\label{fig:gistpp_greedy}
	\end{subfigure} 
	\begin{subfigure}{.3\textwidth}
		\includegraphics[scale=.3]{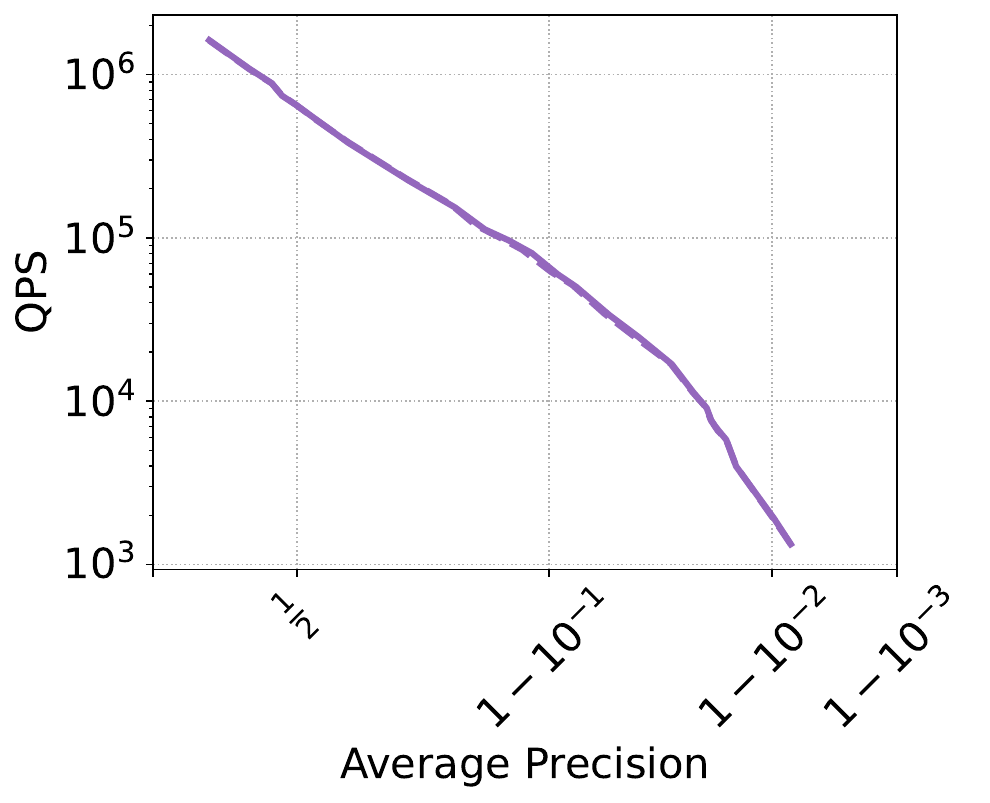}
		\caption{SSNPP-1M}\label{fig:ssnpppp_greedy}
	\end{subfigure}
	\begin{subfigure}{.3\textwidth}
		\includegraphics[scale=.3]{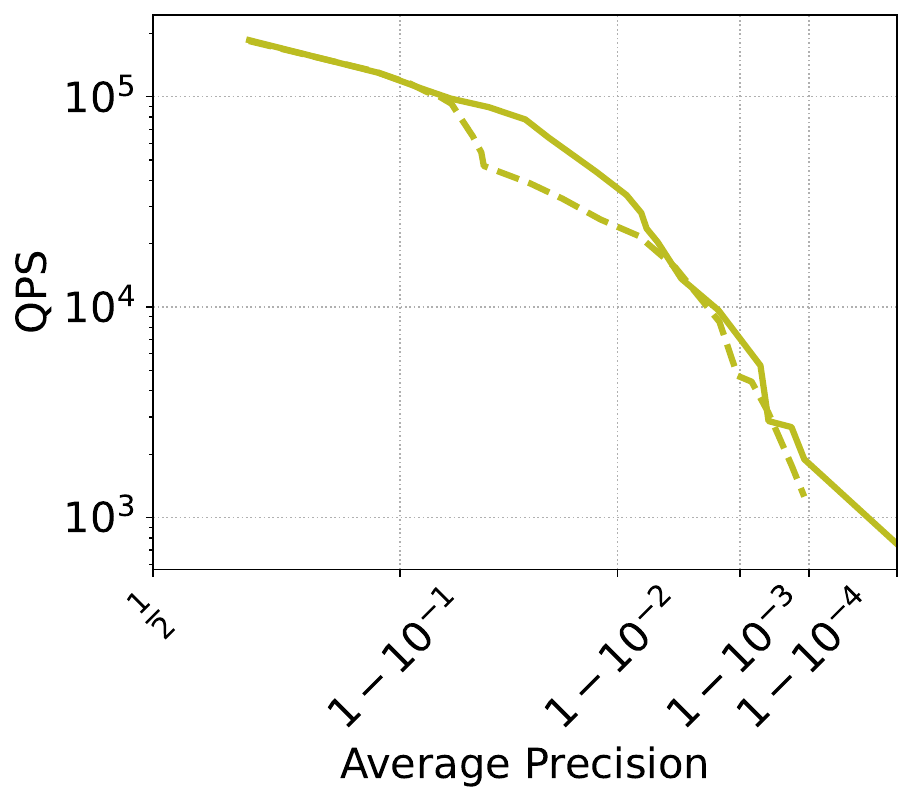}
		\caption{OpenAI-1M}\label{fig:openaipp_greedy}
	\end{subfigure}
	\begin{subfigure}{.3\textwidth}
		\includegraphics[scale=.3]{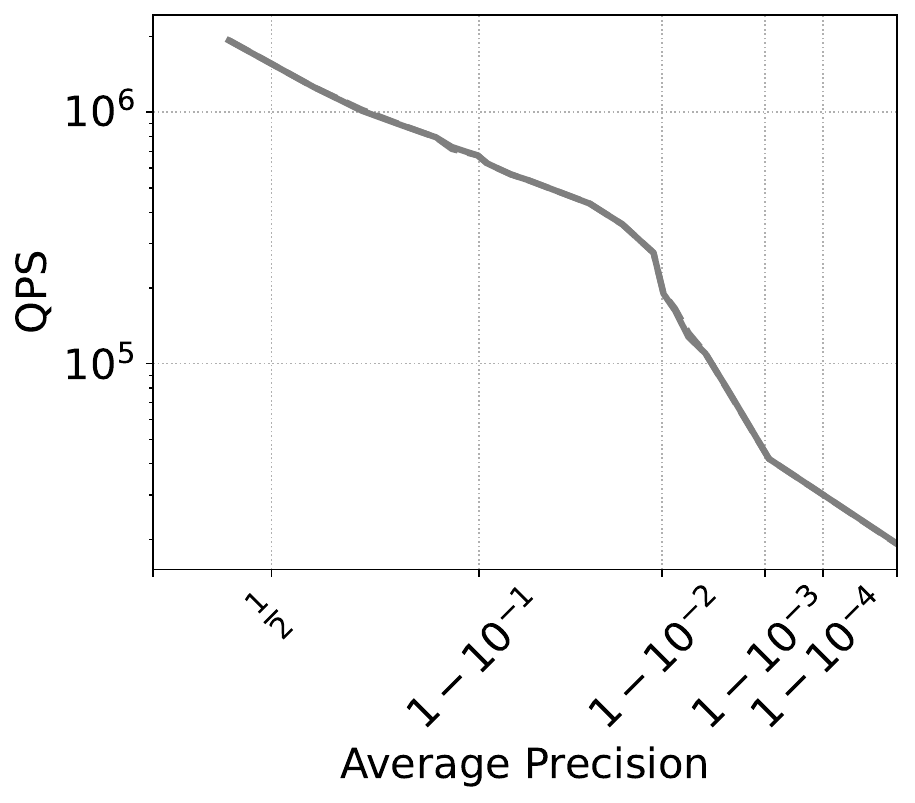}
		\caption{Text2Image-1M}\label{fig:t2ipp_greedy}
	\end{subfigure} 
	\begin{subfigure}{.3\textwidth}
		\includegraphics[scale=.3]{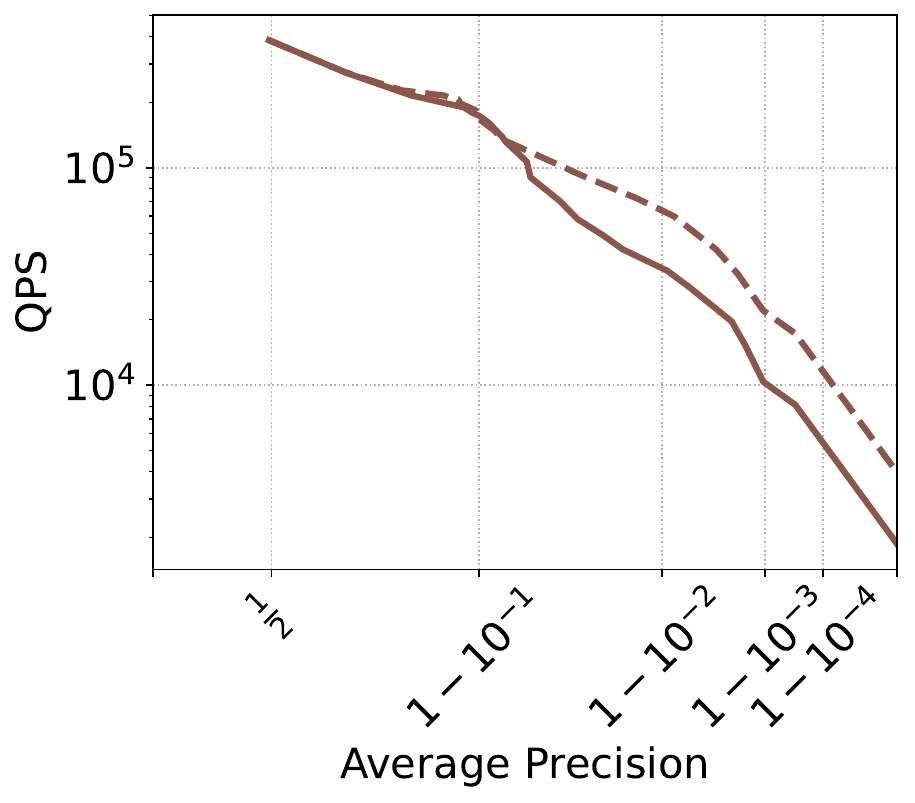}
		\caption{Wikipedia-1M}\label{fig:wikipp_greedy}
	\end{subfigure}
	\begin{subfigure}{.3\textwidth}
		\includegraphics[scale=.3]{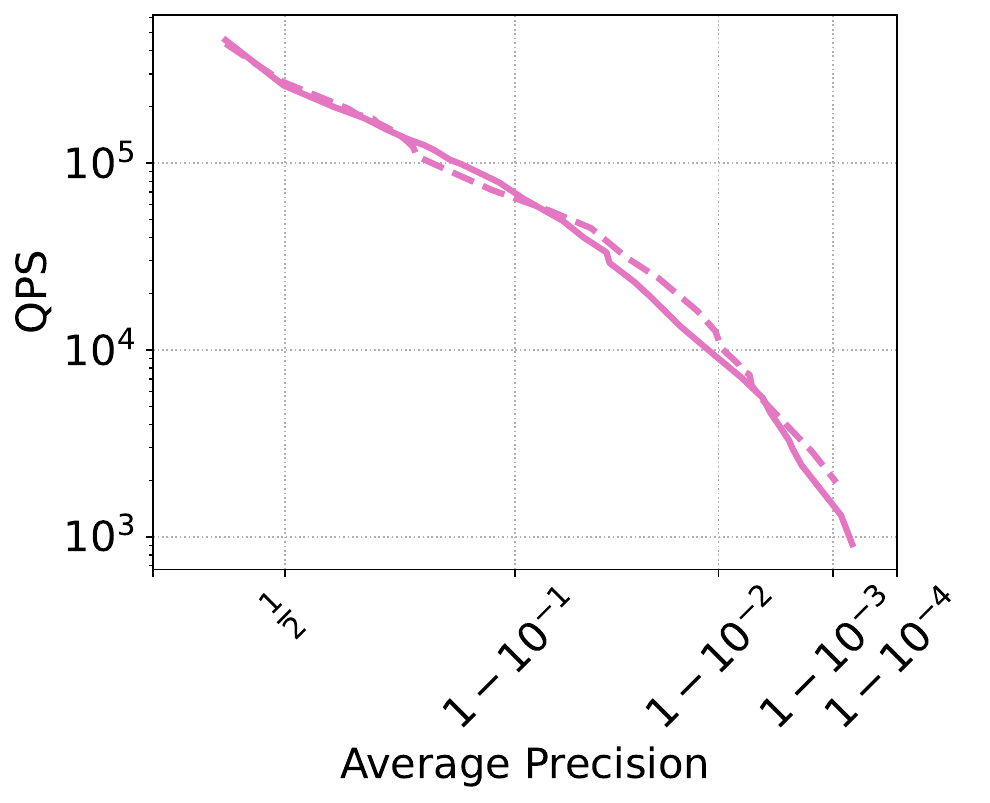}
		\caption{MSMARCOWebSearch-1M}\label{fig:msmarcopp_greedy}
	\end{subfigure}
	\caption{Average precision vs QPS for all nine datasets using greedy search, comparing use of early stopping versus without early stopping.}
	\label{fig:ppgraphearlystoppinggreedy}
\end{figure*}

\begin{figure*}
	\centering
	\includegraphics[scale=.35]{figures/legends/nine_datasets_legend.pdf}
	\includegraphics[scale=.35]{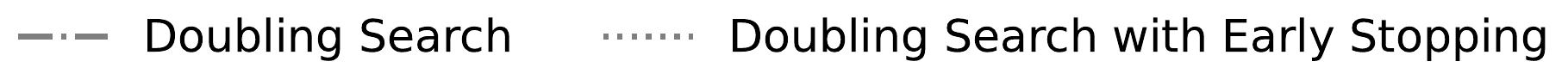} \\
	\begin{subfigure}{.3\textwidth}
		\includegraphics[scale=.3]{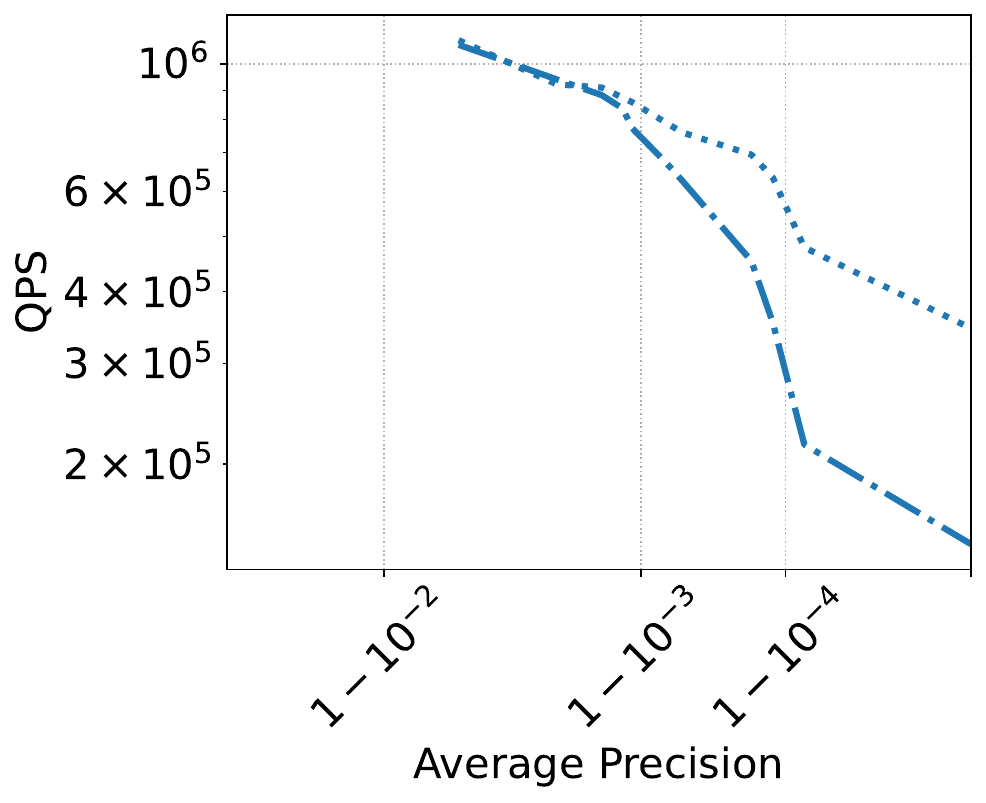}
		\caption{BIGANN-1M}\label{fig:bigannap_doubling}
	\end{subfigure} 
	\begin{subfigure}{.3\textwidth}
		\includegraphics[scale=.3]{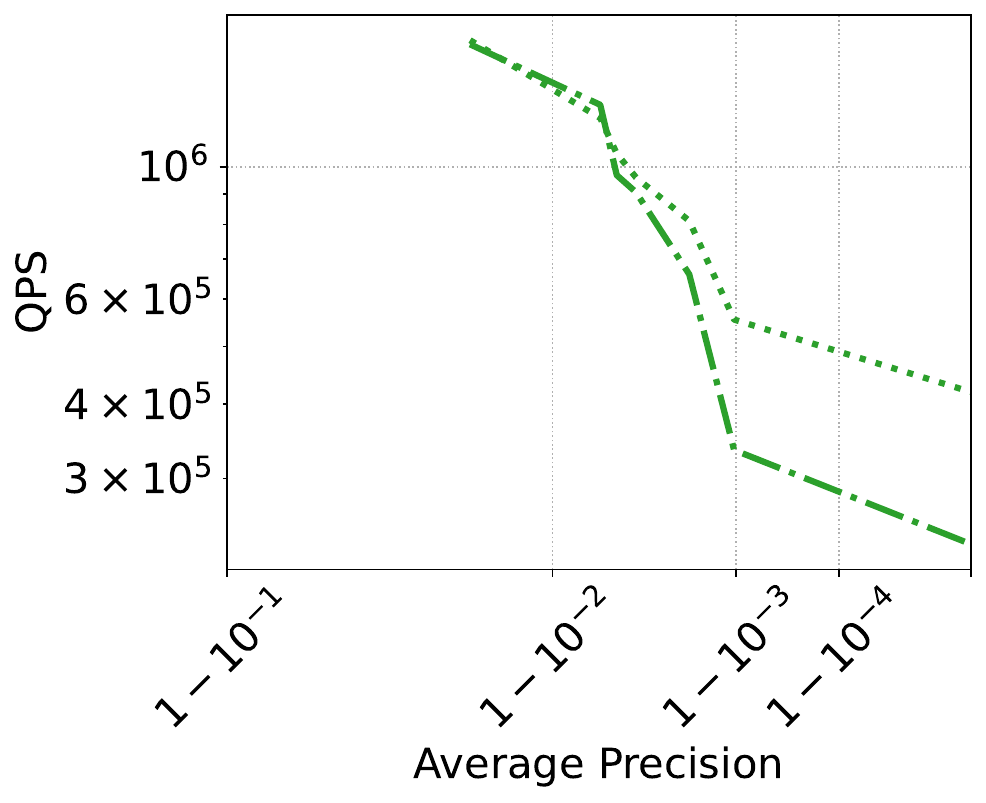}
		\caption{DEEP-1M}\label{fig:deepap_doubling}
	\end{subfigure}
	\begin{subfigure}{.3\textwidth}
		\includegraphics[scale=.3]{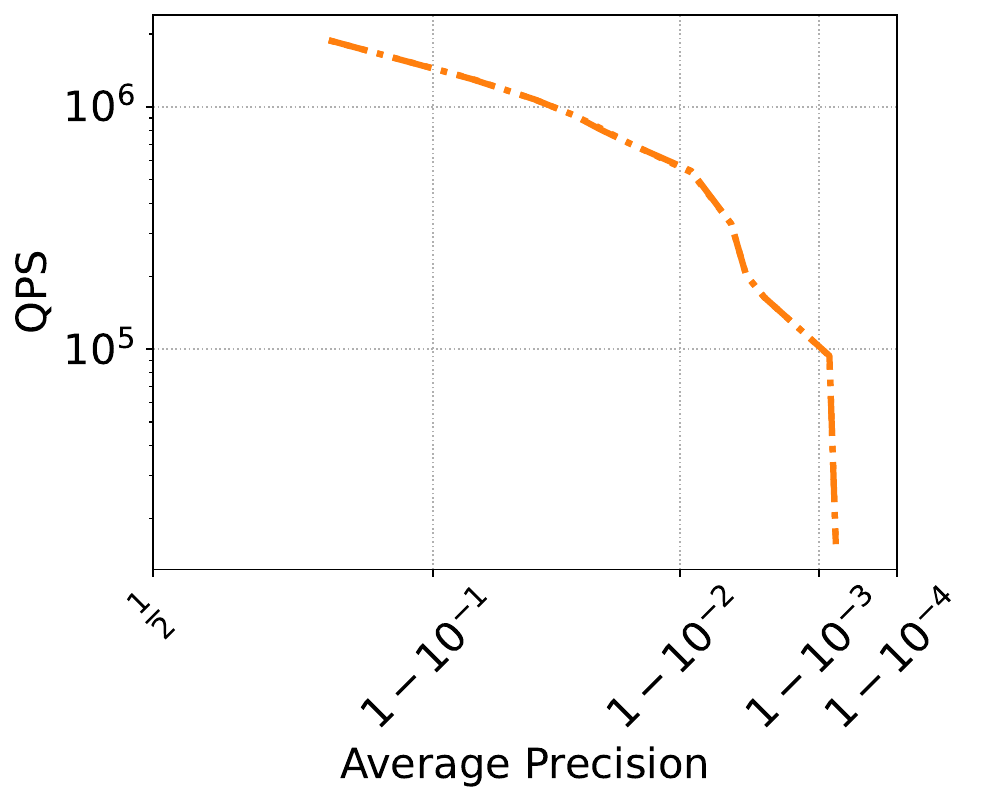}
		\caption{MSTuring-1M}\label{fig:msturingap_doubling}
	\end{subfigure}
	\begin{subfigure}{.3\textwidth}
		\includegraphics[scale=.3]{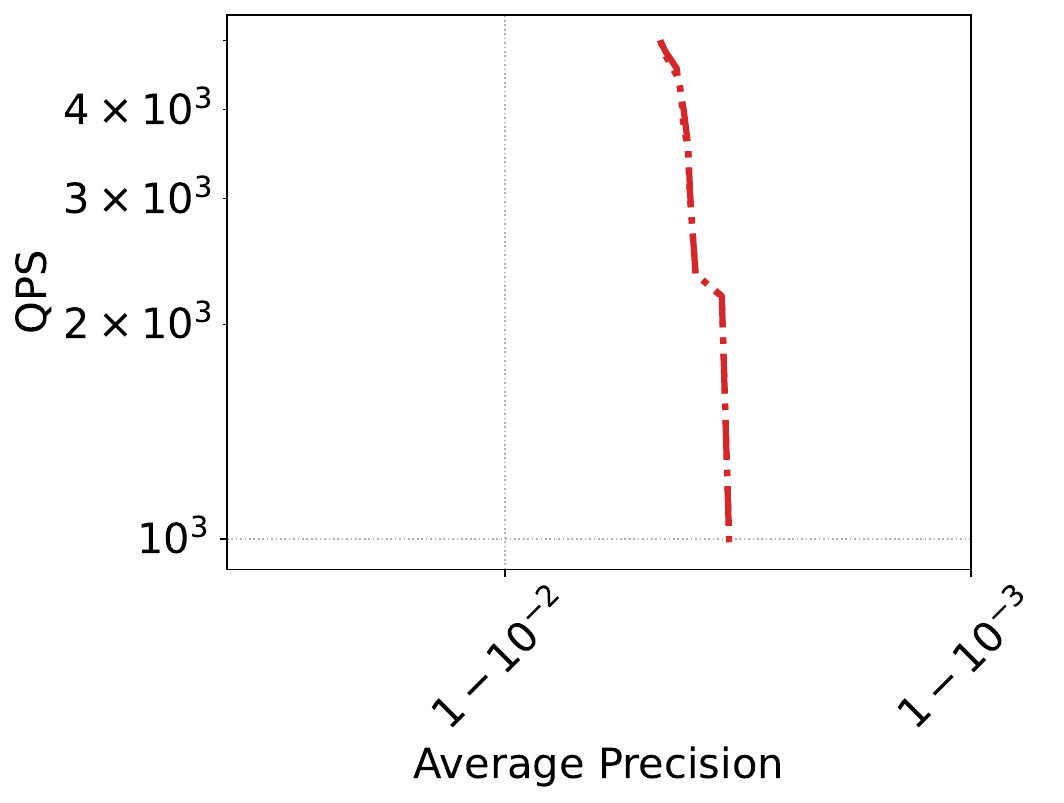}
		\caption{GIST-1M}\label{fig:gistap_doubling}
	\end{subfigure} 
	\begin{subfigure}{.3\textwidth}
		\includegraphics[scale=.3]{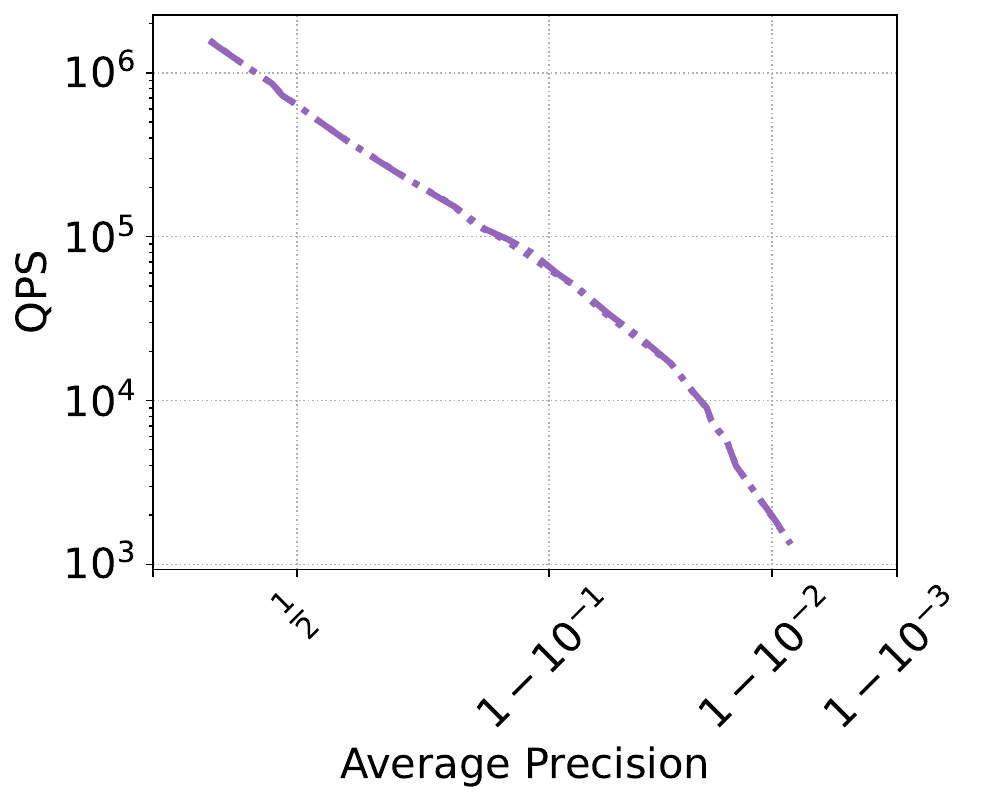}
		\caption{SSNPP-1M}\label{fig:ssnppap_doubling}
	\end{subfigure}
	\begin{subfigure}{.3\textwidth}
		\includegraphics[scale=.3]{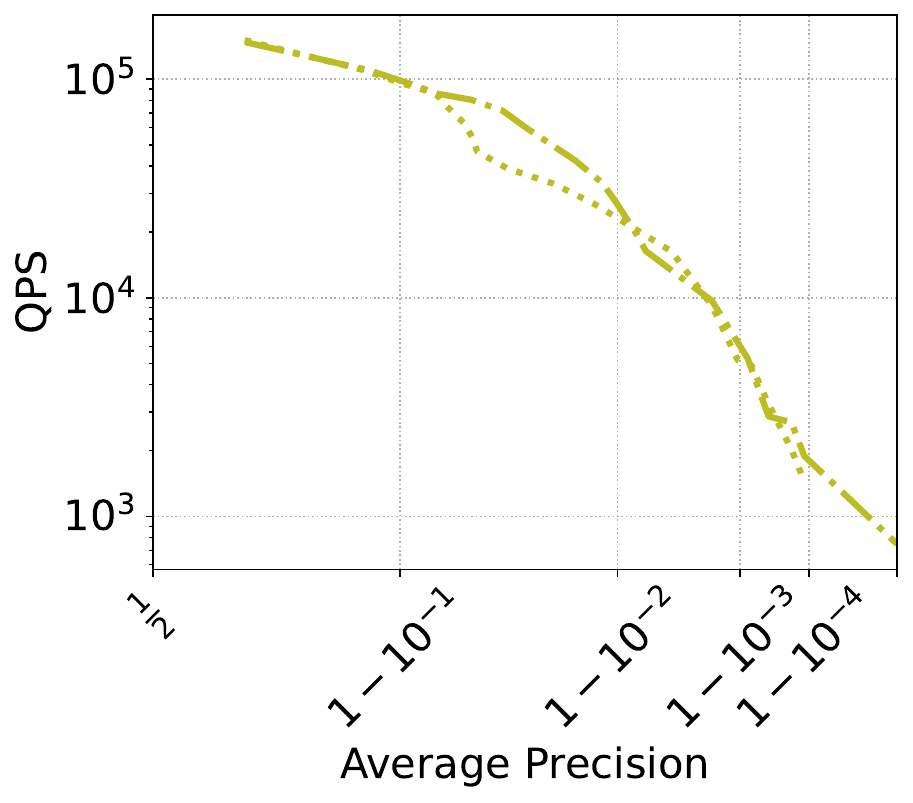}
		\caption{OpenAI-1M}\label{fig:openaiap_doubling}
	\end{subfigure}\hfil
	\begin{subfigure}{.3\textwidth}
		\includegraphics[scale=.3]{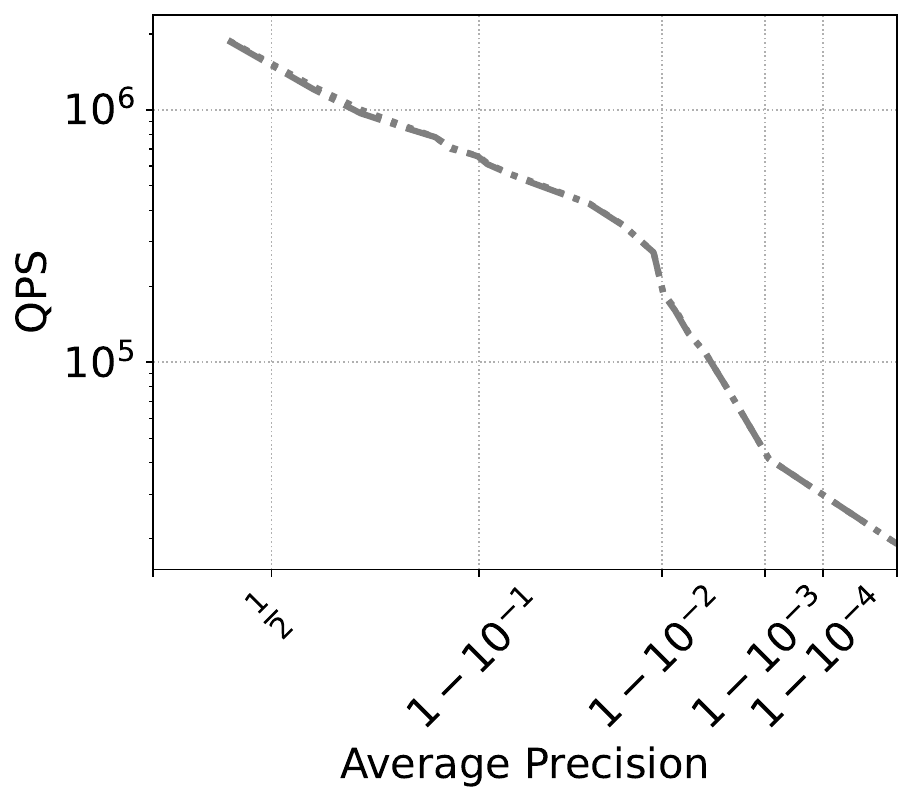}
		\caption{Text2Image-1M}\label{fig:t2iap_doubling}
	\end{subfigure} 
	\begin{subfigure}{.3\textwidth}
		\includegraphics[scale=.3]{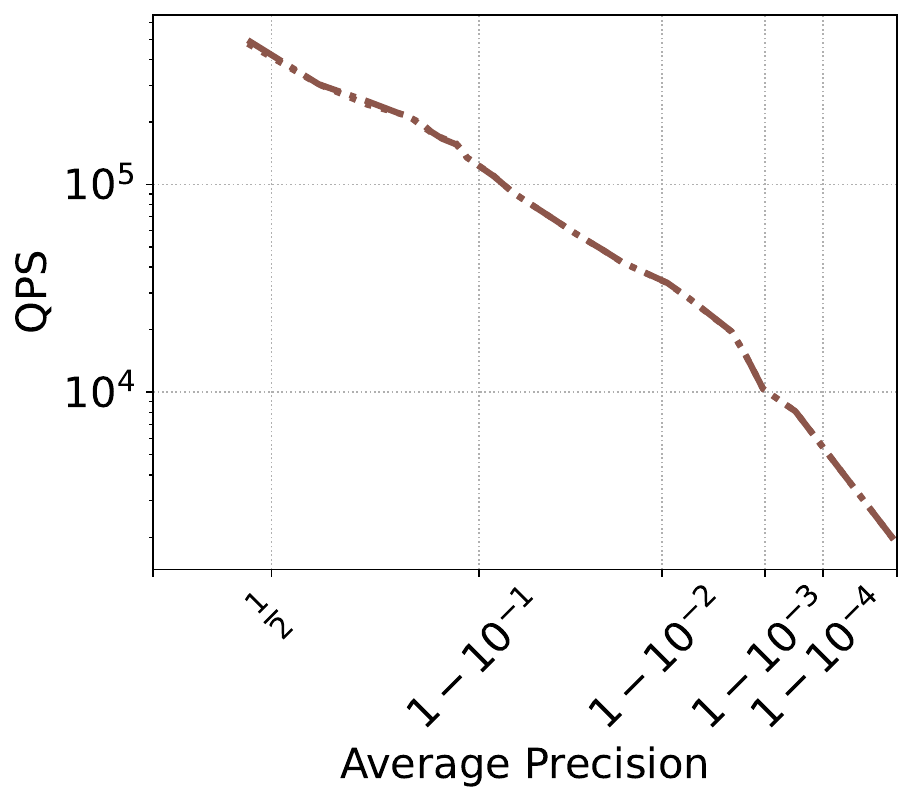}
		\caption{Wikipedia-1M}\label{fig:wikiap_doubling}
	\end{subfigure}
	\begin{subfigure}{.3\textwidth}
		\includegraphics[scale=.3]{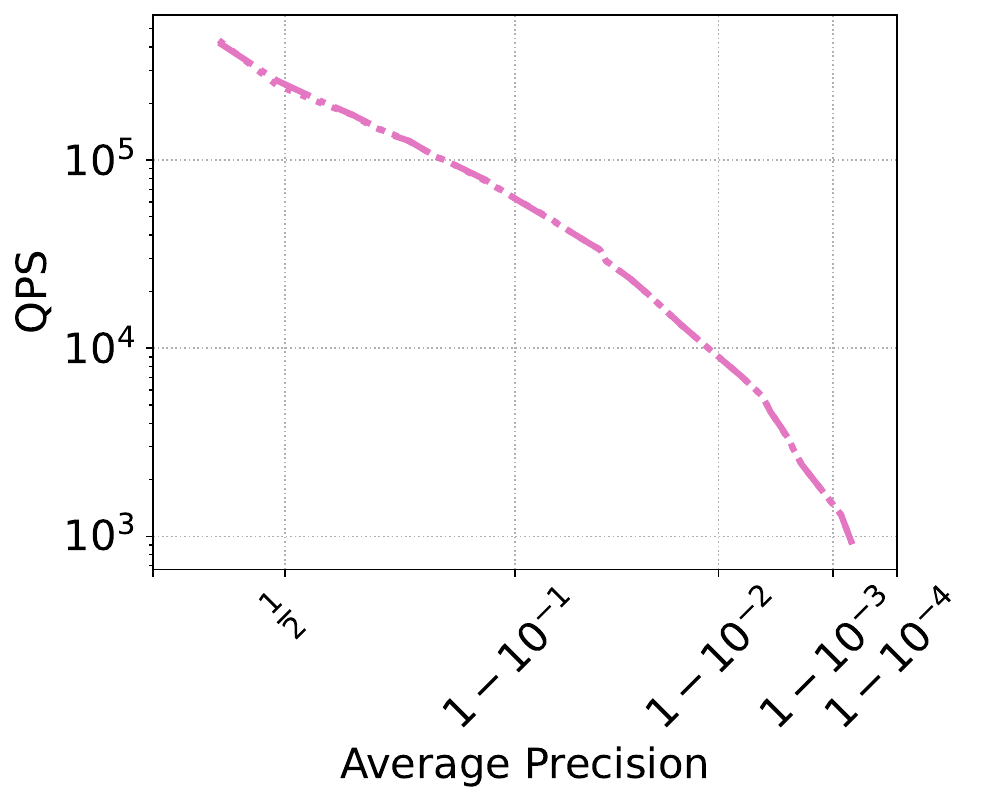}
		\caption{MSMARCOWebSearch-1M}\label{fig:msmarcoap_doubling}
	\end{subfigure}
	\caption{Average precision vs QPS for all nine datasets using doubling search, comparing use of early stopping versus without early stopping.}
	\label{fig:ppgraphearlystoppingdoubling}
\end{figure*}

\subsection{Parameter Values}

\begin{figure*}
	\begin{tabular}{|c|c|}
		\hline
		Dataset & Parameters \\
		\hline
		BIGANN-1M & "OPQ32\_64,IVF65536\_HNSW32,PQ64x4fsr" \\
		\hline
		BIGANN-10M & "OPQ32\_64,IVF1048576\_HNSW32,PQ64x4fsr" \\
		\hline
		BIGANN-100M & "OPQ64\_128,IVF1048576\_HNSW32,PQ128x4fsr" \\
		\hline
		DEEP-1M & "OPQ32\_64,IVF65536\_HNSW32,PQ64x4fsr"\\
		\hline
		MSTuring-1M & "OPQ32\_64,IVF16384\_HNSW32,PQ64x4fsr"\\
		\hline
		GIST-1M & "OPQ32\_64,IVF16384\_HNSW32,PQ64x4fsr" \\
		\hline
	 	SSNPP-1M & "OPQ32\_64,IVF65536\_HNSW32,PQ64x4fsr"\\
	 	\hline
		SSNPP-10M & "OPQ64\_128,IVF65536\_HNSW32,PQ128x4fsr" \\
		\hline
		SSNPP-100M & "OPQ32\_128,IVF1048576\_HNSW32,PQ32" \\
		\hline
		OpenAI-1M & "OPQ32\_64,IVF16384\_HNSW32,PQ64x4fsr" \\
		\hline
		Text2Image-1M & "OPQ64\_128,IVF16384\_HNSW32,PQ128x4fsr" \\
		\hline
		Wikipedia-1M & "IVF16384\_HNSW32,Flat"\\
		\hline
		MSMARCOWebSearch-1M & "IVF16384\_HNSW32,Flat" \\
		\hline
	\end{tabular}
	\caption{Choices of FAISS parameters for each dataset.}
	\label{fig:FAISS_params}
\end{figure*}

In Figure~\ref{fig:FAISS_params}, we show our choice of FAISS parameters for each dataset. For each dataset, we chose a starting set of parameters based on both instructions from the FAISS Wiki~\cite{FAISS_Wiki} and the parameter settings from the NeurIPS'21 Big ANN Benchmarks Competition~\cite{bigann}. In each case, we use two-level clustering to significantly speed up build times. We then experimented with increasing and decreasing the choice of both the size of the PQ codes and the number of centroids, and kept the parameters that gave the optimal QPS/Average Precision curve. For SSNPP-100M, we also made use of the Hamming threshold parameter as suggested from the configurations in the Big ANN Benchmarks repository. The curves were drawn by varying the HNSW \textit{ef\_search} parameter as well as the number of centroids probed. FAISS has a build-in range search method~\cite{FAISS_Wiki}, which we used to conduct searches. For Wikipedia and MSMARCOWebSearch, we used no quantization during range search as the quantization seemed to distort distances for inner product datasets and produced nonsensical results.

\begin{figure*}
	\begin{tabular}{|c|c|c|}
		\hline
		Dataset & Build Parameters & Early Stopping Threshold \\
		\hline
		BIGANN (all sizes) & $R=64,L=128,\alpha=1.15$ & 10000 \\
		\hline
		DEEP-1M & $R=64,L=128,\alpha=1.15$ & .4 \\
		\hline
		MSTuring-1M & $R=64,L=128,\alpha=1.15$ & 1.6 \\
		\hline 
		GIST-1M & $R=64,L=128,\alpha=1.15$ & 2.5 \\
		\hline
		SSNPP (all sizes) & $R=80,L=200,\alpha=1.1$ & 200000 \\
		\hline
		OpenAI-1M & $R=64,L=128,\alpha=1.15$ & .28 \\
		\hline
		Text2Image-1M & $R=64,L=128,\alpha=1.0$ & -1.0 \\
		\hline
		Wikipedia-1M & $R=64,L=128,\alpha=1.0$ & -9.5 \\
		\hline
		MSMARCOWebSearch-1M & $R=64,L=128,\alpha=1.0$ & -57 \\
		\hline
	\end{tabular}
	\caption{Choices of DiskANN parameters for each dataset.}
	\label{fig:diskann_params}
\end{figure*}

In Figure~\ref{fig:diskann_params} we show the choices of build and search parameters for the DiskANN graphs used in our experiments. For the build parameters, we used $R=64,L=128$ wherever possible, although we found it was not sufficient for SSNPP. The $\alpha$ parameter was configured experimentally for each dataset, as was the choice of early stopping threshold.

\fi

\end{document}